\documentclass[12pt]{article}
\usepackage{amsfonts,amsmath}
 \usepackage{amssymb}

\usepackage[hypertex] {hyperref}

\makeatletter \@addtoreset{equation}{section} \makeatother

\def\theequation{\thesection.\arabic{equation}}

\newcommand{\hel}{\mathrm{h},s\,}
\newcommand{\hell}{1,s\,}
\newcommand{\helh}{ \mathrm{{h}}}
  
\newcommand{\dis}{\displaystyle}

\newcommand{\Fursol}{\mathcal{P}}

\newcommand{\ra}{{\mathrm{a}^\prime}}

\newcommand{\la}{\mathrm{{a }}}

\newcommand{\hv}{{f_0}}

\newcommand{\co}{{\rm{  c }}}
\newcommand{\si}{{\rm{  s }}}

\newcommand{\be}{ \begin{equation}}

\newcommand{\etc}{{\it etc}}
\newcommand{\slv}{{}^v{}\mathfrak{sl}_2}
\newcommand{\slh}{{}^h{}\mathfrak{sl}_2}

\newcommand{\bp}{\bar{\partial}}
\newcommand{\has}{l}

\newcommand{\goo}{{w}}
\newcommand{\bm}{\bar{m}}

\newcommand{\NNN}{{\mathcal N}}

\newcommand{\R}{{\mathcal R}}

\newcommand{\KKK}{ {\mathcal{J}}}
\newcommand{\LLLR}{ {\mathcal{J}}}
\newcommand{\LLL}{\widetilde{\mathcal{J}}}%
\newcommand{\LLLL}{ { {\mathcal{J}}_{0,s}}}%
\newcommand{\RRR}{\mathcal{I}}%
\newcommand{\Gg}{\overline{G}}%

\newcommand{\ee}{\end{equation}}
\newcommand{\bn}{\overline{{n}}}%
\newcommand{\uu}{ {{g_0}}}%
\newcommand{\vv}{ {{g}}}

\newcommand{\TTT}{ {I}}
\newcommand{\PPP}{ {J} }

\newcommand{\bee}{\begin{eqnarray}}
\newcommand{\beee}{\begin{array}}
\newcommand{\eee}{\end{eqnarray}}
\newcommand{\eeee}{\end{array}}

\newcommand{\gn}{\nu}

\newcommand{\gm}{\mu}

\newcommand{\gx}{\xi}

\newcommand{\px}{\bar{\xi}}
\newcommand{\gr}{\rho}
\newcommand{\ga}{\alpha}
\newcommand{\pa}{{\ga^\prime}}

\newcommand{\pn}{{\nu^\prime}}
\newcommand{\pmm}{{\mu^\prime}}

\newcommand{\pb}{{\gb^\prime}}
\newcommand{\pga}{{\gamma^\prime}}
\newcommand{\gb}{\beta}

\newcommand{\gga}{\gamma}

\newcommand{\gla}{\lambda}

\newcommand{\M}{{\cal M}}

\newcommand{\FFF}{{ \mathfrak{ F}}}

\newcommand{\FF}{{\Phi}}

\newcommand{\Hh}{{\cal H}}

\newcommand{\rhs}{{\it r.h.s.} }

\newcommand{\ie}{{\it i.e.,} }
\newcommand{\ls}{\!\!\!\!\!\!}

\newcommand{\gd}{\delta}

\newcommand{\gl}{\lambda}

\newcommand{\gep}{\epsilon}
\newcommand{\gvep}{\varepsilon}

\newcommand{\gs}{\sigma}

\newcommand{\go}{\omega}

\newcommand{\by}{{\bar{y}}}

\newcommand{\bxi}{{\bar{\xi}}}

\newcommand{\q}{\,,\qquad}

\newcommand{\mR}{{\mathbb R}}

\newcommand{\bY}{{\overline{Y}}}

\newcommand{\nn}{\nonumber}

\newcommand{\half}{\frac{1}{2}}

\newcommand{\p}{\partial}
\newcommand{\D}{{\cal D}}
\newcommand{\f}{\frac}

\newcommand{\bu}{{\bar{u} }}

\begin{document}

\vskip1.5cm

 \begin{center}
 {\large\bf Unfolded Equations for Current Interactions of $4d$ Massless Fields as a Free System
  in Mixed Dimensions}
 \vglue 0.6  true cm

\vskip0.5cm

 O.A. Gelfond$^1$ and M.A.~Vasiliev$^2$
 \vglue 0.3  true cm

 ${}^1$Institute of System Research of Russian Academy of Sciences,\\
 Nakhimovsky prospect 36-1,
 117218,
 Moscow, Russia

 \vglue 0.3  true cm

 ${}^2$I.E.Tamm Department of Theoretical Physics, Lebedev Physical
 Institute,\\
 Leninsky prospect 53, 119991, Moscow, Russia\\ and\\
 Theory Group, Physics Department, CERN, CH-1211, Geneva 23, Switzerland

 \end{center}

\vskip2cm

 \begin{abstract}
Interactions of massless fields of all spins in four dimensions with
currents of any spin is shown to result from a solution of the
linear problem that describes a gluing between rank-one (massless)
system  and rank-two (current) system in the unfolded dynamics
approach. Since the rank-two system is dual to a free rank-one
higher-dimensional system, that effectively describes conformal
fields in six space-time dimensions, the constructed system can be
interpreted as describing a mixture between linear conformal fields
in four and six dimensions. Interpretation of the obtained results
in spirit of $AdS/CFT$ correspondence is discussed.
  \end{abstract}
\newpage

\tableofcontents
 \newpage

\section{Introduction}

Valery Rubakov has remarkably broad area of scientific interests, ranging
from the theory of fundamental interactions to cosmology. For the volume in honor
of Valery's 60th birthday we contribute a paper which gives
hints on a possible unification of such seemingly different concepts of QFT as conserved
currents in lower dimension and free fields in higher dimension.
Although these days such identification
sounds natural in the context of $AdS/CFT$ correspondence
\cite{Maldacena:1997re,Gubser:1998bc,Witten:1998qj}, the particular realization
suggested in this paper goes beyond the standard setup allowing to interpret
current interactions of $4d$ fields of all spins, including usual fields of spins
$0\leq s\leq 2$, in terms of a linear system mixing free conformal fields in four and
six dimensions. In fact, a part of this work has been
presented some time ago at the seminar headed by Rubakov after which we had a
stimulating discussion with Valery on whether or not it is possible to make fields in
space-times of different dimensions directly interacting in relativistic field theory.
Since then we got more evidences, including those presented in this paper, that
this is not only possible but also can  eventually drive us to a better understanding of
fundamental concepts of QFT including the very concept of space-time. Hence we believe
that this paper is appropriate for the volume in honor of Valery Rubakov.

{

Specifically, we consider  field equations for massless fields of all spins
in four dimensional anti-de Sitter space in the lowest order in interactions
accounting for the contribution of conserved currents built from
bilinears of  the same set of massless fields. The problem is
analyzed in the framework of the covariant first-order unfolded
formulation underlying the known formulation of nonlinear massless field
equations \cite{more,non} (see also \cite{solv} for more details and references).
 Our goal  is to clarify the structure of current interactions in the nonlinear
higher-spin (HS) theory which describes interactions of
massless fields of all spins in four dimensions.

Technically, our approach is based on the correspondence between fields and
currents elaborated in \cite{tens2}, where $Sp(2M)$-invariant field equations
corresponding to rank-$r$ tensor products of the Fock (singleton)
representation of $Sp(2M)$ were studied. These equations were shown to
describe localization on  ``branes'' of different dimensions embedded into
the generalized space-time $\M_M$ with matrix coordinates $X^{AB}=X^{BA}$ with $A,B=1,\ldots,M$
\cite{F,{Bandos:1999qf},BHS, {Vasiliev:2001dc}}. For $M=4$, the indices
$A,B=1,\ldots,4$ can be interpreted as Majorana spinor indices of the
four-dimensional Minkowski space while the space $\M_4$ is ten dimensional.
Minkowski space is a subspace of $\M_4$ with local coordinates\footnote{(Un)primed
indices from the beginning of the
Greek alphabet take two values $\ga,\gb = 1,2$ and $\pa,\pb =  {1}^\prime,  {2}^\prime$.
The two-component indices are raised and lowered as
follows $ A^\ga = \gvep^{\ga \gb}A_\gb\,,$ $A_\ga =
\gvep_{\gb\ga}A^\gb\,,$ $\gvep_{\gb\ga} = -\gvep_{\ga\gb}
\,,$ $\gvep_{12}=1 $ and analogously for primed indices.} 
$x^{\ga\,\pa}$ in two-component spinor
notations. The relation with the tensor notations is
$ x^{\ga\pb}=
x^n \sigma_n^{\ga\pb}\,,$
where $\sigma_n^{\ga\pb}$
($n = 0,1,2,3$)  are four independent Hermitian $2\times 2$ matrices.

  The conserved currents built from bilinears of the rank-one fields in $\M_M$
were shown \cite{tens2} to obey the field equations of the rank-two fields in
$\M_M$. More generally,  it was shown that products of $r$ rank-one fields
obey the rank-$r$ field equations. On the other hand, a rank-$r$ field in
$\M_M$ was interpreted as a ``compactification" of an ``elementary" rank-one
field in $M_{rM}$. This correspondence is in spirit of  $AdC/CFT$
\cite{Maldacena:1997re,Gubser:1998bc,Witten:1998qj} with a field in
higher-dimensional (bulk) space-time identified with the current in a
lower-dimensional (boundary) space-time.  We believe that this phenomenon has
far-reaching consequences, partially discussed already in \cite{BHS}. In
particular, from this perspective, the very notion of space-time dimension
acquires dynamical origin \cite{Vasiliev:1990bu,Vasiliev:2001dc}.

Genuine massless fields in $d=4$ are rank-one fields in the ten-dimensional
space $\M_4$ \cite{BHS}.} In \cite{tens2,gelcur}, it was  shown that, for
$M=4$, the realization of a rank-two field in terms of bilinears of rank-one
fields gives rise to the full list of conformal gauge invariant
 conserved currents of all spins in four dimensions \cite{GSV}, which generalize
 the so-called generalized Bell-Robinson currents constructed by
 Berends, Burgers and van Dam \cite{Berends:1985xx}.

On the other hand, the rank-two field in $\M_4$ can be identified with the
elementary rank-one field in $\M_8$ that gives rise to usual conformal fields
in six dimensions \cite{Bandos:1999qf,Vasiliev:2001dc,Bandos:2005mb}, which,
in accordance with the general results of \cite{Siegel,Mets}, are  the
mixed-symmetry fields described by various two-row rectangular Young diagrams. It
should be noted that the idea that currents realized as bilinears of
elementary fields  behave as fields in higher dimension is not  new
 and was discussed for example in \cite{Das:2003vw,Koch:2010cy}
(see also references therein). However, in the framework of HS
theories that describe infinite towers of massless fields of all spins this
idea gets particularly neat realization.

This correspondence suggests the idea that the current interaction in four
dimensions can be interpreted as a mixture between linear rank-one and
rank-two fields in $\M_4$, where the latter field is only assumed to satisfy
the rank-two unfolded field equations. This implies that the seemingly
nonlinear interaction of massless fields in four dimensions with the currents
(that can be constructed from the same fields) results from a solution of the
linear problem that describes a gluing between rank-one and rank-two  fields
in the unfolded dynamics approach. As mentioned above an interesting
interpretation of this system is that it mixes massless fields in four
space-time dimensions with conformal fields in six space-time dimensions
interpreted as currents in the four-dimensional space.

In  this paper we show how this works in practice. Namely, we present a
linear unfolded system of equations that glues the unfolded equations of
rank-one and rank-two fields in such a way that, upon realization of the
rank-two fields in terms of bilinears of the rank-one fields, the usual field
equations for massless fields receive  corrections that just describe the
contribution of currents to the field equations. It is interesting to note that the same mechanism
brings Yukawa interactions to the field equations of massless fields of spins
0 and 1/2.

The rest of the paper is organized as follows.
 In   Section \ref{Preliminaries}, we recall the unfolded form of $4d$ free
 HS  field equations in $AdS_4$ proposed in
\cite{4dun,Ann} and their flat limit. In   Section \ref{currents},
 the constructions of conserved currents in the flat space
 of \cite{tens2, gelcur} is recalled and its generalization to $AdS_4$
 is given.
The nontrivial current deformation of the rank-one unfolded system with the
rank-two unfolded system is presented in Section \ref{Current deformation}.
In Section \ref{contribution}, it is shown in detail
 how the   deformed unfolded equations affect the form of dynamical
equations for massless fields bringing currents to their right-hand sides.
Section \ref{conc} contains summary of obtained results and discussion of
 further research directions.
Appendices A, B, C  and D collect technical details of the calculations.

\section{Preliminaries}
 \label{Preliminaries}
\subsection{Higher-spin gauge fields in $AdS_4$}
\label{Higher spin gauge fields}
 In this section, we recall the unfolded form of $4d$ free
 HS  field equations proposed in
\cite{4dun,Ann}. It is based on the frame-like approach to HS gauge
fields \cite{frame,Fort1} where a spin-$s$ HS gauge field is
described by the set of one-forms \be\nn \go_{\ga_1\ldots
\ga_k, \ga^\prime_1\ldots \ga^\prime_l}= dx^n \go_{n\,\ga_1\ldots
\ga_k, \ga^\prime_1\ldots \ga^\prime_l}\q k+l=2(s-1)\, \ee
 and the set of zero-forms
$C{}_{\alpha_1\ldots\alpha_n}{}_,{}_{ \gb^\prime_1 \ldots \gb^\prime_m}(x)$ with
$n-m =2s$ along with their conjugates
$\overline{C}{}_{\alpha_1\ldots\alpha_n}{}_,{}_{ \gb^\prime_1
\ldots \gb^\prime_m}(x)$ with $m-n =2s$.
  The HS gauge fields are self-conjugate
$\overline{\go_{\ga_1\ldots \ga_k\,,\gb^\prime_1\ldots\gb^\prime_l}}=
\go_{\gb_1\ldots\gb_l\,, \ga^\prime_1\ldots  \ga^\prime_k}.
$ This set is equivalent to the real
one-form $\go_{A_1\ldots A_{2(s-1)}}$
symmetric in the  Majorana spinor indices $A=1,\ldots 4$, that
carries an irreducible module of the  $AdS_4$
symmetry algebra $\mathfrak{sp}(4,{\mathbb R})\sim \mathfrak{o}(3,2)$.

$AdS_4$  is described by the Lorentz connection $\goo^{\ga\gb}$,
$\overline{\goo}^{\pa\pb}$ and vierbein $e^{\ga\pa}$.
Together, they form the $\mathfrak{sp}(4,{\mathbb R})$ connection $w^{AB}=w^{BA}$
that satisfies the $\mathfrak{sp}(4,{\mathbb R})$ zero curvature conditions
\be
\label{ads}
R^{AB}=0 \,,\qquad R^{AB} = dw^{AB} +w^{AC}\wedge w_C{}^B\,,
\ee
where indices are raised and lowered by a $\mathfrak{sp}(4,\mR)$
invariant form $\mathrm{C}_{AB}=-\mathrm{C}_{BA}$
\be
\label{Cind}
A_B =A^A \mathrm{C}_{AB}\q A^A = \mathrm{C}^{AB} A_B\q \mathrm{C}_{AC}\mathrm{C}^{BC} = \delta_A^B\,.
\ee

In terms of Lorentz components $ w^{AB} = (\goo^{\ga\gb},
\overline{\goo}^{\pa\pb}, \gla e^{\ga\pb}) $ where $\lambda^{-1}$ is
the  $AdS_4$ radius, the  $AdS_4$ equations (\ref{ads}) read as \be
\label{adsfl} R_{\ga\gb}=0\,,\quad \overline{R}_{\pa\pb}=0\,, \quad
R_{\ga\pa}=0\,, \ee where \be \label{nR} R_{\alpha
\gb}=d\goo_{\alpha \gb} +\goo_{\alpha}{}^\gamma \wedge \goo_{\gb
\gamma} +\lambda^2\, e_{\alpha}{}^{{\delta}^\prime} \wedge e_{\gb
{\delta}^\prime}\,, \ee \be
\nn
\overline{R}_{{\pa} {\pb}}
=d\overline{\goo}_{{\pa}
{\pb}} +\overline{\goo}_{{\pa}}{}^{{\gamma}^\prime}
\wedge \overline{\goo}_{{\pb} {\gga}^\prime} +\lambda^2\,
e^\gamma{}_{{\pa}} \wedge e_{\gamma {\pb}}\,,
\ee
\begin{equation}
\label{nr}
R_{\alpha {\pb}} =de_{\alpha{\pb}} +\goo_\alpha{}^\gamma \wedge
e_{\gamma{\pb}} +\overline{\goo}_{{\pb}}{}^{{\delta}^\prime}
\wedge e_{\alpha{\delta}^\prime}\,.
\end{equation}

The unfolded equations of motion of a spin-$s$ massless  field
are \cite{Ann}
\bee \label{CON1}
 D^{ad}\go(y,{\bar{y}}|x) =
   \overline{H}^{\pa\pb}
\f{\p^2}{\p \overline{y}^{\pa} \p \overline{y}^{\pb}}
\overline{C}(0,\overline{y}\mid x) +  H^{\ga\gb}
\f{\p^2}{\p {y}^{\ga} \p {y}^{\gb}}
C (y,0\mid x)\,,\eee
 \bee
\label{CON2}
D^{tw}C(y,{\bar{y}}|x) =0\,,
\eee
where
\be
\label{H}
H^{\ga\gb} = e^{\ga}{}_\pa \wedge e^\gb{}^\pa\q
\overline {H}^{\pa\pb} = e_{\ga}{}^\pa\wedge e^{\ga\pb}\,,
\ee
$y^\ga$ and $\by^\pb$ are
auxiliary commuting conjugate two-component spinor variables,
 $1-$form  $\go(y,{\bar{y}}|x)$ and $0-$form $C(y,{\bar{y}}|x)$ 
have the form
$$\go(y,{\bar{y}}|x)=\sum_{m,n\ge0}
\omega{}_{\alpha_1\ldots\alpha_n}{}_,{}_{{\gb^\prime}_1
\ldots{\gb^\prime}_m}(x)y^{\alpha_1}\dots y^{\alpha_n} \by^{\gb^\prime_1}\dots
\by^{\gb^\prime_m}\,
$$ with $n+m=2(s-1)$ (for $s\geq 1$)\,,
 $$C(y,{\bar{y}}|x)=\sum_{m,n\ge0}C{}_{\alpha_1\ldots\alpha_n}{}_,{}_{{\gb^\prime}_1
\ldots{\gb^\prime}_m}(x) y^{\alpha_1}\dots y^{\alpha_n}
\by{}^{\gb^\prime_1}\dots \by{}^{\gb^\prime_m}$$
 with $n-m=2s$,\, $\overline{C}(y,{\bar{y}}|x)$ is complex conjugated to
 $C(y,{\bar{y}}|x),$ 
and
 \be
\label{ad}D^{ad}\go(y,{\bar{y}}|x) =
D^L \go (y,\bar{y}| x) -
 \lambda e^{\ga\pb}\Big (y_\ga \frac{\partial}{\partial \bar{y}^\pb}
+ \frac{\partial}{\partial {y}^\ga}\bar{y}_\pb\Big )
\go (y,\bar{y} | x) \,,\quad (D^{ad})^2=0\,,
 \ee
 \be
\label{tw}
D^{tw} C(y,{\bar{y}}|x) =
D^L C (y,{\bar{y}}|x) +\lambda e^{\ga\pb}
\Big (y_\ga \bar{y}_\pb +\frac{\partial^2}{\partial y^\ga
\partial \bar{y}^\pb}\Big ) C (y,{\bar{y}}|x)\,,\quad (D^{tw})^2=0\,,
\ee
where  the Lorentz covariant derivative $D^L$ is
\be
\label{dlor} D^L A (y,{\bar{y}}|x) = d A (y,{\bar{y}}|x) - \Big
(\goo^{\ga\gb}y_\ga \frac{\partial}{\partial {y}^\gb} +
\overline{\goo}^{\pa\pb}\bar{y}_\pa \frac{\partial}{\partial
\bar{y}^\pb} \Big ) A (y,{\bar{y}}|x)\,. \ee
%
  $x^{\ga\pb}=x^n \sigma_n^{\ga\pb}\,$ are Minkowski coordinates
where $\sigma_n^{\ga\pb}$ are four Hermitian $2\times 2$ matrices.

As explained in \cite{Ann,gol,33}, the dynamical massless fields are
\begin{itemize}
\item
$C(x)$ and $\overline{C}(x)$ for two spin-zero fields,
\item
$C_\ga(x)$ and $\overline{C}_\pa (x)$ for a massless spin-$1/2$ field,
\item
$\omega_{\ga_1\ldots\ga_{s-1}, \ga^\prime_1\ldots \ga^\prime_{s-1}}(x)$
for an integer spin-$s\geq 1 $ massless field,
\item
 $\omega_{\ga_1\ldots\ga_{s-3/2}, \ga^\prime_1\ldots \ga^\prime_{s-1/2}}(x)$
and its complex conjugate
 $\omega_{\ga_1\ldots\ga_{s-1/2}, \ga^\prime_1\ldots \ga^\prime_{s-3/2}}(x)$
for a half-integer spin-$s\geq 3/2$ massless field.
\end{itemize}
All other fields  are auxiliary, being
expressed via derivatives of the dynamical massless fields
by the equations  (\ref{CON1}) and (\ref{CON2}).

The equations (\ref{CON2})  are
independent of (\ref{CON1}) for spins $s=0$ and $s=\half$ and partially independent
for  $s=1$ but become consequences
of (\ref{CON1}) for  $s>1$.
The equations  (\ref{CON1}) express the holomorphic
and antiholomorphic components of spin-$s\geq 1$ zero-forms
$C(y,\bar{y}|x)$ via derivatives of the massless field
gauge one-forms described by $\go(y,\bar{y}|x)$. This identifies
the spin-$s\geq 1$  holomorphic
and antiholomorphic components of the zero-forms
$C(y,\bar{y}|x)$ with the Maxwell tensor,
on-shell Rarita-Schwinger curvature, Weyl tensor and their HS
generalizations. In addition, the equations (\ref{CON1}) impose
the standard field equations on the spin-$s>1$ massless gauge fields.
The dynamical equations
for  $s\leq 1$ are  contained in the equations (\ref{CON2}).

\subsection{$\sigma_-$--cohomology}
In the unfolded dynamics approach, {\it dynamical
fields}  , their {\it differential gauge symmetries}
({\it i.e.,} those that are not Stueckelberg (\ie shift) symmetries)
and {\it differential field equations} ({\it i.e.,} those that are
not constraints) are characterized by the so-called
$\sigma_-$--cohomology.

 Let us briefly recall the  $\sigma_-$--cohomology analysis
 following to \cite{33}.
A space $V_0$, where
zero-forms $C$, $\overline{C}$ are valued, is endowed with
the   grading $G_0$
 \be
\label{G0}
G_0=\half (   n+\bar{n})  \q
n=y^\gb \f{\p}{\p y^\gb}\q \bar{n}=\by^\pb \f{\p}{\p \by^\pb}\,.
\ee
This gives
\be \label{Gtw} D^{tw} = D^L
+\gl\sigma^{tw}{}_- +\gl\sigma^{tw}{}_+\,, \ee
where
\be\nn
\sigma^{tw}{}_- = e^{\ga\pa} \f{\p^2}{\p y^\ga \p \bar{y}^\pa}\q
\sigma^{tw}{}_+ = e^{\ga\pa}  {  y_\ga   \bar{y}_\pa}\,.
\ee
We have $
[G_0\,,\sigma^{tw}{}_\pm ] =
\pm \sigma^{tw}{}_\pm\,,
$  \,\,$
[G_0\,,\D^L ]= 0\,,$
 and
$ (\sigma^{tw}{}_\pm)^2=0\,.
$

A space $V_1$, where
one-forms $\go$  are valued, is endowed with
the   grading $G_1$
\be
\label{Gnbn}
G_1=   \half\Big |     n-\bar{n}  \Big | .
\ee
This gives
 \be \label{Gad} D^{ad} = D^L
-\gl\sigma{}^{ad}{}_- -\gl\sigma{}^{ad}{}_+\,, \ee
where
\bee\label{gsp}
\sigma{}^{ad}{}_- =
   \gr_-\, \theta(n-\bar{n}-2)
+\overline{\gr}_-  \,\theta(\bar{n}-n-2) \q
\gs^{ad}{}_+=\gr_-\,
 \theta(\bar{n}-n)
+\overline{\gr}_-\,
 \theta(n-\bar{n} )\,,
\eee
\be\label{theta,ro}
\gr_-=e^{\ga\pb}\f{\p}{\p y^\ga} \by_\pb \q \overline{\gr}_- =e^{\ga\pb}\f{\p}{\p \bar{y}^\pb } {y}_\ga\q
\theta (m)=1\,(0)\q m\geq 0\,\, (m<0)\,.
\ee
We have   $ [G_1\,,\sigma^{ad}{}_\pm ] =\pm \sigma^{ad}{}_\pm\,,$ \,\,\,
$[G_1\,,\D^L ]= 0$.
Although $\rho_-$ and $\overline{\rho}_-$
do not anticommute, $ (\sigma^{ad}{}_-)^2=0\,$
  because $(\rho_-)^2=(\overline{\rho}_- )^2=0$
and the step functions guarantee that the parts of $\gs_-$
associated with $\rho_-$ and $\overline{\rho}_-$ act in different spaces.

Setting \be\label{sigma-prime}
\gs_- =\gs^{tw}{}_-+\gs^{ad}{}_-\,,\ee where $\gs^{tw}{}_-$
 acts on zero-forms, while
 $\gs^{ad}{}_-$ acts on one-forms, cohomology of $\sigma_- $ determines
the dynamical content of the dynamical
system at hand.  Namely,
 from the level-by-level analysis of the equations (\ref{CON1}) and (\ref{CON2})
it follows that all fields, that do not belong to $Ker\,\sigma_- $,
are auxiliary, being expressed by (\ref{CON1}) and (\ref{CON2})
via derivatives of the lower grade fields.
 (For more details  see e.g. \cite{solv,33}.)
In the case of massless fields, the nontrivial cohomology of $\gs_- $ is concentrated in
the subspaces with $G_j =0 $ and $\pm 1/2$ \cite{33}.
In particular, the nontrivial cohomology of $H^0(\gs{} _-)$ 
appears in the subspaces of grades $G_1=0$ or $1/2$, where $\gs_-$
acts trivially because of the step functions in (\ref{gsp}).

Field equations contained in the sector of $(p+1)$-form curvatures
are characterized by $H^{p+1}(\gs{} _-)$ which describes
those parts of the generalized curvatures that contain nontrivial
gauge invariant combinations of derivatives of dynamical fields.
Since massless
equations for bosons and fermions are, respectively, of second and
first order, the respective cohomologies have levels two and one. As
anticipated, there are as many nontrivial field equations as
components of the Fronsdal fields. In particular, in the bosonic
case, dynamical equations for a spin-$s$ field are described by the
traceless symmetric tensors of ranks $s$ and $s-2$ (for $s\geq 2$).
For example, in the case of gravity, these include the traceless
part of Ricci tensor and scalar curvature. In this paper, we
only consider conformal HS currents that are generated by
generalized HS stress tensors that in the tensor notation are described
by traceless tensors. This means that in this paper we will only
study those current deformations of the massless field equations
that contribute to the rank--$s$ traceless part of the HS field
equations.

 \subsection{Flat limit}\label{flatlim}
To take the flat limit it is necessary to perform certain rescalings.
 To this end, it is useful to introduce notations \cite{33}
$A_\pm$ and $A_0$ so that the spectrum of the operator
$
 \left (y^\ga\f{\p}{\p y^\ga} - \overline{y}^\pa
\f{\p}{\p \overline{y}^\pa}\right )
$
is positive on  $A_+(y,\overline{y}\mid x) $,
negative on  $A_-(y,\overline{y}\mid x) $ and zero on
$A_0(y,\overline{y}\mid x)$.
Having the decomposition
\be
\label{dec}
A(y,\overline{y}\mid x) =
A_+(y,\overline{y}\mid x) + A_-(y,\overline{y}\mid x)
+A_0(y,\overline{y}\mid x)\,,
\ee
the rescaled fields are introduced as follows
\bee
\label{resc}
\tilde{A}(y,\overline{y}\mid x)=
A_+(\lambda^{\half} y,\lambda^{-\half}\overline{y}\mid x) +
 A_-(\lambda^{-\half}y,\lambda^{\half} \overline{y}\mid x)
+A_0(\lambda^{\half} y,\lambda^{-\half}\overline{y}\mid x)\,,\\ \nn
\tilde{\tilde{A}}(y,\overline{y}\mid x)=
A_+(\lambda^{\half}y,\lambda^{\half}\overline{y}\mid x)
+ A_-(\lambda^{\half}y,\lambda^{\half}\overline{y}\mid x)
+A_0(\lambda^{\half} y,\lambda^{\half}\mid x)\,.
 \eee
(Note that $A_0(\lambda y,\overline{y}\mid x)=
A_0(y,\lambda \overline{y}\mid x)$).
For the rescaled variables, the flat limit $\lambda \to 0$ of
the adjoint and twisted adjoint covariant derivatives (\ref{ad})
and (\ref{tw}) gives
\be
\label{adfl}
D_{fl}^{ad}\tilde{A}(y,\bar{y} \mid x)
= D^L \tilde{A} (y,\bar{y} \mid x) -
 e^{\ga\pb}\Big (y_\ga \frac{\partial}{\partial \bar{y}^\pb}
\tilde{A}_-(y,\bar{y} \mid x)
+ \frac{\partial}{\partial {y}^\ga}\bar{y}_\pb \tilde{A}_+(y,\bar{y} \mid x)
\Big ) \,,
\ee
\be
\label{fltw}
D_{fl}^{tw} \tilde{\tilde{A}}(y,\bar{y} \mid x) =
D^L \tilde{\tilde{A}}(y,\bar{y} \mid x) + e^{\ga\pb}
\frac{\partial^2}{\partial y^\ga\partial \bar{y}^\pb}
\tilde{\tilde{A}}(y,\bar{y} \mid x)\,.
\ee
The flat limit of the unfolded massless equations results from
(\ref{CON1}) and (\ref{CON2}) via the substitution of
$D^L$ and $e^{\ga\pa}$ of Minkowski space and the
replacement of $D^{ad}$ and $D^{tw}$
by $D^{ad}_{fl}$ and $D_{fl}^{tw}$, respectively.
The resulting field equations
describe free HS fields in Minkowski space. Let us stress that
the flat limit prescription (\ref{resc}),
that may look somewhat unnatural
in the two-component spinor notation,
is designed just to give rise to the theory of
Fronsdal \cite{fronsdal_flat} and Fang and Fronsdal
 \cite{FF} (for more details see \cite{33}).

Note that, although  the contraction $\lambda\to 0$
with the rescaling (\ref{resc}) is consistent with the free
HS field equations, it turns out to be inconsistent
in the nonlinear HS theory because negative powers
of $\lambda$ survive in the full nonlinear equations
 upon the rescaling (\ref{resc}), not allowing the
flat limit in the nonlinear theory. This is why
the Minkowski background is unreachable in the
nonlinear HS gauge theories of \cite{FV1,more,non}.

\subsection{Unfolded equations in matrix spaces $\M_M$}
As observed in \cite{BHS}, the massless equations (\ref{CON2})
can be promoted to a larger space $\M_4$ with matrix coordinates $X^{AB}=X^{BA}$
by extending the system (\ref{CON2}) to
\be
\label{r1}
d X^{AB} \Big (\frac{\partial}{\partial  X^{AB}} \pm
\frac{\partial^2}{\partial Y^A \partial Y^B}\Big ) C_\pm(Y|X)=0\,,
\ee
where the $\pm$ sign is introduced for the future convenience.
This extension makes the
$Sp(8)$ symmetry of the tower of massless fields of all spins,
observed originally by Fronsdal \cite{F},
geometrically realized on the Lagrangian Grassmannian that was shown
in \cite{F} to be a minimal $Sp(8)$ invariant space that contains
Minkowski space as a subspace. (Note that in \cite{Bandos:1999qf} it was also
observed that the tower of $4d$ massless fields of all spins is naturally realized
in $\M_4$.)

That $Sp(8)$ is a symmetry of both the system (\ref{CON2}) and
(\ref{r1}) follows from the general property of unfolded equations
that any subalgebra of  $End\, V$, where $V$ is the module where
zero-forms $C$ are valued, forms a symmetry of the free system (for
more details see e.g. \cite{33} and references therein).
$\mathfrak{sp}(8)$ is the algebra of various bilinears of $Y^A$ and
$\f{\p}{\p Y^A}$ that act on the space $V$ of functions $C(Y)$.
Conformal algebra $\mathfrak{su}(2,2)$ is the subalgebra of
$\mathfrak{sp}(8)$ spanned by those bilinears that commute with the
helicity operator $H\in \mathfrak{sp}(8)$
\be\label{hel1} H=y^\ga
\f{\p}{\p y^\ga}-\by^\pga \f{\p}{\p \by^\pga}\,, \ee
which associates helicities of fields
to its eigenvalues. More
precisely, the centralizer of $H$ in $\mathfrak{sp}(8)$ is
$\mathfrak{su}(2,2)\oplus \mathfrak{u}(1)$ where $\mathfrak{u}(1)$
is generated by $H$ while $\mathfrak{su}(2,2)$
is the conformal algebra. Thus, in the zero-form sector, massless
equations of fields of different spins are conformal.

The system (\ref{r1}) extends to $\M_4$ the  $4d$ massless equations in Minkowski background
formulated in Cartesian  coordinates. Its extension to a $AdS$-like version of $\M_4$,
which is the group manifold $Sp(4)$ \cite{BHS}, is also available
\cite{Didenko:2003aa}  in any coordinate system. Note that more recently
the one-form sector of HS  equations (\ref{CON1}) was also extended to
$\M_4$ in \cite{33}.
By general properties of unfolded equations, the equations (\ref{r1}) are equivalent
to the flat limit of the  $4d$ HS equations (\ref{CON2}). Interesting details of
this correspondence were worked out in \cite{Vasiliev:2001dc,Bandos:2005mb}.

In \cite{tens2}, the  equation (\ref{r1}) was extended to so-called
rank-$r$ systems of the form
\be
\label{co}
d X^{AB} \Big (\frac{\partial}{\partial  X^{AB}}\pm\eta^{ij}
\frac{\partial^2}{\partial Y^{iA} \partial Y^{jB}}\Big ) C^r_\pm(Y|X)=0\,,
\ee
where $i,j=1,\ldots, r$ and $\eta^{ij}=\eta^{ji}$ is some nondegenerate metric.
The following comments on the properties of higher-rank systems are most
relevant to the analysis of this paper.

Higher-rank systems inherit all symmetries of the lower-rank system
from which they are built simply because they correspond to the tensor
product of the lower-rank representation of one symmetry or another.
In particular, this means that higher-rank systems are conformal once
the underlying lower-rank systems were.

In the basis where $\eta^{ij}$
is diagonal, the higher-rank
equations (\ref{co}) are satisfied by the products of rank-one fields
\be
C^r (Y_i|X) = C_1(Y_1|X)C_2(Y_2|X)\ldots C_r(Y_r|X)\,.
\ee

The  rank-$r$ systems in $\M_M$ can further be extended to
a rank-one system (\ref{r1}) in the larger space $\M_{rM}$
with coordinates $X^{AB}_{ij}$ via reinterpretation of the
twistor coordinates
\be
Y^A_i \to Y^{\widetilde A}\q \widetilde A = 1,\ldots, rM\,.
\ee
The diagonal embedding $\M_M$ into $\M_{rM}$ is
\be
X^{AB}_{11}= X^{AB}_{22}=\ldots = X^{AB}_{rr} =X^{AB}\,.
\ee

On the other hand, as shown
in \cite{Bandos:1999qf,Vasiliev:2001dc,Bandos:2005mb}
the rank-one fields in $\M_M$ with
higher $M$ describe conformal fields in diverse  space-time dimensions. In particular,
a rank-one field in $\M_8$ describes all conformal fields
in the six-dimensional Minkowski space. This
implies that conformal currents in four space-time dimensions, that were shown in \cite{gelcur}
to be described by rank-two fields in $\M_4$, are equivalent to conformal fields
in six space-time dimensions. More precisely we should say that the $4d$ currents
are dual to the $6d$ conformal fields. The reason is that the space
of states of higher-dimensional fields are represented
by the product of $C_-$ fields in (\ref{r1}) while the currents are represented
by the product of $C_+$ and $C_-$, where $C_+$ and $C_-$ describe, respectively,
particles and anti-particles, \ie the space of single-particle states and its
dual.\footnote{Strictly speaking this interpretation requires
an additional factor of $i$ in the second term of (\ref{r1}), skipped in this paper.
For more details on these issues we refer the reader to \cite{gelcur}.}
In this paper we will loosely identify the currents with the fields.

Now we are in a position to explain how rank-two equations give rise
to conserved currents considering for simplicity  the reduction of
$\M_4$ to usual Minkowski space.

\section{Conserved currents}

\label{currents}
\subsection{Minkowski case }
\label{currentsflat}
The reduction of the rank-two field equations of \cite{gelcur} to Minkowski space
gives
\bee\label{Dtw2flj}
D^{tw}_{fl}{}_2 \PPP{}(y^\pm,\by^\pm|x)= \Big(D^L +   e^{\ga\pb}
\Big( \frac{\partial^2}{\partial y^+{}^\ga\partial \bar{y}^-{}^\pb}\,
  + \frac{\partial^2}{\partial y^-{}^\ga\partial \bar{y}^+{}^\pb}\,\Big)\,\Big)\PPP{}
   (y^\pm,\by^\pm|x)=0\,.
\eee
Let $ \PPP{}(y^\pm,\by^\pm|x)$, that satisfies Eq.~(\ref{currentsflat}),
 be called rank-two current field.
 Introducing basis three-forms
\be\label{Hh}
 {\Hh}^{\ga \delta^\prime}=-\f{1}{3} e^{\ga}{}_\pa \wedge e^\gb{}^\pa
\wedge e_{\gb}{}^{\delta^\prime}\,
\ee
and using relations
\be\label{HhREL}
e^{\gamma\rho^\prime}\wedge {\Hh}^{\ga \delta^\prime}=\frac{1}{4}
\epsilon^{\gamma\ga}\epsilon^{\rho^\prime \delta^{\prime}}
e_{\eta\sigma^\prime}\wedge {\Hh}^{\eta \sigma^\prime}
\,,
\ee
it is easy to check  that the  three-forms
\bee\label{concur1}
\Omega_{- }( \PPP{})&=& \Hh^{\ga\pa}\f{\p}{\p y^-{}^\ga}\f{\p}{\p \by^-{}^\pa}
  \PPP{}(y^\pm,\by^\pm|x)\Big|_{{y^\pm=\by^\pm=0}}\,
 \,\q
\\\label{concur1++}
\Omega_{+}( \PPP{})&=& \Hh^{\ga\pa}\f{\p}{\p y^+{}^\ga}\f{\p}{\p \by^+{}^\pa}
 \PPP{}(y^\pm,\by^\pm|x)\Big|_{{y^\pm=\by^\pm=0}}\,
 \,
 \,\q\\ \label{concur1+-}\Omega_{\pm}( \PPP )&=&\Hh^{\ga\pa}\left(\f{\p}{\p y^-{}^\ga}\f{\p}{\p \by^+{}^\pa}
-\f{\p}{\p y^+{}^\ga}\f{\p}{\p \by^-{}^\pa}
\right) \PPP{}(y^\pm,\by^\pm|x)\Big|_{{y^\pm=\by^\pm=0}}\,
\eee
are closed provided that  $ \PPP{}(y^\pm,\by^\pm|x)$ satisfies
(\ref{currentsflat}).

 To define symmetry parameters that produce more conserved currents, consider  the {\it adjoint} covariant derivative
 \bee\label{Dcurad2fl}\!\!
  D{}_{fl\,2}  =  \!D^L+
  e^{\ga\pb} \Big( \! u_-{}_\ga\frac{\partial }{ \partial \bar{y}^+{}^\pb}\!
  +  \!\bu_-{}_\pb\frac{\p }{\p  y^+{}^\ga}\,\Big) \,,\quad
\eee
resulting from $ D^{tw}_{fl\,2}$ via
the substitution
\bee\label{FUR}y^-{}^\ga\to - \frac{\p }{\p  u_-{}_\ga}\q\by^-{}^\pa\to
-\frac{\p }{\p  \bu_-{}_\pa}\q
 \frac{\p }{\p  y^-{}^\ga}\to  u_-{}_\ga \q \frac{\p }{\p  \by^-{}^\pa}\to \bu_-{}_\pa\,,\eee
 that formally coincides with the ``half Fourier transform" of \cite{gelcur}.
 Since the covariant derivative (\ref{Dcurad2fl}) is of the first order, the space
of  regular solutions of the equation
  \bee\label{cur2fl}\!\!
  D{}{}{}_{fl\,2}^{tw}\eta(y^+,\by^+,u_- ,\bu_- |x)=  0\qquad
\eee
forms a commutative algebra $\Fursol_{fl}{}{}$.
Evidently, $\Fursol_{fl}{}{}$ is generated by the elementary solutions
 \be \label{parfl}\ls
  u _-{}_\gb\,,\,\,\quad\,\,
   y^+{}^\ga  -  x{}^{\ga\pb}  \bu_-{}_\pb \,
 \,, \,\quad\,\,\,
\, \bu _-{}_\pb
 \,,\,\,\,\quad\,\,
     \by^+{}^\pa -  x{}^{\gb\pa}   u_-{}_\gb .\ee

By the substitution inverse to  (\ref{FUR})  
\bee\label{FURin}u_-{}_\ga\to \frac{\p }{\p  y^-{}^\ga}\q\bu_-{}_\pa\to \frac{\p }{\p  \by^-{}^\pa}
\q
 \frac{\p }{\p  u_-{}_\ga}\to -y^-{}^\ga \q \frac{\p }{\p  \bu_-{}_\pa}\to-\by^-{}^\pa\, \eee
the algebra $\Fursol_{fl}{}$ is mapped to the
algebra $\R_{fl}$ of differential operators
$\eta (\xi{}_-{}_\gb\,,\bxi{}_-{}_\pb \,,\,  \xi^{+}{}^\ga \, ,\, \bxi^{+}{}^\pa )$
 generated by
   \be \label{parflop}
  {\xi}{}{}_-{}_\ga=\frac{\p }{\p  y^-{}^\ga}\,,\,\,
  \bar{\xi}{}{}{}_-{}_\pb=\frac{\p }{\p  \by^-{}^\pb}
 \,,\,\,
 {\xi}{}{}{}^+{}^\ga= y^+{}^\ga \,-  \,x{}^{\ga\pb}  \frac{\p }{\p  \by^-{}^\pb} \, ,\,\,\
\bar{\xi}{}{}{}^+{}^\pa= \by^+{}^\pa \,-  \,x{}^{\gb\pa}  \frac{\p }{\p  y^-{}^\gb} \,. \ee

Since any  $\eta (\xi{}_-{}_\gb\,,\bxi{}_-{}_\pb \,,\, \xi^{+}{}^\ga
\, ,\, \bxi^{+}{}^\pa )$ $\in \R_{fl} $ satisfies (\ref{cur2fl}),
 it follows that
\bee\label{Dtw2fljeta}
D{}^{tw}_{fl\,2}\,   \PPP{}(y^\pm,\by^\pm|x) =0\quad\Longrightarrow\quad
D{}^{tw}_{fl\,2}\, \,\big (\eta (\gx,\px)\,\,\PPP{}(y^\pm,\by^\pm|x)\big )  =0\,.
\eee
Hence, the three-form
\bee\label{concur2}
\Omega_-(\eta  \PPP{})= \Hh^{\ga\pa}\f{\p}{\p y^-{}^\ga}\f{\p}{\p \by^-{}^\pa}
\eta (\gx,\px)  \PPP{}(y^\pm,\by^\pm|x)\Big|_{{y^\pm=\by^\pm=0}}\,
 \,
\eee
is closed.  Thus any element of $\R_{fl} $ generates a conservation law.
As explained in more details in \cite{gelcur}, $\R_{fl} $ matches
 the space of HS global symmetry parameters  of \cite{GSV}.

The relation with usual currents is due to the fact  that
Eq. (\ref{Dtw2flj}) is solved
by the  bilinear  \cite{tens2}
   \be\label{bilinear}
   \PPP(y^\pm\,\by^\pm|x)= 
   C_+(y^++y^-,\by^++\by^-| x )C_-(y^+-y^-,\by^+-\by^- | x)\,\ee
of   rank-one fields $C_\pm(y\,\by|x)$ that solve  the rank-one equations
 \be
\label{fltwh}
 D^L C_\pm(y\,\by|x) \pm   e^{\ga\pb}
\frac{\partial^2}{\partial y^\ga\partial \bar{y}^\pb}C_\pm(y\,\by|x) =0\,,
\ee
which coincide with the Minkowski reduction of the equation (\ref{r1})
and, up to a sign, with the flat limit of Eq.~(\ref{CON2}).
 The resulting currents reproduce the
lower-spin and HS conserved currents built from massless fields, originally obtained
in \cite{Berends:1985xx}.

The change of minuses  to pluses in the ``half Fourier transform"
(\ref{FUR}) gives another set of operators
   \be \label{parflop+}
  {\chi}{}{}_+{}_\ga=\frac{\p }{\p  y^+{}^\ga}\,,\,\,
  \bar{\chi}{}{}{}_+{}_\pb=\frac{\p }{\p  \by^+{}^\pb}
 \,,\,\,
 {\chi}{}{}{}^-{}^\ga= y^-{}^\ga \,- \,x{}^{\ga\pb}  \frac{\p }{\p  \by^+{}^\pb} \, ,\,\,\
\bar{\chi}{}{}{}^-{}^\pa= \by^-{}^\pa \,- \,x{}^{\gb\pa}  \frac{\p }{\p  y^+{}^\gb} \,,
 \ee
  that commute with $D^{tw}_{fl\, 2}$
(\ref{Dtw2flj}), hence also generating symmetry  parameters and
conserved currents. Generally, the following set of
closed three-forms can be written with an arbitrary parameter
$g(\gx,\px,\chi,\bar{\chi})$
 \bee\nn
\Omega_{- }( g \PPP{})&=& \Hh^{\ga\pa}\f{\p}{\p y^-{}^\ga}\f{\p}{\p \by^-{}^\pa}
g (\gx,\px,\chi,\bar{\chi})  \PPP{}(y^\pm,\by^\pm|x)\Big|_{{y^\pm=\by^\pm=0}}\,
 \,\q
\\\nn
\Omega_{+}( g \PPP{})&=& \Hh^{\ga\pa}\f{\p}{\p y^+{}^\ga}\f{\p}{\p \by^+{}^\pa}
g (\gx,\px,\chi,\bar{\chi}) \PPP{}(y^\pm,\by^\pm|x)\Big|_{{y^\pm=\by^\pm=0}}\,
 \,
 \,\q\\ \nn
 \Omega_{\pm}(g  \PPP )&=&\Hh^{\ga\pa}\left(\f{\p}{\p y^-{}^\ga}\f{\p}{\p \by^+{}^\pa}
-\f{\p}{\p y^+{}^\ga}\f{\p}{\p \by^-{}^\pa}
\right)g (\gx,\px,\chi,\bar{\chi}) \PPP{}(y^\pm,\by^\pm|x)\Big|_{{y^\pm=\by^\pm=0}}\,
\eee
 However, most of these forms turn out to be exact giving rise to
zero charges. As  will be shown in the forthcoming publication
\cite{fut}, both  in Minkowski and $AdS_4$ cases, nontrivial
charges (\ie current cohomology) are fully represented by the
following closed three-forms
\bee\label{concur2hel}
\Hh^{\ga\pa}\f{\p}{\p y^-{}^\ga}\f{\p}{\p \by^-{}^\pa}
\eta (\gx,\px, H_1-H_2) \PPP{}(y^\pm,\by^\pm|x)\Big|_{{y^\pm=\by^\pm=0}}\,,
 \,
\\\nn
\Hh^{\ga\pa}\f{\p}{\p y^+{}^\ga}\f{\p}{\p \by^+{}^\pa} \eta (\chi,\bar{\chi},
H_1-H_2) \PPP{}(y^\pm,\by^\pm|x)\Big|_{{y^\pm=\by^\pm=0}}\,,
 \,
\eee where
$$H_j=y^j{}^\ga\f{\p}{\p  y^j{}^\ga}-\by^j{}^\pa\f{\p}{\p \by^j{}^\pa}\,.$$
Note that  $ (H_1-H_2)J=4(h_+-h_-)J$ for bilinear currents $J$ (\ref{bilinear})
with the fields $C_\pm $ of helicities $h_\pm$.

\subsection{$AdS_4$}
\label{currentsAdS}
In the case of  $AdS_4$\,,
the rank-two unfolded equations, \ie ``current equations", are
\bee\label{newtwistjads}
  D^{tw}_2{}  \PPP(y^\pm,\by^\pm|x)\,=0
 \,,
\eee
where \bee\label{Dtw2}
D^{tw}_2{}= D^L +\gl e^{\ga\pb}
\Big(y^+{}_\ga\,\bar{y}^-{}_\pb+
y^-{}_\ga\,\bar{y}^+{}_\pb+
\frac{\partial^2}{\partial y^+{}^\ga\partial \bar{y}^-{}^\pb}\,
  + \frac{\partial^2}{\partial y^-{}^\ga\partial \bar{y}^+{}^\pb}\,\Big)\,.
\eee
Again, the current equations (\ref{newtwistjads}) imply that, being evaluated at
${{y^\pm=\by^\pm=0}}$,  three-forms (\ref{concur1}) - (\ref{concur1+-}) are closed.

\subsubsection{The Howe dual algebra}
\label{Howe}
To sort out different solutions  of the rank-two equation (\ref{newtwistjads})
we observe that the operators
\bee\label{slv}\ls
f_+=y^+{}^\gn y^-{}_\gn -\f{\p^2}{\p  \by^+{}^\pn \by^-{}_\pn}\,,\quad
f_-=-\f{\p^2}{\p y^+{}^\gga\p y^-{}_\gga}+\by^+{}^\pga \by^-{}_\pga\,,\quad
\\\nn
 \hv{} =y^+{}^\ga\f{\p}{\p y^+{}^\ga}
+ y^-{}^\ga\f{\p}{\p y^-{}^\ga}-\by^+{}^\pa\f{\p}{\p  \by^+{}^\pa}-\by^-{}^\pa\f{\p}{\p  \by^-{}^\pa}\q
\,\quad
 \eee
 and
\bee\label{slh}
\vv_+=y^+{}^\ga\f{\p}{\p y^-{}^\ga}
 -\by^+{}^\pa\f{\p}{\p  \by^-{}^\pa}\q
 \vv_-= y^-{}^\ga\f{\p}{\p y^+{}^\ga}- \by^-{}^\pa\f{\p}{\p  \by^+{}^\pa}
\,,\\ \nn \uu=
 y^+{}^\ga\f{\p}{\p y^+{}^\ga}
 +\by^+{}^\pa\f{\p}{\p  \by^+{}^\pa} - y^-{}^\ga\f{\p}{\p y^-{}^\ga}- \by^-{}^\pa\f{\p}{\p  \by^-{}^\pa}
  \eee
commute with $D_2^{tw}$.
These operators form two mutually commuting $\mathfrak{sl}_2$ algebras %
 with the nonzero   commutation relations
\bee\nn
[f_+,f_-]=\hv{}\q [\hv{},f_\pm]=\pm 2f_\pm
\,;
\\ \nn
[\vv_+\,,\vv_-]=\uu{}\q [\uu{},\vv_\pm]=\pm 2\vv_\pm
\,.
\eee
The algebras (\ref{slv}) and (\ref{slh}) will be referred to as
 vertical $\slv$  and horizontal $\slh$, respectively.
The Cartan operator $\hv\in\slv$ (\ref{slv}) will be referred to as
{\it rank-two helicity   operator}.

 It is easy to see that
\bee\label{concurepm}
 \!\!\Hh^{\ga\pa}\f{\p^2}{\p y^-{}^\ga\by^-{}^\pa}
\,  f_-\, \PPP(y^\pm,\by^\pm|x)\Big|_{{y^\pm=\by^\pm=0}}\,=
\f{1}{2\gl}d \Big (H^{\ga\gb}\f{\p^2}{\p y^-{}^\ga \p y^-{}^\gb}
\PPP(y^\pm,\by^\pm|x)\Big|_{{y^\pm=\by^\pm=0}}\Big )
\q \\ \nn
 \Hh^{\ga\pa}\f{\p^2}{\p y^-{}^\ga \p \by^-{}^\pa}\!
   f_+  \PPP(y^\pm,\by^\pm|x)\Big|_{{y^\pm=\by^\pm=0}} =
-\f{1}{2\gl}d \Big ( \overline{H}^{\pa\pb}\!\f{\p^2}{\p \by^-{}^\pa
\p \by^-{}^\pb}\PPP(y^\pm,\by^\pm|x)\Big|_{{y^\pm=\by^\pm=0}}
\Big )
 \,,\qquad
\eee
provided that $\PPP$ satisfies (\ref{newtwistjads}). Recall, that the  two-forms
$H^{\ga\gb}$  and
$\overline {H}^{\pa\pb}$ are defined in (\ref{H}), while
the three-form
$\Hh^{\ga\pa}$  is defined in (\ref{Hh}).


The system of equations (\ref{newtwistjads}) decomposes into a set
of subsystems
associated with different elements of $\slh\oplus \slv$-modules realized by
rank-two fields.
Let \be\label{Y}
Y=y^+{}^\ga y^-{}_\ga\q \bY=\by^+{}^\pa \by^-{}_\pa.
\ee
Any polynomial  $P(y^\pm) $   can be
represented in the form
\be\nn
P(y^\pm)=\sum_{n,m,k=0}^\infty   Y{}^n\,
C^{n,m,k}_{\ga(m+k) }y^+{}^{\ga(m)}y^-{}^{\ga(k)}
 ,
\ee
where multispinors
$C^{n,m,k}_{\ga(m+k)}$ are   symmetric.
It is easy to see that
 \be\nn
\frac{\partial^2}{\partial y_\gga^- \partial y^{+\gga}}
\Big ( Y^n\,
C^{n,m,k}_{\ga(m+k) }y^+{}^{\ga(m)}y^-{}^{\ga(k)}\Big)
=n(n+1+m+k) \,Y^{n-1}\,C^{n,m,k}_{\ga(m+k)}y^+{}^{\ga(m)}y^-{}^{\ga(k)}\,.
\ee
{}From this relation it follows that
  lowest vectors $F_m$ of the vertical $\slv$ (\ref{slv}), satisfying
  $\dis{f_-F_m=0}$, have the form
   \bee\label{howelowmF}
  F_m(y,\by,Y,\bY)=f^m(y,\by ,\,\bY) \, \sum_{n=0}^\infty
Y^n \bY{}^n\f{1}{n!\,(1+m+n)!}\q
\eee
 where $f^m(y,\by ,\,\bY)$ is an  arbitrary  function that satisfies
the conditions
\bee\label{howelowm}\f{\p^2}{\p y^+{}^\gga\p y^-{}_\gga}
f^m(y,\by ,\,\bY)=0\q
\Big(y^+{}^\gga\f{\p}{\p y^+{}^\gga}+y^-{}^\gga
\f{\p}{\p y^-{}_\gga}\Big)f^m(y,\by ,\,\bY)=
m\,f^m(y,\by ,\,\bY)
.\eee
Note that $F_{m }(y,\by,Y,\bY)$ (\ref{howelowmF}) satisfies
$$
\left(\Big(Y\f{\p}{\p Y} +y^j{}^\ga\f{\p}{\p y^j{}^\ga}+1\Big)\f{\p}{\p Y}-\bY\right)
F_{m}(y,\by,Y,\bY)=0\,\q
$$
where the derivatives over $Y$ and $y$ are treated as independent.

Since $\overline{f}_+ =f_-$,\,\, highest vectors are complex conjugate
to the lowest ones.
 Therefore   singlets $F_{m,m}$ of  the vertical $\slv$ (\ref{slv})  have the form
\bee\label{howesingletm}
F_{m,m}(y ,\by ,Y,\bY)=    s^m(y, \by )
\, \sum_{n\ge0}^\infty
 Y{}^n  \bY{}^n\f{1}{(1+m+n)!n!}\,,
\eee
 where polynomials $s^m(y, \by )$ satisfy (\ref{howelowm})
along with the  conjugate  conditions
\bee\nn
 \f{\p^2}{\p \by^+{}^\pga\p \by^-{}_\pga}s^m(y, \by )( \by )=0\q
 \Big(\by^+{}^\pa\f{\p}{\p \by^+{}^\pa}+\by^+{}^\pa\f{\p}{\p \by^+{}^\pa}\Big )
 s^m(y, \by )=m s^m(y, \by )\,.
\eee

It is easy to see, that lowest vectors $F_-$ and highest vectors $F_+$
of the  horizontal  $\slh$
(\ref{slh}) have the form
$$
F_- \big(y^-,\by^-,\,\,(y^+{}_\ga\,\bar{y}^-{}_\pb+y^-{}_\ga\,\bar{y}^+{}_\pb )\big)\q
F_+ \big(y^+,\by^+,\,\,(y^+{}_\ga\,\bar{y}^-{}_\pb+y^-{}_\ga\,\bar{y}^+{}_\pb) \big)\,,
$$
while $\slh$   singlets are
$$ G  \big(y^+{}_\ga\,\bar{y}^-{}_\pb+y^-{}_\ga\,
\bar{y}^+{}_\pb\big)\,,
$$
where $F_\pm$ and $G$ are arbitrary functions of their arguments.

Note that $f_0$ and the algebra $\slh$ (\ref{slh}) commute with
$D_{fl}^{tw}{}_2$ while the flat limit of the operators $f_\pm$
gives the following mutually commuting operators
\be\label{slvflat} f{}_+{}_{fl} =
 -\f{\p^2}{\p  \by^+{}^\pn \by^-{}_\pn}\,,\quad
f{}_-{}_{fl}=-\f{\p^2}{\p y^+{}^\gga\p y^-{}_\gga} \,,\quad
\ee
that commute with $D_{fl}^{tw}{}_2$.

\subsubsection{Symmetry parameters  of $AdS_4$ currents}
\label{symparam}

Proceeding as in the  Minkowski case, to find  symmetry parameters  of
$AdS_4$ currents we have to solve  the equation
 \bee\label{cur2ad} D{}_2^{ad}\,\eta(y^+,\by^+,u_- ,\bu_- |x)=0\q \eee
\be\nn  D{}_2^{ad}=  D^L+
\gl e^{\ga\pb} \Big( - y_+{}_\ga\frac{\partial }{ \partial \bar{u}^-{}^\pb}\,
  -  \by_+{}_\pb\frac{\p }{\p  u^-{}^\ga}\,
  + u_-{}_\ga\frac{\partial }{ \partial \bar{y}^+{}^\pb}\,
  +  \bu_-{}_\pb\frac{\p }{\p  y^+{}^\ga}\,\Big) ,
   \ee
where $ D{}_2^{ad}$ is again related to $D_2^{tw}$ via
(\ref{FUR}).

As in the Minkowski case, the space of solutions  of the first-order
system of partial differential equations (\ref{cur2ad})
forms a commutative algebra that  possesses two gradings
 \be
\label{Gi}
G^+=\half \Big(   y^+{}^\ga\f{\p}{\p y^+{}^\ga} +\bu_-{}_\pa\f{\p}{\p  \bu_-{}_\pa}\Big)  \q
G_-=\half \Big( u_-{}_\pa\f{\p}{\p   u_-{}_\ga} + \by^+{}^\pa\f{\p}{\p \by^+{}^\pa} \Big)
\,.
\ee
Since the compatibility of the equation (\ref{cur2ad}) is guaranteed by the
flatness condition (\ref{adsfl}), the space of   solutions of (\ref{cur2ad}) is
isomorphic to the space of
arbitrary functions of $y^+,\by^+,u_- ,\bu_- $, \ie $\xi(y^+,\by^+,u_- ,\bu_- |x)$
is reconstructed via its values at any given point $x=x_0$. Since the equation (\ref{cur2ad}) is homogeneous in the variables
$y^+,\by^+,u_- ,\bu_- $ its solutions can also be chosen to be homogeneous.
Moreover, it is enough to find a complete set of solutions of minimal
grades with respect to the both gradings (\ref{Gi}),
  hence, linear either in $y^+$ and $\bu_- $ or
in $u_-$ and $ \by^+ $.

To this end let us introduce
Killing spinors $\co^\gb(x)$ and $\si^\pb(x) \, $
 that satisfy  the equations
\bee\label{resh}
D^L  \co{}^\ga(x)  + \gl e^{\ga\pb}   \si{}_\pb(x)   =0
\q
 D^L  \si^\pb(x)  +\gl e^{\ga}{}^{\pb}     \co_\ga{}(x) =0\,.\quad
\qquad\eee
Let
a basic of this system be formed by four independent pairs
of spinors
$(\co_\la{}^\gb(x),\si_\la{}^\pb(x))$ and $(\co_\ra{}^\gb(x),\si_\ra{}^\pb(x))$
labeled by indices $\la=1,2$ and $\ra =1,2$. For example, basic
solutions of (\ref{resh}) can be chosen to obey the following
initial conditions at  $x=0$
$$\co_\la{}^\gb(0)=\gd_\la{}^\gb\q \si_\la{}^\pb(0)=0\q\co_\ra{}^\gb(0)=0\q \si_\ra{}^\pb(0)=\gd_\ra{}^\pb\,. $$
{}From these conditions it follows that
$$\overline{\co_\la{}^\gb(x)}=\si_\ra{}^\pb(x)\q
\overline{\si_\la{}^\pb(x)} =\co_\ra{}^\gb(x).$$
A particular form of solutions
$\co_\la{}^\gb(x)\,,\,\,\,\si_\la{}^\pb(x)\,,\,\,
\,\co_\ra{}^\gb(x)\,,\,\,\,\si_\ra{}^\pb(x)$
 depends on a chosen coordinate system.

 Evidently, the fundamental solutions
\bee\label{paramads}
\varrho{}_\la(u_- ,\by^+|x) = \co_\la{}^\gn(x)u_-{}_\gn+ \si_\la{}_\pn(x)\by^+{}^\pn\q
   \gep{}_\la(y^+ ,\bu_- |x) = \co_\la{}_\gb(x)y^+{}^\gb+ \si_\la{}{}^\pb(x)\bu_-{}_\pb\q
   \\ \nn
   \overline{\varrho}{}_\ra(\bu_-  ,y^+|x)
    = \si_\ra{}^\pb(x)\bu_-{}_\pb+ \co_\ra{}_\gn(x) y^+{}^\gn\q
   \overline{\gep}{}_\ra(\by^+ , u_- |x)
   = \si_\ra{}_\pb(x)\by^+{}^\pb+ \co_\ra{}{}^\gb(x) u_-{}_\gb\qquad  \eee
 generate the commutative algebra $\Fursol_{AdS}$ of solutions  of
 (\ref{cur2ad}) of the form
\be\label{paramad}
\eta^\prime {}(y^+,\by^+,u_- ,\bu_- |x)= P(\varrho{}_\la,\gep{}_\la\,,\overline{\varrho}{}_\ra\,,
\overline{\gep}{}_\ra).
\ee
As in the Minkowski case, the
 substitution (\ref{FURin}) maps  $\Fursol_{AdS}$
to  the  commutative  algebra  $\R_{AdS}$ of differential operators   generated by\footnote{
$\bp_\pm $ and $\p_\pm $ are \,shorthand \, notations for
$\frac{\p }{\p  \by^\pm}
$ and $\frac{\p }{\p   y^\pm }$, respectively.}
\bee\label{curparAds} \varrho{}_\la(\p_-  ,\by^+|x)  \,,  \,\,
\gep{}_\la(y^+ ,\bp_-   |x)   \,,  \,\,
  \overline{\varrho}{}_\ra(\bp_-  , y^+|x)   \,,  \,\,
   \overline{\gep}{}_\ra (\by^+ ,\p_- |x) .\eee
 Again it follows that
 \bee\nn
D_2^{tw} \big (\eta\, \PPP(y^\pm,\by^\pm|x)\big ) =0
\eee
provided that $\eta\in \R_{AdS}$  and $ \PPP(y^\pm,\by^\pm|x)$ satisfies  (\ref{newtwistjads}).

The commutative algebra $ \R_{AdS}$ of the current parameters 
  is a
 representation of the vertical   $\slv$ (\ref{slv}).
  In particular,
\bee\nn[\varrho{}_\la\big( \p_-   ,\by^+|x\big),f_+]=
                             \gep_\la\big(y^+ ,\bp_- |x\big) \q\,\,&&
                              [\gep{}_\la\big(y^+ ,\bp_- |x\big),f_+]=0\q \,\,\\ \nn
[\overline{\gep}{}_\ra\big( \p_-   ,\by^+|x\big),f_+]=
                      \overline{\varrho}{}_\ra\big(\bp_-  , y^+|x\big) \q&&
                      [\overline{\varrho}{}_\ra\big(\bp_-  , y^+|x\big),f_+]=0,\,\,
\etc.\eee
 On the other hand, the parameters (\ref{curparAds}) are highest vectors
 of the horizontal  $\slh$ (\ref{slh})
\bee\nn[\varrho{}_\la\big( \p_-   ,\by^+|x\big),\vv_+]=
                               [\gep{}_\la\big(y^+ ,\bp_- |x \big),\vv_+]
=[\overline{\gep}{}_\ra\big( \p_-   ,\by^+|x\big),\vv_+]
                      =[\overline{\varrho}{}_\ra\big(\bp_-  , y^+|x\big),\vv_+]=0,\,\,
 \eee
 while $\vv_-\in\slh$ maps them to new parameters
\bee\label{par*}
 [\varrho{}_\la\big(\p_-   ,\by^+|x\big),\vv_-]
    =\varrho{}_\la\big(\p_+   ,\by^-|x\big)\q
 [ \gep{}_\la(y^+ ,\bp_- |x),\vv_-] =
-\gep{}_\la(y^- ,\bp_+ |x),\,\, \etc.\qquad \eee
which result  from the original ones via exchange of
pluses and minuses.

\newcommand{\nv}{\,n}
\newcommand{\kv}{\,k}
\newcommand{\mv}{\,m}
\newcommand{\mh}{\,\widehat{m}}
\newcommand{\nh}{\,\widehat{n}}
\newcommand{\kh}{\,\widehat{k}}

Since   $\slh$ commutes with $D_2^{tw}$, the new oscillators also
 commute with $D_2^{tw}$. The full list of covariantly constant
 spinors can be packed into the form
 \be
 \label{packed}
 \varrho_\la^{{\nv \nh}}\q \overline{\varrho}{}_\ra^{{\nv \nh}}\,,
 \ee
where $\nv=+,-$ and $\nh=+,-$ are indices of the doublet
representations of $\slv$ and $\slh$, respectively.
Namely,
\bee\nn\!\!\!
\! \varrho{}_\la\big( \p_-   ,\by^+|x\big)=-\varrho{}_\la^{+-}\!\!\q\!\!
\gep{}_\la\big(y^+ ,\bp_- |x \big)=\varrho{}_\la^{++}\!\q\!
 \overline{\varrho}{}_\ra\big( \bp_-  , y^+|x\big)=\overline{\varrho}{}_\la^{++}\!\q \!\!\!
  \overline{\gep}{}_\ra\big( \by^+,\p_-   |x\big) =-\overline{\varrho}{}_\ra^{+-}\!\!\q
\\ \nn  \!\!
\overline{\varrho}{}_\ra\big(\bp_+  , y^-|x\big) =\overline{\varrho}{}_\la^{-+}\q\!
 \overline{\gep}{}_\ra\big(  \by^-,\p_+|x\big) =\overline{\varrho}{}_\ra^{--}\q\!
\varrho{}_\la\big( \p_+   ,\by^-|x\big)=\varrho{}_\la^{--}\q\!
\gep{}_\la\big(y^- ,\bp_+ |x \big)=\varrho{}_\la^{-+}
 .\quad\quad\eee
Since all oscillators (\ref{packed}) are covariantly constant,
they have $x$--independent commutation relations
  \bee \label{sootnoshenie}
[\varrho_\gb^{\nv  \kh} \,,\,\varrho_\ga^{\mv \nh}]=
 \gvep^{\nv \mv  } \gvep^{\kh\nh}\gvep_{\gb\ga}\,\q
 [\overline{\varrho}_\pb^{\nv  \kh} \,,\,\overline{\varrho}_\pa^{\mv \nh}]=
 \gvep^{\nv \mv  } \gvep^{\kh\nh}\gvep_{\pb\pa}\q
 \gvep^{-+}=1\,.
\eee
In fact, as  will be explained in more details in \cite{fut},
the covariantly constant spinors (\ref{packed}) are related to
supergenerators of (conformal) SUSY.

The full set of parameters  belongs to the space
$ P $  of arbitrary functions of the oscillators
(\ref{packed}). This space is much bigger than the space
of HS global symmetry parameters. As will be shown in \cite{fut},
most of the currents associated to elements of $P$
are exact, hence generating no nontrivial charges, while
the nontrivial currents are represented by the current
cohomology (\ref{concur2hel}) with $\xi$, $\chi$ replaced by $\varrho$ and $  \gvep$ (\ref{packed}), respectively. (Note that the ambiguity in
the dependence on $H_1-H_2$ in (\ref{concur2hel}),
is physically trivial, expressing
the ambiguity in the normalization of the rank-one fields
in the formula (\ref{bilinear}).)

To introduce currents  bilinear in rank-one fields  it is convenient to
 consider the operators $D^{tw}_\pm$, that  differ from $D^{tw}$ (\ref{tw}) by a
  sign in front of $\gl$ so that the corresponding rank-one
  equations  are
  \be
\label{twpm}
D^{tw}_{\pm } C{}_\pm (y,{\bar{y}}|x) =
D^L C{}_\pm (y,{\bar{y}}|x) \pm\lambda e^{\ga\pb}
\Big (y_\ga \bar{y}_\pb +\frac{\partial^2}{\partial y^\ga
\partial \bar{y}^\pb}\Big ) C{}_\pm  (y,{\bar{y}}|x)\,.
\ee
Analogously to the Minkowski  case,
for any parameter $\eta \in\R_{AdS}$, Eq. (\ref{newtwistjads}) is solved
by the  bilinears
   \be \label{bilinearAdS}
   \PPP(y^\pm\,\by^\pm|x)= \eta\,C_+(y^++y^-,\by^++\by^-| x )C_-(y^+-y^-,\by^+-\by^- | x) \ee
of rank-one fields $C_\pm(\sqrt{2}y\,,\sqrt{2}\by|x)$  that solve the
equations  (\ref{twpm}).

Now we are in a position to consider a deformation
 of the system (\ref{CON1}), (\ref{CON2}) combined with
 the rank-two equations (\ref{newtwistjads}). We will show
 in particular that, upon the bilinear substitution (\ref{bilinear}),
 the constructed deformed system leads to the Maxwell equations with
 nonzero current and to the linearized Einstein equations with a nonzero
 stress tensor.

   \section{Current deformation}
\label{Current deformation}
To describe the current interactions of $4d$ massless fields we
look for a nontrivial deformation of the combination of the
rank-one and rank-two unfolded systems (\ref{CON1}), (\ref{CON2})
and (\ref{newtwistjads}).
The  form of the deformation is fixed by its formal consistency.
The problem is solved in two steps.
 First,   we consider the zero-form sector to find a
 gluing of the rank-two current module to the rank-one Weyl module.
The  result is presented in Section \ref{Current def0} while details of
derivation are given in Appendix A. Second,  the  result for
the gluing in the one-form sector is presented in Section
\ref{Current def1}, while details are given in Appendices B, C  and D.

\subsection{Current deformation in the zero-form sector}
\label{Current def0}
The deformation in the zero-form sector is independent of that in the
one-form sector. On the other hand, because of the $C$--dependent
part of the equation (\ref{CON1}), the form of the deformation
in the zero-form sector affects the deformation in the one-form
sector.

A most general consistent deformation of  the  equations
(\ref{CON2}) by  rank-two fields has the form
\bee\label{newtwC}
D^{tw} C(y,\by|x)
 + e^{ \ga  }{}^{\pa} F \big( \NNN_\pm \,,\overline{\NNN}_\pm\big)
   y^j{}_\ga\bar{\p}_j{}_\pa
 \PPP(y^\pm,\by^\pm|x)\,\Big|_{{y^\pm=\by^\pm=0}}
\qquad\quad\\ \rule{0pt}{22pt}\nn +
 e^{ \ga  }{}^{\pa}\FF\big( \NNN_\pm \,,\overline{\NNN}_\pm\big)
   \by^j{}_\pa{\p}_j{}_\ga
    \TTT{}  (y^\pm,\by^\pm|x)\,\Big|_{{y^\pm=\by^\pm=0}}
 =0
 ,\quad
\eee
where $D^{tw}$ is defined in (\ref{tw})\,, 
    $\PPP(y^\pm\,,\by^\pm)$ and $\TTT(y^\pm\,,\by^\pm)$ are
rank-two fields satisfying unfolded field equations (\ref{newtwistjads}).
 The form of the gluing operators $F$ and $\FF$ is determined by 
 the consistency of Eq.~(\ref{newtwC}) analyzed in detail in
 Appendix A, which is the condition that application of
$D^{tw}$ to (\ref{newtwC}) leads to identity $0=0$ provided that the
current fields $\PPP(y^\pm,\by^\pm|x)$ and $\TTT(y^\pm,\by^\pm|x)$
satisfy the current equation. Here  we use the following notations
 \bee \label{Npm}
   a^j b_j= a^+ b_+-a^- b_- \q
\NNN_\pm=y^\ga \p_\pm{}_\ga \q\quad
\overline{\NNN}_\pm=\bar{y}^\pa \bp_\pm{}_\pa
\,.
 \eee

  The final result is
 \bee   \label{resultFpm}
{F}  \big(\NNN_\pm\,,\overline{\NNN}_\pm\big)= \sum_{m\ge0}
\sum_{ n=0}  ^{m} a_{n,m}
 \FFF {}^{n ,m-n } \big( \NNN_\pm , \overline{\NNN}_\pm \big)
\,,\\
   \label{resultFpmcc}
\FF \big(\NNN_\pm\,,\overline{\NNN}_\pm\big)=
\sum_{m\ge0} \sum_{ n=0}  ^{m}  b_{n,m}
  \overline{\FFF} {}^{n ,m-n } \big( \NNN_\pm , \overline{\NNN}_\pm \big)
  \,,\eee
where $a_{n,m}$ and $b_{n,m} $
are arbitrary coefficients and
\bee \label{GENF_K1}
 \FFF{}^{{n_+} ,{n_-} } \big( \NNN_\pm , \overline{\NNN}_\pm \big)=    \big( \NNN_+\big)^{ {n_+}}
   \big( \NNN_-\big)^{ {n_-}  }
  \sum_{m\ge0\, }
\f{\big(\overline{\NNN}_+\,\NNN_-+ \overline{\NNN}_-\,\NNN_+\big)^{m}}
 { \,m! (m+ {n_+}+{n_-}+1 )!}\, \q\\\nn
\overline{ \FFF}{}^{{n_+} ,{n_-} } \big( \NNN_\pm , \overline{\NNN}_\pm \big)=
\big( \overline{\NNN}_+\big)^{ {n_+} }
   \big( \overline{\NNN}_-\big)^{ {n_-}   }
  \sum_{m\ge0\, }
\f{\big(\overline{\NNN}_+\,\NNN_-+ \overline{\NNN}_-\,\NNN_+\big)^{m}}
 { \,m! (m+ {n_+}+{n_-}+1 )!}\, .\qquad
 \eee

As shown in Appendix D, the fields of the form
  $\PPP=f_-\PPP^\prime$ and $\TTT=f_+\TTT^\prime$,
 give a $D^{tw}-$exact  deformation (\ref{newtwC}) which can be removed by a
 local field redefinition.

 Note that  the functions (\ref{GENF_K1})  express via the
regular Bessel functions  (see, e.g., \cite{bessel})
\be
\label{bess}
  I_{k+1}(2 {x}^{\half})=   {x} ^{ \f{k+1}{2} }\sum_m  \f{ {x} ^{ m }}{m!(m+k+1)!}
\ee
as follows
\bee \nn\FFF {}^{n  ,m}
=
\f{\big( \NNN_+\big)^{  {n}  }  \big( \NNN_-\big)^{m    }}{
\Big(\overline{\NNN}_+\,\NNN_-+ \overline{\NNN}_-\,\NNN_+\Big)
{}^{\f{{ {n} +m +1} }{2}}}\,\,
I_{ {n}+m+1 } \Big(2
\big(\overline{\NNN}_+\,\NNN_-+ \overline{\NNN}_-\,\NNN_+\big)^{\half}\Big) .
\eee

To see the origin of the ambiguity associated with the coefficients
$a_{n,m}$ and $b_{n,m}$ we use that
\bee\label{slhN}
[\hv{}\,,\, N_\pm]=-  N_\pm\q
[\hv{}\,,\, \overline{N}_\pm]=   \overline{N}_\pm\q
[\uu{}\,,\, \overline{N}_\pm]= \mp \overline{N}_\pm\q
[\uu{}\,,\,  {N}_\pm]= \pm  {N}_\pm\q
\\ \nn
[\vv_\pm ,\, N_\pm]=-N_\mp  \q\quad [\vv_\mp,\, N_\pm]=0\q\quad 
[\vv_\pm ,\, \overline{N}_\pm]=\overline{N}_\mp \q \quad
[\vv_\mp ,\, \overline{N}_\pm]=0\,,
\eee
{}from where it follows that
\bee\label{gpmFF}
\Big[g_\pm\,,
  \overline{\NNN}_+\,\NNN_-+ \overline{\NNN}_-\,\NNN_+ \Big]=\Big[\uu \,,
  \overline{\NNN}_+\,\NNN_-+ \overline{\NNN}_-\,\NNN_+ \Big]=\Big[\hv\,,
  \overline{\NNN}_+\,\NNN_-+ \overline{\NNN}_-\,\NNN_+ \Big]=0
 \,,\quad
\eee\bee\label{actingdlhF}
\Big[g_-\,, \FFF {}^{n_+ ,n_-} \Big]=-n_-\FFF {}^{n_++1 ,n_--1} \q&&
\Big[g_+\,, \FFF {}^{n_+ ,n_-} \Big]=-n_+\FFF {}^{n_+-1 ,n_++1}
 \q\,\,\\ \nn
\Big[g_-\,, \overline{\FFF} {}^{\bar{n}_+ ,\bar{n}_-} \Big]= \,\bar{n}_-
\overline{\FFF} {}^{\bn_++1 ,\bn_--1}
\q\,\,\,\,&&
\Big[g_+\,, \overline{\FFF} {}^{\bar{n}_+ ,\bar{n}_-} \Big]= \bar{n}_+
\overline{\FFF }{}^{\bn_+-1 ,\bn_-+1}
.\qquad\qquad
\eee
Here $f_a$ and $g_b$ are generators of $\slv$ (\ref{slv}) and $\slh$ (\ref{slh}), respectively.

On the other hand, the $J$ and $I$--dependent  terms
  of (\ref{newtwC}) are invariant under    the action of
$f_0$ and $g_{j}$ on the variables $y^\pm$ and $\bar y^\pm$
simply because the result   is zero at
$y^\pm =\bar y^\pm=0$.
  (However, this is not the case
for the operators $f_\pm$ which contain second derivatives in
$y^\pm$ and $\bar y^\pm$.)
This means that
 the action of the rank-two helicity   operator   $\hv$ on the gluing functions
is equivalent up to a sign to their action on $J$ and $I$, shifted to $\mp 2$
respectively, since $$[\hv, y^j{}_\ga\bar{\p}_j{}_\pa]=2y^j{}_\ga\bar{\p}_j{}_\pa\q
[\hv,\by^j{}_\pa {\p}_j{}_\ga]=-2\by^j{}_\pa {\p}_j{}_\ga\,.\qquad
$$
For example,
\bee\label{helF} 
 0= \big( \hv  {\FFF} {}^{k ,n }y^j{}_\ga\bar{\p}_j{}_\pa
 \PPP(y^\pm,\by^\pm|x)\,\big)\Big|_{{y^\pm=\by^\pm=0}}
     =\qquad\\ \nn
    = (2-k-n){\FFF} {}^{k,n }  y^j{}_\ga\bar{\p}_j{}_\pa  \PPP(y^\pm,\by^\pm|x)  \Big|_{{y^\pm=\by^\pm=0}}
 +  {\FFF} {}^{k,n } y^j{}_\ga\bar{\p}_j{}_\pa
 \hv   \PPP(y^\pm,\by^\pm|x)  \Big|_{{y^\pm=\by^\pm=0}}
\,.\qquad\eee

 Analogously,
 the action of the horizontal operators  $g_{j}$ on the gluing functions
is equivalent up to a sign to their action on $J$ and $I$ since
the operators $ y^j{}_\ga\bar{\p}_j{}_\pa$
and their complex conjugate  $ \by^j{}_\pa {\p}_j{}_\ga$ are
invariant under $\slh$, namely, for example,
\bee\label{akthor0}\!\!\!
 0= \Big( \vv_\pm\,\, {\FFF} {}^{k ,n }y^j{}_\ga\bar{\p}_j{}_\pa
 \PPP(y^\pm,\by^\pm|x)\,\Big)\Big|_{{y^\pm=\by^\pm=0}}
      =\qquad\qquad\qquad\qquad\\ \nn
    =    [\vv_\pm\,,{\FFF} {}^{k,n }]
    \,y^j{}_\ga\bar{\p}_j{}_\pa  \PPP(y^\pm,\by^\pm|x) \,\Big|_{{y^\pm=\by^\pm=0}}
  + {\FFF} {}^{k,n }y^j{}_\ga\bar{\p}_j{}_\pa
 \vv_\pm   \PPP(y^\pm,\by^\pm|x)  \Big|_{{y^\pm=\by^\pm=0}}\,,
  \\ \nn
 0= \big( \uu  {\FFF} {}^{k ,n }y^j{}_\ga\bar{\p}_j{}_\pa
 \PPP(y^\pm,\by^\pm|x)\big)\,\Big|_{{y^\pm=\by^\pm=0}}
 =\qquad\qquad\qquad\qquad\\ \nn
    =(-k+n  ) {\FFF} {}^{k,n }   y^j{}_\ga\bar{\p}_j{}_\pa  \PPP(y^\pm,\by^\pm|x) \,\Big|_{{y^\pm=\by^\pm=0}}
   + {\FFF} {}^{k,n } y^j{}_\ga\bar{\p}_j{}_\pa
 \uu\,  \PPP(y^\pm,\by^\pm|x) \,\Big|_{{y^\pm=\by^\pm=0}}.
 \eee

Since
$\varphi(f,g)J$ and $\psi(f,g)I$ satisfy the rank-two equation, hence
providing new conserved currents for any functions $\varphi(f,g)J$
and $\psi(f,g)I$, the general deformation (\ref{newtwC}) realizes
a representation of  $\mathfrak{gl}_2$ formed by $\hv$ and $\slh$. Application of
$\hv$ and {$g_{ j}$} to the deformation transforms the coefficients
as finite-dimensional spin-$\half (n+k)$ representations of
$\mathfrak{gl}_2.$
Indeed, the deformation (\ref{newtwC}) for
a spin $s$ rank-one field with currents obeying
\be
\label{cij}
\hv J^{s-1} =  2(s-1)J^{s-1}\q
\hv I^{-s+1} =- 2(s-1)I^{-s+1}\,\q
\ee
is
\bee\label{newtwCs}
D^{tw} C^s(y,\by|x)
 + e^{ \ga  }{}^{\pa} \sum_{m=0}^{2s} a_{m, 2s}{\FFF} {}^{m,2s-m } \big( \NNN_\pm \,,\overline{\NNN}_\pm\big)
  y^j{}_\ga\bar{\p}_j{}_\pa
    \PPP^{s-1}_{(2s-2m)} (y^\pm,\by^\pm|x)\,\Big|_{{y^\pm=\by^\pm=0}}
 =0\q \\ \rule{0pt}{22pt} \label{newtwC-s}
 D^{tw} \overline{C}^{-s}(y,\by|x)
 + e^{ \ga  }{}^{\pa}\sum_{m=0}^{2s} \bar{a}_{m,2s} \overline{{\FFF}} {}^{m,2s-m }\big( \NNN_\pm \,,\overline{\NNN}_\pm\big)
  \by^j{}_\pa{\p}_j{}_\ga
    \TTT^{-s+1}_{(2s-2m)}{}  (y^\pm,\by^\pm|x)\,\Big|_{{y^\pm=\by^\pm=0}}
 =0\,\qquad
  \eee
for $s>0$
and
\bee\label{newtwC0}
D^{tw} C^0(y,\by|x)
 + e^{ \ga  }{}^{\pa}  a_{0,0} {\FFF} {}^{0,0} \big( \NNN_\pm \,,\overline{\NNN}_\pm\big)
  y^j{}_\ga\bar{\p}_j{}_\pa
     {\PPP}^{ -1}_{(0)} (y^\pm,\by^\pm|x)\,\Big|_{{y^\pm=\by^\pm=0}}
+\\ \nn+    e^{ \ga  }{}^{\pa} \bar{a}_{0,0} \overline{{\FFF}} {}^{0,0}\big( \NNN_\pm \,,\overline{\NNN}_\pm\big)
  \by^j{}_\pa{\p}_j{}_\ga
     {\TTT}^1{}_{(0)}  (y^\pm,\by^\pm|x)\,\Big|_{{y^\pm=\by^\pm=0}}
 =0
 \,, \qquad
\eee
for $s=0$.
Here $\PPP^{p}_{(k)}$ satisfies
${\uu\PPP}^{p}_{(k)}=k\,\PPP^{p}_{(k)}$, $\uu\in \slh$ (\ref{slh}),
and  $a_{i,j}$ - arbitrary coefficients.

Since the deformation coefficients form finite-dimensional $\mathfrak{gl}_2$--modules, it suffices to
consider the problem for any element of these modules.
       In  Section \ref{Current def1}
and examples of Section \ref{contribution} we consider ``$\slh$-highest deformations"
 with \,\,\be\label{higslh0}{a_{m, 2s-m}=\gd_m^0 a_{0, 2s}}\q \bar{a}_{m, 2s-m}=\gd_m^0 \bar{a}_{0, 2s} .\ee
For the future convenience we   set
 $a_{0, 2s}=\bar{a}_{0, 2s}=2s+1.$

To define flat limit of the deformed equations (\ref{newtwCs}),  (\ref{newtwC-s}) and  (\ref{newtwC0})
it is necessary to introduce the appropriate $\gl$-depended coefficients
of added deforming terms.
It is evident, that the terms
\bee\label{rescdefs}
  e^{ \ga  }{}^{\pa}{{\FFF}} {}^{m,2s-m }
   y^j{}_\ga\bar{\p}_j{}_\pa \big(f^+\big)^n
   {\PPP}^{s-1}_{(2s-2m)} (y^\pm,\by^\pm|x)\,\Big|_{{y^\pm=\by^\pm=0}}
\eee and
\bee\label{rescdef-s}
  e^{ \ga  }{}^{\pa}{\overline{{\FFF}}} {}^{m,2s-m }
   \by^j{}_\pa{\p}_j{}_\ga \big(f^-\big)^n
    {\TTT}^{1-s}_{(2s-2m)} (y^\pm,\by^\pm|x)\,\Big|_{{y^\pm=\by^\pm=0}}
\eee
require some coefficient $a(\gl^n)$\, to yield
  the coefficient $a(1)$ after the
rescaling (\ref{resc}) in the flat limit $\gl\to 0$.

\subsection{Current deformation in the one-form  sector}
\label{Current def1}

Since zero-forms contribute to the right-hand-sides of the equations (\ref{CON1}),
their formal consistency in presence of the deformation (\ref{newtwC})
requires an appropriate deformation in the one-form sector.
Since the analysis of the deformation in the one-form sector is more
complicated due to the gauge ambiguity, instead of consideration of
the problem in  full generality we use an appropriate Ansatz,
 that not only guarantees formal consistency but also gives rise to
 correct current deformation of the dynamical equations.

The problem is considerably simplified by using the $\mathfrak{gl}_2=f_0\cup\slh$
symmetry acting on the gluing coefficients in (\ref{resultFpm}) and
(\ref{resultFpmcc})  of the deformation (\ref{newtwC}). Indeed it allows us first
to find the  deformation
in the one-form sector in the particular case of
$\slh$ (\ref{slh}) highest-weight coefficients    of the form (\ref{higslh0}) in  (\ref{newtwC}),
 then extending the result to arbitrary gluing
coefficients  by the action
of $\slh$ 
on the gluing functions.

 Here we present the final results of the "highest-weight"
 deformation. Details of their derivation are quite
 complicated and are presented in Appendices B  and C (see pp. \pageref{Appendix 2}, \pageref{Appendix 3} ).

First, for a given spin-$s$ we introduce ``seed current fields"
$\LLLR_{\hel}$ that
 solve
Eq. (\ref{newtwistjads}) and obey  the conditions
\bee\label{primfield}
\hv{}\,\LLLR_{\hel}   (y^\pm\,,\by^\pm|x)
\!\!&=&2 \helh \,\LLLR_{\hel}   (y^\pm\,,\by^\pm|x)
\,,\quad\\\nn
\rule{0pt}{20pt}
\uu
\LLLR_{\hel}
(y^\pm\,,\by^\pm|x)\!\!&=& -2 s \,\LLLR_{\hel}(y^\pm\,,\by^\pm|x) \q
 \eee
where $\hv{}$ (\ref{slv}) is the  rank-two helicity
 operator, $\uu{}$ (\ref{slh}) is the  Cartan operator of
$\slh$,
      $\helh =0$ for integer $s$ and
$ \helh =\pm\half$ for half-integer $s$\,.
 The reality condition requires $\LLLR_{\hel}=\overline{\LLLR}_{-\hel }\,$.

Given an integer spin-$s\ge2$ and a seed current field $\LLLL$,
 the  deformed equation in the one-form sector is
  \bee  \label{CON1new}
 D^{ad}\go(y,{\bar{y}}|x) -   \overline{H}^{\pa\pb}
\bar{\p}{}_\pa\bar{\p}{}_\pb
\overline{C}(0,\by\mid x) -  H^{\ga\gb}
 {\p}{}_\ga {\p}{}_\gb
C (y,0\mid x)\,=\qquad\\\nn=
  {H}^{\ga \gb}{\p}_-{}_\ga {\p}_-{}_\gb
 \sum_{k=0}^{s-2}  \f{ \big(  \NNN_- \big)^{s-k-2}   \,
  \big(\overline{\NNN}_-\big)^{ s+k}}{ (s+k)!  }   \left( f_-\right)^k\LLLL
  \,
  \big|_{y^\pm=\by^\pm=0}+\qquad
 \\ \nn{}+
  \overline{H}^{\pa \pb}\bar{\p}_-{}_\pa\bar{\p}_-{}_\pb 
 \sum_{k=0}^{ s-2 } \f{   \big(  \NNN_- \big)^{  s+k}   \,
 \big(\overline{\NNN}_-\big)^{s-k-2}
}{  (s+k)! }
 \left( f_+\right)^k\LLLL\, \big|_{y^\pm=\by^\pm=0} \,,
 \qquad%
 \eee
  where $f_\pm \in \slv$ (\ref{slv}).

 The associated deformation in the zero-form sector  is
  \bee
 \label{CON2new}
 D^{tw}C(y,\by\mid x)
  +\gl(2s+1)e^{ \gm  }{}^{\pb}
       \FFF {}^{0 ,2s } y^j{}_\ga\bar{\p}_j{}_\pb
    \Big( f_+\Big)^{s- 1} \LLLL \,\big|_{y^\pm=\by^\pm=0}
     &=&0\q
 \,\\\nn
 D^{tw}\overline{C} (y,\by\mid x)
+  \gl(2s+1) e^{ \gm  }{}^{\pb}
     \overline{\FFF }{\,}^{0 ,2s }  {\p}_j{}_\gm \by^j{}_\pb
 \Big( f_-\Big)^{s- 1} \LLLL\,\big|_{y^\pm=\by^\pm=0}
  &=&0\, \q
\eee
where
$ {\FFF }{\,}^{0 ,2s }$ and $\overline{\FFF }{\,}^{0 ,2s }$
are defined in (\ref{GENF_K1}).

Given a half-integer   spin $s=\has+\half$ and seed current fields $\LLLR_{\pm\hell}$,
 the deformed equation in the one-form sector is
    \bee \label{CON1newh}
\!\!\! D^{ad}\go(y,{\bar{y}}|x) =   \overline{H}^{\pa\pb}
 \bar{\p}{}_\pa\bar{\p}{}_\pb
\overline{C}(0,\by\mid x) +  H^{\ga\gb}
{\p}{}_\ga {\p}{}_\gb
C (y,0\mid x)\, \qquad\qquad\qquad\qquad
 \\\nn+\rule{0pt}{28pt}
  {H}^{\ga \gb} {\p}_-{}_\ga {\p}_-{}_\gb\Big\{
 \sum_{k=0}^{ \has-2}  \f{  \big(  \NNN_- \big)^{ \has-k-2}   \,
  \big(\overline{\NNN}_-\big)^{   \has+1+k}}{ (  \has+1+k)! }
     \left( f_-\right)^k\LLLR_{-\hell}+\sum_{k=0}^{ \has-1 }  \f{ \big(  \NNN_- \big)^{ \has-1-k}   \,
  \big(\overline{\NNN}_-\big)^{   \has +k}}{\has (  \has +k)! }
      \left( f_-\right)^k\LLLR_{\hell}
 \Big\}
  \,
  \big|_{y^\pm=\by^\pm=0} \quad
 \\ \nn{}
+  \overline{H}^{\pa \pb}\bar{\p}_-{}_\pa\bar{\p}_-{}_\pb
 \Big\{
\! \sum_{k=0}^{   \has-1}  \f{   \big(  \NNN_- \big)^{   \has+k}
  \big(\overline{\NNN}_-\big)^{  \has -k-1}
}{ \has ( \has+k)!}
  \left( f_+\right)^k\!\LLLR_{-\hell}
  +\!\sum_{k=0}^{(  \has -2)}  \f{   \big(  \NNN_- \big)^{   \has+1+k}
 \big(\overline{\NNN}_-\big)^{  \has -k-2}
}{  ( \has+1+k)!}
  \left( f_+\right)^k\!\LLLR_{\hell}\Big\} \big|_{y^\pm=\by^\pm=0} \,.
  \eee
  The associated
  deformation in the zero-form sector  is
  \bee\label{CON2newh}
D^{tw}C(y,\by\mid x)
 +  \gl(2s+1) e^{ \gm  }{}^{\pb}
     {\FFF }{\,}^{0 ,2s } y^j{}_\ga \bar{\p}_j{}_\pb
      \Big\{
        \Big( f_+\Big)^{ \has-1}\LLLR_{\hell}
  +\f{ 1}{  \has  }\Big( f_+\Big)^{ \has}\LLLR_{-\hell}\Big\}\big|_{y^\pm=\by^\pm=0}
  &=&0
 \,,\qquad\qquad\\\nn
 D^{tw}\overline{C} (y,\by\mid x)
 + \gl(2s+1) e^{ \gm  }{}^{\pb}
     \overline{\FFF }{\,}^{0 ,2s }  {\p}_j{}_\gm \by^j{}_\pb \Big\{
         \,
 \Big( f_-\Big)^{ \has- 1} \LLLR_{ -\hell } \,+    \f{ 1 }{    \has  }       \,
 \Big( f_-\Big)^{ \has} \LLLR_{\hell} \,
\Big\}
\big|_{y^\pm=\by^\pm=0} &=&0\, .\qquad\qquad
\eee

Note that these deformations are  nontrivial provided that the seed current fields
 $\LLLR_{\hel}$ (\ref{primfield})
satisfy
$\LLLR_{\hel}(y^\pm\,,\by^\pm|x)\big|_{y_+=\by_+=0}\ne 0.$

\section{Current contribution to dynamical equations}
\label{contribution}
Let us explain how the deformed  unfolded equations affect the form of dynamical
equations for massless fields.
To obtain usual current interactions, the rank-two
fields should be realized as bilinears of massless fields  
\bee\label{X}
\KKK_0=  C_+\Big( {y^++y^-} , {\by^++\by^-} \Big| x \Big)
 C_-\Big( {y^+-y^-} , {\by^+-\by^-} \Big|  x \Big)\,,
 \eee
 where $C_\pm( \f{1}{\sqrt{2}}y\,,\f{1}{\sqrt{2}}\by|x)$
 solve the rank-one equations  (\ref{twpm}).
For the future convenience we will use the following decompositions
 \bee\label{jdecom}
  A(y^\pm,\by^\pm| x)= \sum_{m,\bm } A^{m\,,\bm}(y^\pm,\by^\pm| x)\q
     B(y ,\by | x)= \sum_{m,\bm } B^{m\,,\bm}(y,\by| x)\,,
  \eee
  where
  \bee \nn
\Big( y^+{}^\gb \f{\p}{\p y^+{}^\gb}+ y^-{}^\gb \f{\p}{\p y^-{}^\gb} \Big)  A^{m\,,\bm}(y^\pm,\by^\pm| x)=m A^{m\,,\bm}(y^\pm,\by^\pm| x)\,,
\qquad
 \\ \nn
\Big( \by^+{}^\pb \f{\p}{\p \by^+{}^\pb}+\by^-{}^\pb \f{\p}{\p \by^-{}^\pb}\Big) A^{m\,,\bm}(y^\pm,\by^\pm| x)=\bm A^{m\,,\bm}(y^\pm,\by^\pm| x)\,,
\qquad
\\\nn\Big( y^\gb \f{\p}{\p y^\gb}\Big)  B^{m\,,\bm}(y,\by| x)=m B^{m\,,\bm}(y,\by| x)\,
 \q
\Big( \by^\pb \f{\p}{\p \by^\pb}\Big)   B^{m\,,\bm}(y,\by| x)=\bm B^{m\,,\bm}(y,\by| x)\eee

\renewcommand{\PPP}{{ \overline{J}}}
\renewcommand{\TTT}{{J}}

  \subsection{Spin zero}
\label{examps0j}
Using (\ref{cij}), consider such $\TTT$ that $\hv{}\TTT = 2\TTT$.
Eq.~(\ref{newtwC0}) with \,\,$ a_{0, 0} =\bar{a}_{0, 0} =1 $
 gives
  \bee\label{CON21j}
  D^{L}{}_{\ga\pa}C(0\,,0| x) +\gl
  C_{\ga\pa}(0\,,0| x) &=&0\,\,,\qquad\\
  \rule{0pt}{22pt} \nn D^{L}{}_{\ga\pa}C_{\gb\pb}(0\,,0| x)  +\gl
  C_{\ga\gb\pa\pb}(0\,,0| x)  +\gl \gvep_{\pa\pb}\gvep_{\ga\gb}C(0\,,0| x)&{}&\qquad\\
  \rule{0pt}{22pt}\nn -
  \f{\gvep_{\pa\pb}}{2}  \Big(\f{\p^2}{\p y^+{}^\gb \p y^-{}^\ga}
  -\f{\p^2}{\p y^-{}^\gb\p y^+{}^\ga}\Big) \TTT{}(y^\pm,0| x)\big|_{y^\pm=\by^\pm=0}\,
  &{}&\\ \nn\rule{0pt}{22pt} -
\f{ \gvep_{\ga\gb}}{2}     \Big( \f{\p^2}{\p \by^+{}^\pb \p \by^-{}^\pa}
   - \f{\p^2}{\p \by^-{}^\pb \p \by^+{}^\pa}\Big)  {\PPP}(0,\by^\pm| x )\big|_{y^\pm=\by^\pm=0}
    &=&0.
 \eee
Hence
  \bee\label{KG}
 D^{L}_{\ga\pa}D^{L}{}^{\ga}{}^{\pa}C(0\,,0| x) =4\gl^2C(0\,,0| x)
 -4  \f{\p^2}{\p y^+{}_\ga\p y^-{}^\ga}\TTT{}(y^\pm,0| x)
  -4  \f{\p}{\p^2 \by^+{}_\pa \p \by^-{}^\pa} {\PPP}(0,\by^\pm| x )
   \,. \eee
{}From  (\ref{X}) we obtain
 \bee\label{KGbe}
 D^{L}_{\ga\pa}D^{L}{}^{\ga}{}^{\pa}C(0\,,0| x) =4\gl^2C(0\,,0| x)
 +4   \overline{C}_+{}_\pa( x )\overline{C}_-{}^\pa( x )
  +4  C_+{}_\ga(x  )C_-{}^\ga(x )
   \,. \eee

Remarkably, in the spin-zero sector, the proposed unfolded construction just
reproduces  Yukawa interaction since $C_{\pm\ga}(x)$ are dynamical spin-$1/2$
fields.  Note that a $C^2$ deformation, that one might naively expect in the
spin-zero  sector, does not appear in agreement with the fact that the
construction of this paper is conformal, while the $C^2$ deformation is not
conformal in four dimensions.

  \subsection{Spin   $1/2$}
\label{exampshalfj}
  Let $\hv{}\TTT = \TTT$.
    Eq. (\ref{newtwCs}), (\ref{newtwC-s}) with \,\,$\dis{a_{m, 2s-m}=
    \bar{a}_{m, 2s-m}} =2\gd_m^0 $
  give
   \bee\label{CON22j1/2}
D^{L}{}_{\ga\pa}C_\gm(0\,,0| x) +\gl
C_{\gm\ga\pa}(0\,,0| x)+  \gvep_{\gm\ga}  \f{\p}{\p \by^-{}^\pa}{\PPP}(0,\by^-|x )\big|_{\by^-=0} =0 \,\,,
\\ \nn
 D^{L}{}_{\ga\pa}\overline{C}{}_\pmm(0\,,0| x) +\gl
 \overline{C}{}_{ \ga\pmm\pa}(0\,,0| x) +  \gvep_{\pmm\pa}
 \f{\p}{\p  y^-{}^\ga}{\TTT{}}(y^-\,,0 |x)\big|_{y^-=0} =0 \,.
\eee
    From (\ref{CON22j1/2}) it follows
\bee\label{dirj}
D^{L}{}_{\ga\pa}C^\ga(0\,,0| x)  - 2   \f{\p}{\p \by^-{}^\pa}{\PPP}(0,\by^- |x)\big|_{\by^-=0}  =0
\q \\ \nn
 D^{L}{}_{\ga\pa}\overline{C}{}^\pa(0\,,0| x)-  2   \f{\p}{\p  y^-{}^\ga}{\TTT{}}(y^-\,,0|x )\big|_{ y^-=0} =0\,.\qquad
\eee
Substitution of bilinear
  $\PPP\,$ and $\TTT\,$ (\ref{X}), built   from     fermions    and bosons,   gives
\bee\label{dirjbe}\!\!
&&D^{L}{}_{\ga\pa}C^\ga(  x)  -\!   \sqrt{2}  \overline{{C}}_+{}_\pa(x ) \overline{{C}}_- (x )+
\!   \sqrt{2}\overline{{C}}_+ (x ) \overline{{C}}_-{}_\pa(x )  =0
\q  \\ \nn&&\rule{0pt}{22pt}
 D^{L}{}_{\ga\pa}\overline{C}^\pa(   x)-     \sqrt{2}C_+{}_\ga( x ) C_-( x )+
     \sqrt{2}  C_+( x ) C_-{}_\ga( x )   =0\,,
\eee
which is  Yukawa interaction in the spin-$1/2$ sector.

\subsection{Maxwell equations}
  \label{s=1}
 Let  $\hv{}\TTT =  0$. Then the reality condition  requires $\PPP=\TTT.$
  Eq.
      (\ref{CON1}) still reads as
\bee\label{CON11}
 D^{ad}\go( x) =
   \overline{H}^{\pa\pb}
\overline{C}_{\pa\pb}( x) +  H^{\ga\gb}
C_{\ga\gb}( x)\, .\eee
This identifies $C_{\ga\gb}(x)$ and $\overline{C}_{\pa\pb}( x)$
with selfdual and anti-selfdual parts of the Maxwell field strength.
The consistency conditions of   (\ref{CON11}) imply the Bianchi identities
 \bee \label{Max1}
D^{ad}
\big(
 {H}^{\ga \gb}{C }_{\ga \gb}(x)+\overline{H}^{\pa \pb}
\overline{C}_{\pa \pb}(x) \big)&=&0\,.
\eee
Deformed  equation (\ref{CON2new}) for $s=1$ at $y=\by=0$ gives
   \bee\label{CON22j}
D^{L}{}_{\ga\pa}C_{\gm\gn}(0\,,0| x)\, +\,\gl
C_{\gm\gn\ga\pa}(0\,,0| x)\qquad\qquad\qquad\qquad\\ \nn
+
\Big( \gvep_{\gm\ga}  \f{\p^2}{\p \by^-{}^\pa \p y^-{}^\gn }{\TTT} (y^\pm,\by^\pm|x) +
  \gvep_{\gn\ga}  \f{\p^2}{\p \by^-{}^\pa \p y^-{}^\gm }{\TTT} (y^\pm,\by^\pm|x)
\Big)\big|_{y^\pm=\by^\pm=0}\,
 =0\,.\quad\eee
 {}From (\ref{CON22j})    it follows that, in accordance with the decompositions (\ref{jdecom}),
\bee\label{CON22jrez}
D^{L}{}^{\gm}{}_{\pa}C_{\gm\gn}(0\,,0| x)
+ 3\gl \f{\p^2}{\p \by^-{}^\pa \p y^-{}^\gn }{\TTT}^{1,1}(y^\pm,\by^\pm|x)
 =0\,.\eee
 By virtue of (\ref{CON22jrez}) along with the  identities
\be\label{iden2-3}
H^{\ga\gb}\wedge e^{\mu \pmm} = \epsilon^{\ga\mu} \Hh^{\gb\pmm}+
\epsilon^{\gb\gm} \Hh^{\ga\pmm}\q
\overline{H}^{\pa\pb}\wedge e^{\gm \pmm} =
-\epsilon^{\pa\pmm} \Hh^{\gm\pb}-
\epsilon^{\pb\pmm} \Hh^{\gm\pa}\,,
\ee
we have
 \bee \nn
  {H}^{\ga \gb}e^{\gn\pn}D^{L}{}_{\gn\pn}{C }_{\ga \gb}(x)=2{\Hh}^{\gb \pn}
  D^{L}{}^{\ga}{}_{\pn}{C }_{\ga \gb}=
-6 \gl{\Hh}^{\gb \pn} \f{\p^2}{\p  y^-{}^\gb \p y^-{}^\pn }{\TTT}^{1,1}(y^\pm,\by^\pm|x).
\eee
 Analogously,
 \bee \nn
  \overline{H}^{\pa \pb}D^{L}
\overline{C}_{\pa \pb}(x)  =
6\gl {\Hh}^{\gb \pn} \f{\p^2}{\p  y^-{}^\gb \p y^-{}^\pn }{\TTT}^{1,1}(y^\pm,\by^\pm|x).
\eee
Hence it follows that, as anticipated, the Bianchi identities (\ref{Max1}) are respected and
  \bee \label{Max2}
D^{L}
\left(
 {H}^{\ga \gb}{C}_{\ga \gb}(x)-\overline{H}^{\pa \pb}
\overline{C}_{\pa \pb} (x)\right)&=&-12\gl {\Hh}^{\gb \pn}
\f{\p^2}{\p  y^-{}^\gb \p y^-{}^\pn }{\TTT}^{1,1}(y^\pm,\by^\pm|x)\,.
\eee
This just reproduces the  Maxwell equations with a nonzero current.

  For $\TTT\,$ (\ref{X}) built from scalars and spinors we have,
  respectively,
\bee \nn
 {\Hh}^{\gb \pn}\f{\p^2}{\p  y^-{}^\gb \p y^-{}^\pn }
 \TTT\,^{1,1}(y^\pm\,,\by^\pm|x)
 =
\f{1}{3\gl}  {\Hh}^{\gb \pn}
 \Big( - C_-( x)\f{\p}{\p x^\gb{}^\pn}{}C_+{}(x )
+C_+(x )\f{\p}{\p x^\gb{}^\pn}{}C_-{} ( x)\, \Big)\,   ,
\eee
 \bee \nn
 {\Hh}^{\gb \pn}
 \f{\p^2}{\p  y^-{}^\gb\p y^-{}^\pn }
 \TTT\,^{1,1}(y^\pm\,,\by^\pm|x) =
\f{-1}{3\gl} {\Hh}^{\gb \pn}C_+{}_\gb(x )\overline{C}_-{}_\pn(x)
 \,,
 \eee
    which are the standard expressions for spin-one currents.

\renewcommand{\KKK}{ {\mathcal{J}_{+}}}
\renewcommand{\RRR}{ {\mathcal{J}_{-}}}%

   \subsection{Spin $3/2$}
\label{examps=1.5j}
Using the decomposition (\ref{jdecom}), from Eq.~(\ref{CON1newh})
 we have 
 \bee
 \label{level-01.5}
D^L \go^{0\,,1}(0,\by)-\gl e^{\gb\pb} \by_\pb\f{\p}{\p y^\gb} \go^{1\,,0}(y,0|x)
   =\qquad\qquad\\ \nn=
\overline{H}^{\pa\pb}\f{\p^2}{\p \by^\pa \p \by^\pb  }
   \overline{C}(0\,,\by|x)+  2  H^{\ga\gb}
\by^\pb\f{\p}{\p \by^-{}^\pb}\f{\p^2}{\p {y}^-{}^{\ga} \p {y}^-{}^{\gb} }
 \KKK^{2,1}(y^\pm,\by^\pm|x)
\,,\\
   \label{level+01.5}
D^L \go^{1\,,0}(y,0|x)-\gl e^{\gb\pb}  y_\gb\f{\p}{\p \by^\pb} \go^{0\,,1}(0,\by|x)
=\qquad\qquad\\ \nn=
H^{\ga\gb}\f{\p^2}{\p  y^\ga \p  y^\gb  }  C(y\,,0|x)
+2    \overline{H}^{\pa\pb}y^\gb\f{\p}{\p y^-{}^\gb}
\f{\p^2}{ \p \by^-{}^{\pa} \p \by^-{}^{\pb}}
 \RRR^{1, 2}(y^\pm,\by^\pm|x)\,.
\quad\eee
 Substituting $$\go^{j\,,k}{}=e^\ga{}^{\pb}\go^{j\,,k}{}_\ga{}_{\pb} 
 $$
into (\ref{level-01.5}), (\ref{level+01.5}) we obtain spin-$ 3/2$
 massless equations in $AdS_4$ in the form
\bee
 \label{RarSh1}
     D^L{}_{\gb\pb} \go^{0\,,1}{}_\ga{}^{\pb}(0,\by)
  -\gl   \by_\pb\f{\p}{\p y^\gb} \go^{1\,,0}{}_\ga{}^{\pb}(y,0|x)  = 2
\by^\pb\f{\p}{\p \by^-{}^\pb}\f{\p^2}{\p {y}^-{}^{\ga} \p {y}^-{}^{\gb}  }
 \KKK^{ 2,1}(y^\pm,\by^\pm|x) \,,\quad\\
   \nn
   D^L{}_{\gb\pb} \go^{1\,,0}{}^\gb{}_{\pa}(y,0|x)
 -\gl    y_\gb\f{\p}{\p \by^\pb} \go^{0\,,1}{}^\gb{}_{\pa}(0,\by|x)
 =2 y^\gb\f{\p}{\p y^-{}^\gb}
\f{\p^2}{ \p \by^-{}^{\pa} \p \by^-{}^{\pb}}
 \RRR^{1, 2}(y^\pm,\by^\pm|x)
 \,.
 \quad\eee
Substitution of the bilinear current
$\KKK=\overline{\RRR}\,$ 
 (\ref{X})   gives
 \bee\label{RarSh1bil}
    \f{\p}{\p {\by}^-{}^{\pn}}
     D^L{}_{\ga\pb} \go^{0\,,1}{}_\ga{}^{\pb}(0,\by)
+\gl    \f{\p}{\p y^\ga} \go^{1\,,0}{}_\ga{}_{\pn}(y,0|x)   \\ \nn=
\sqrt{2}
\Big(-C^{2,0}_+{}_{\ga\ga}\big( 0 , 0 \big| x \big)
 \overline{C}^{0,1}_-{}_{\pn}\big( 0 , 0 \big| x \big)  -
  C^{0,0}_+{}_{\ga\pn}\big( 0 , 0 \big| x \big)
 C^{1,0}_-{}_{\ga}\big( 0 , 0 \big| x \big)\Big)
 \,+( +  \leftrightarrow- ),\quad\\
   \nn
  \f{\p}{\p {y}^-{}^{\gn}}
 D^L{}_{\gb\pb} \go^{1\,,0}{}^\gb{}_{\pa}(y,0|x)
 +\gl     \f{\p}{\p \by^\pb} \go^{0\,,1}{}_\gn{}_{\pa}(0,\by|x)
 \\ \nn=
\sqrt{2}
\Big(-\overline{C}^{0,2}_+{}_{\pa\pa}\big( 0 , 0 \big| x \big)
 C^{1,0}_-{}_{\gn}\big( 0 , 0 \big| x \big)  -
  \overline{C}^{0,0}_+{}_{\gn\pa}\big( 0 , 0 \big| x \big)
 \overline{C}^{0,1}_-{}_{\pa}\big( 0 , 0 \big| x \big)\Big)
+( +  \leftrightarrow- ).\eee
This is the Rarita-Schwinger equation with the  super-current built
from a scalar and spinor.
\renewcommand{\KKK}{ {\mathcal{J}_{0}}}
\renewcommand{\RRR}{\mathcal{I}}%

   \subsection{Spin two}
\label{examps=20j}
In  the case of $s=2$
from the conditions (\ref{primfield})  and (\ref{cij})
it follows that $\hv{}\,\KKK=0  $ and
 $\big(  y^-{}^\ga{\p_-{}_\ga}+ \by^-{}^\pa{\bar{\p}_-{}_\pa}-4\big)\KKK
(y^\pm\,,\by^\pm|x)\big|_{y_+=\by_+=0}= 0
$.
  From Eq.~(\ref{CON1new}), we hence  obtain
\bee\label{CON1news2}
&&  D^{ad}\go(y,{\bar{y}}|x) =
   \overline{H}^{\pa\pb}
\f{\p^2}{\p \by^{\pa} \p \by^{\pb}}
\overline{C}(0,\by\mid x) +  H^{\ga\gb}
\f{\p^2}{\p {y}^{\ga} \p {y}^{\gb}}
C (y,0\mid x)\,+\\\nn&+&\f{1}{2}
    \overline{H}^{\pa\pb}
\f{\p^2}{\p \by^-{}^{\pa} \p \by^-{}^{\pb}}
    \big(\NNN_-\big)^{2 }
\KKK (y^\pm,\by^\pm|x) \big|_{    y^\pm=\by^\pm=0}
+  \f{1}{2} H^{\ga\gb}
\f{\p^2}{\p {y}^-{}^{\ga} \p {y}^-{}^{\gb}}
 \big(  \overline{\NNN}_- \big)^{2}  \KKK (y^\pm,\by^\pm|x)\big|_{    y^\pm=\by^\pm=0}
\,.\eee
In accordance with the decompositions   (\ref{jdecom}), this gives
\bee\label{level0j}
D^L \go^{1\,,1}(y,\by|x)&=& \gl e^{\ga\pb} \by_\pb\f{\p}{\p y^\ga} \go^{2\,,0}(y,0|x)+\gl
e^{\ga\pb}{y}_\ga\f{\p}{\p \bar{y}^\pb }  \go^{0\,,2}(0,\by|x)\,,\\
 \label{level-1j}
 D^L \go^{0\,,2}(0,\by)&=&\gl e^{\ga\pb} \by_\pb\f{\p}{\p y^\ga} \go^{1\,,1}(y,\by|x)
 +\overline{H}^{\pa\pb}\f{\p^2}{\p \by^\pa \p \by^\pb  }   \overline{C}(0\,,\by|x)
+
\\\nn&+&
   H^{\ga\gb}\by^{\pa}  \by^{\pb}
\f{\p^2}{\p \by^-{}^{\pa} \p \by^-{}^{\pb}}
\f{\p^2}{\p {y}^-{}^{\ga} \p {y}^-{}^{\gb}}
   \,\KKK^{2,2}(y^\pm,\by^\pm|x)
\,,\\
   \label{level+1j}
D^L \go^{2\,,0}(y,0|x)&=&\gl e^{\ga\pb}  y_\ga\f{\p}{\p \by^\pb} \go^{1\,,1}(y,\by|x)\,
+H^{\ga\gb}\f{\p^2}{\p  y^\ga \p  y^\gb  } C(y\,,0|x)
+\\\nn&+&
    \overline{H}^{\pa\pb} y^{\ga}  y^{\gb}\f{\p^2}{\p {y}^-{}^{\ga} \p {y}^-{}^{\gb}}
\f{\p^2}{\p \by^-{}^{\pa} \p \by^-{}^{\pb}}
    \,  \,\KKK^{2,2} (y^\pm,\by^\pm|x).\eee
Introducing
 $\go^{j\,,k}{}=e^\ga{}^{\pb}\go^{j\,,k}{}_\ga{}_{\pb} $,{} from Eq.~(\ref{level0j})
 we obtain
\bee \label{spin2ur1j}
 D^L{}_{\gb\pb} \go^{1\,,1}{}_{\gb}{}^{\pb}(y,\by|x)=
 \gl   \by_\pb\f{\p}{\p y^\gb} \go^{2\,,0}{}_{\gb}{}^{\pb}(y,0|x)&+&\gl
 {y}_\gb\f{\p}{\p \bar{y}^\pb }  \go^{0\,,2}{}_{\gb}{}^{\pb}(0,\by|x)\,,
 \eee
 \bee \label{spin2ur2j}
    D^L{}_{\gb\pb} \go^{1\,,1}{}^{\gb}{}_{\pb}(y,\by|x)=
 \gl  \by_\pb\f{\p}{\p y^\gb} \go^{2\,,0}{}^{\gb}{}_{\pb}(y,0|x)&+&\gl
 {y}_\gb\f{\p}{\p \bar{y}^\pb }  \go^{0\,,2}{}^{\gb}{}_{\pb}(0,\by|x)
\,.
 \eee
   Eq.~(\ref{level-1j})    gives (omitting the arguments)
\bee\!\!\!
 \label{spin2ur3j}
  \!D^L{}_{\gb\pb} \go^{0\,,2}{}_{\gb}{}^{\pb}&=&  \by^{\pa}  \by^{\pb}
\f{\p^2}{\p \by^-{}^{\pa} \p \by^-{}^{\pb}}
\f{\p^2}{\p {y}^-{}^{\gb} \p {y}^-{}^{\gb}}
   \,\KKK^{2,2}
\!
+\gl   \by_\pb\f{\p}{\p y^\gb}
 \go^{1\,,1}{}_{\gb}{}^{\pb}
   \,,\quad\\
   \label{spin2ur4j}\!\!\!
\!D^L{}_{\gb\pb} \go^{2\,,0}{}^{\gb}{}_{\pb}
&=&
      y^{\ga}  y^{\gb}\f{\p^2}{\p {y}^-{}^{\ga} \p {y}^-{}^{\gb}}
\f{\p^2}{\p \by^-{}^{\pb} \p \by^-{}^{\pb}}
    \,  \,\KKK^{2,2} 
\!+\gl   y_\gb
\f{\p}{\p \by^\pb} \go^{1\,,1}{}^{\gb}{}_{\pb}
\,.
\quad
\eee

 The equations (\ref{spin2ur1j}) and (\ref{spin2ur2j}) express the Lorentz connection $\go^{2,0}$ and $\go^{0,2}$
via derivatives of the vierbein $\go^{1,1}$ while the equations (\ref{spin2ur3j})
and(\ref{spin2ur4j})  contain the
Bianchi identities for Eq. (\ref{level0j})
 \bee
   \label{spin2ur4jj}&&\f{\p^2}{\p \by^{\pn} \p \by^{\pn}}
D^L{}_{\gb\pb} \go^{0\,,2}{}_{\gb}{}^{\pb}(0,\by|x)
  = \f{\p^2}{\p {y}^{\gb} \p {y}^{\gb}}
D^L{}_{\gn\pn} \go^{2\,,0}{}^{\gn}{}_{\pn}(y,0|x) \,,\quad
 \eee
 and the linearized Einstein equations
 \be
 \label{spin2ur3jj}
 \f{\p^2}{\p \by^{\pn} \p \by^{\pn}}
D^L{}_{\gb\pb} \go^{0,2}{}_{\gb}{}^{\pb}(0,\by|x)
- 2\gl   \f{\p^2 }{\p \by^{\pn} \p y^\gb}
 \go^{1 ,1}{}_{\gb}{}_{\pn}(y,\by|x)
  = 2 
\f{\p^2}{\p \by^-{}^{\pn} \p \by^-{}^{\pn}}
\f{\p^2}{\p {y}^-{}^{\gb} \p {y}^-{}^{\gb}}
  \KKK^{2,2}(y^\pm,\by^\pm|x)
  , \, \ee
which contain the contribution of the stress tensor.

Substitution of the bilinear $\KKK\,$
 (\ref{X})   gives linearized Einstein equations
\bee\nn
 \f{\p^2}{\p \by^{\pn} \p \by^{\pn}}
D^L{}_{\gb\pb} \go^{0,2}{}_{\gb}{}^{\pb}(0,\by)
- 2\gl   \f{\p }{\p \by^{\pn}  }
\f{\p}{\p y^\gb}
 \go^{1 ,1}{}_{\gb}{}_{\pn}(y,\by|x)
  =\qquad\qquad\\ \nn = 2
   \Big(C^{2,0}_+{}_{\ga\ga}\big( 0 , 0 \big| x \big)
   \overline{C}^{0,2}_-{}_{\pa\pa}\big( 0 , 0 \big| x \big)
+C^{1,0}_+{}_{\ga\ga \pa}( 0 , 0 \big| x \big)\overline{C}^{0,1}_-{}_{\pa}
\big( 0 , 0 \big| x \big)
+\\ \nn
+ C^{0,0}_+{}_{\ga\pa}\big( 0 , 0 \big| x \big)
\overline{C}^{0,0}_-{}_{\ga\pa}\big( 0 , 0 \big| x \big)
 +\big( + \leftrightarrow- \big)\Big) \,
 \eee
with  the stress tensor of  massless fields of spins $0,\,\,1/2$
and 1 (recall that $C^{2,0}_+{}_{\ga\ga}\big( 0 , 0 \big| x \big)$ and
$\overline{C}^{0,2}_-{}_{\pa\pa}\big( 0 , 0 \big| x \big)$ describe the selfdual
and anti-selfdual combination of the spin-one field strength).
 \subsection{  Higher  spins }
\label{examps=s0j}
 \subsubsection{Integer spins}
For any integer $s\geq 2$ and real seed current field $\KKK{}=\overline{\KKK }{}$,
 we should obtain equations for the
components $\go_{\ga\pa}{}^{m\,,n}$ of  $\go ^{m\,,n}=e^{\ga\pa}\go_{\ga\pa}{}^{m\,,n}$.
In particular, for $m=s-1-k$, $n=s-1+k$ \,,   $k=-1,0,1$, using the
decomposition (\ref{jdecom}) for $\go$, it follows  from (\ref{CON1new}) that
 \bee\label{levels0}
D^L \go^{s-1\,,s-1}(y,\by|x)
=
 \gl e^{\ga\pb} \by_\pb\f{\p}{\p y^\ga} \go^{s\,,s-2}(y,\by|x) \!+\!\gl
e^{\ga\pb}{y}_\ga\f{\p}{\p \bar{y}^\pb }  \go^{s-2\,,s}(y,\by|x)
\,,
\\ \label{level+1s}
D^L \go^{s\,,s-2}(y,\by|x)=\gl
e^{\ga\pb}{y}_\ga\f{\p}{\p \bar{y}^\pb }  \go^{s-1\,,s-1}(y,\by|x)
 + \qquad\qquad\qquad\qquad \qquad\\ \nn+
 \gl e^{\ga\pb} \by_\pb\f{\p}{\p y^\ga} \go^{s+1\,,s-3}(y,\by|x)
+  \overline{H}^{\pa\pb}
\f{\p^2}{\p \by^-{}^{\pa} \p \by^-{}^{\pb}}
\f{{1}}{ s!   } \big(  \NNN_- \big)^{s }   \,
 \big(\overline{\NNN}_-\big)^{ s-2}\,   \KKK{}^{s,s}(y^\pm,\by^\pm|x)
\,,\\ \nn\\
 \label{level-1s}\!
D^L \go^{s-2\,,s}(y,\by|x)=\gl e^{\ga\pb} \by_\pb\f{\p}{\p y^\ga} \go^{s-1\,,s-1}(y,\by|x)
 +\qquad\qquad\qquad\qquad\qquad \\\nn+
 \gl e^{\ga\pb}{y}_\ga\f{\p}{\p \bar{y}^\pb }  \go^{s-3\,,s+1}(y,\by|x)\!+  H^{\ga\gb}
\f{\p^2}{\p {y}^-{}^{\ga} \p {y}^-{}^{\gb}}
   \f{{1}}{ s!    } \big(  \NNN_- \big)^{s-2}   \,
 \big(\overline{\NNN}_-\big)^{s }
     \KKK{}^{s,s}(y^\pm,\by^\pm|x)\, .
\quad\eee

{}From here it follows that (omitting the arguments)
\bee\label{Levs0}
e^{\mu\pmm}e^{\nu\pn}D^L_{\mu\pmm} \go^{s-1\,,s-1}{}_{\nu\pn}
=
 \gl e^{\ga\pb}e^{\nu\pn} \by_\pb\f{\p}{\p y^\ga} \go^{s\,,s-2}{}_{\nu\pn}  \!+\!\gl
e^{\ga\pb}e^{\nu\pn}{y}_\ga\f{\p}{\p \bar{y}^\pb }  \go^{s-2\,,s}{}_{\nu\pn}
\,,
\\ \label{Lev+1s}
e^{\mu\pmm}e^{\nu\pn}D^L_{\mu\pmm}  \go^{s\,,s-2}{}_{\nu\pn} =\gl
e^{\ga\pb}e^{\nu\pn}{y}_\ga\f{\p}{\p \bar{y}^\pb }  \go^{s-1\,,s-1}{}_{\nu\pn}
 + \qquad\qquad\qquad\qquad \qquad\\ \nn+
 \gl e^{\ga\pb}e^{\nu\pn} \by_\pb\f{\p}{\p y^\ga} \go^{s+1\,,s-3}{}_{\nu\pn}
+  \overline{H}^{\pa\pb}
\f{\p^2}{\p \by^-{}^{\pa} \p \by^-{}^{\pb}}
\f{{1}}{ s ! {}  } \big(  \NNN_- \big)^{s}   \,
 \big(\overline{\NNN}_-\big)^{ s-2}\,   \KKK{}^{s,s}
\,,\\ \nn\\
 \label{Lev-1s}\!
e^{\mu\pmm}e^{\nu\pn}D^L_{\mu\pmm}   \go^{s-2\,,s}{}_{\nu\pn} =\gl e^{\ga\pb} \by_\pb\f{\p}{\p y^\ga} \go^{s-1\,,s-1}
 +\qquad\qquad\qquad\qquad\qquad \\\nn+
 \gl e^{\ga\pb}e^{\nu\pn}{y}_\ga\f{\p}{\p \bar{y}^\pb }  \go^{s-3\,,s+1}{}_{\nu\pn} \!+  H^{\ga\gb}
\f{\p^2}{\p {y}^-{}^{\ga} \p {y}^-{}^{\gb}}
   \f{{1}}{ s!  {}  } \big(  \NNN_- \big)^{s-2}   \,
 \big(\overline{\NNN}_-\big)^{s}
     \KKK{}^{s,s}\, .
\quad\eee
Hence
\bee\label{Luvs0}
  D^L_{\ga\pmm} \go^{s-1\,,s-1}{}_{\ga}{}^{\pmm}
=
 \gl   \by_\pb\f{\p}{\p y^\ga} \go^{s\,,s-2}{}_{\ga}{}^{\pb}  \!+\!\gl
 {y}_\ga\f{\p}{\p \bar{y}^\pb }  \go^{s-2\,,s}{}_{\ga}{}^{\pb}
\,,\\ \nn
 D^L_{\mu\pb} \go^{s-1\,,s-1}{}^{\mu}{}_{\pb}
=
 \gl  \by_\pb\f{\p}{\p y^\ga} \go^{s\,,s-2}{}^{\ga}{}_{\pb}  \!+\!\gl
 {y}_\ga\f{\p}{\p \bar{y}^\pb }  \go^{s-2\,,s}{}^{\ga}{}_{\pb}
\,,
\eee

\bee \label{Luv+1s}
 D^L_{\ga\pmm}  \go^{s\,,s-2}{}_{\ga}{}^{\pmm} =\gl
 {y}_\ga\f{\p}{\p \bar{y}^\pb }  \go^{s-1\,,s-1}{}_{\ga}{}^{\pb}
 +
 \gl  \by_\pb\f{\p}{\p y^\ga} \go^{s+1\,,s-3}{}_{\ga}{}^{\pb}\,,\quad
\\ \nn
 D^L_{\mu\pb}   \go^{s-2\,,s}{}^{\mu}{}_{\pb}
=\gl  \by_\pb\f{\p}{\p y^\ga} \go^{s-1\,,s-1}{}^{\ga}{}_{\pb}
 +
 \gl {y}_\ga\f{\p}{\p \bar{y}^\pb }  \go^{s-3\,,s+1}{}^{\ga}{}_{\pb}\,
\,,\quad\eee

\bee
 \label{Luv-1s}\!
 D^L_{\ga\pmm}   \go^{s-2\,,s}{}{}_{\ga}{}^{\pmm}
=\gl   \by_\pb\f{\p}{\p y^\ga} \go^{s-1\,,s-1}{}_{\ga}{}^{\pb}
 +
 \gl  {y}_\ga\f{\p}{\p \bar{y}^\pb }  \go^{s-3\,,s+1}{}_{\ga}{}^{\pb}+  \\\nn+
\f{\p^2}{\p {y}^-{}^{\ga} \p {y}^-{}^{\ga}}
   \f{{1}}{s!   {}  } \big(  \NNN_- \big)^{s-2}   \,
 \big(\overline{\NNN}_-\big)^{s}
     \KKK{}^{s,s}\, ,
\quad\\
 D^L_{\mu\pb}  \go^{s\,,s-2}{}^{\mu}{}_{\pb} =\gl
 {y}_\ga\f{\p}{\p \bar{y}^\pb }  \go^{s-1\,,s-1}{}^{\ga}{}_{\pb}
 +
 \gl   \by_\pb\f{\p}{\p y^\ga} \go^{s+1\,,s-3}{}^{\ga}{}_{\pb}
 +\\ \nn
+\f{\p^2}{\p \by^-{}^{\pb} \p \by^-{}^{\pb}}
\f{{1}}{ s!  {}  } \big(  \NNN_- \big)^{s }   \,
 \big(\overline{\NNN}_-\big)^{ s-2}\,   \KKK{}^{s,s}
\,. \nn
\eee
Substitution of the bilinear $\KKK $
 (\ref{X})
   gives
 \bee
  \label{bilis}\!
 D^L_{\ga\pmm}   \go^{s-2\,,s}{}{}_{\ga}{}^{\pmm}
=\gl   \by_\pb\f{\p}{\p y^\ga} \go^{s-1\,,s-1}{}_{\ga}{}^{\pb}
 +
 \gl  {y}_\ga\f{\p}{\p \bar{y}^\pb }  \go^{s-3\,,s+1}{}_{\ga}{}^{\pb}
 +\qquad\\\nn+
\f{\p^2}{\p {y}^-{}^{\ga} \p {y}^-{}^{\ga}}
   \f{{\big(  \NNN_- \big)^{s-2}
 \big(\overline{\NNN}_-\big)^{s}}}{s!  {}  }\!\!
 \sum_{p ,\,\,  n+m=s-p}    \!\!  \Big(C^{p+n, n }_+{}  \big(  y_- , \by_-\big| x \big)
   {C}^{m, p+m}_-{}  \big( -y_- , -\by_- \big| x \big)
  +cc\Big)\big|_{y_-=\by_-=0} \,\, ,
\quad\\
\label{bilisc}
 D^L_{\mu\pb}  \go^{s\,,s-2}{}^{\,\,\mu}{}_{\pb} =\gl
 {y}_\ga\f{\p}{\p \bar{y}^\pb }  \go^{s-1\,,s-1}{}^{\ga}{}_{\pb}
 +
 \gl   \by_\pb\f{\p}{\p y^\ga} \go^{s+1\,,s-3}{}^{\ga}{}_{\pb}
 +\qquad\\\nn+ \f{\p^2}{\p \by^-{}^{\pb} \p \by^-{}^{\pb}}
\f{ \big(  \NNN_- \big)^{s}
 \big(\overline{\NNN}_-\big)^{ s-2}}{s!   }\!\!
   \sum_{p ,\,\,  n+m=s-p}   \!\!   \Big(C^{p+n, n }_+{}  \big(  y_- , \by_-\big| x \big)
   {C}^{m, p+m}_-{}  \big( -y_- , -\by_- \big| x \big)
  +cc
\Big)\big|_{y_-=\by_-=0} \,.
\quad
 \nn \eee

To obtain the dynamical spin-$s$ equations with the current corrections
it remains to project out the terms, that contain $\omega^{s-3,s+1}$ and
$\omega^{s+1,s-3}$. This is achieved by the contraction of free indices in
(\ref{bilis}) with $y^\ga y^\ga$ and in (\ref{bilisc}) with $\bar{y}^\pb \bar{y}^\pb$.
The resulting equations describe the contribution of HS currents of
\cite{Berends:1985xx} to the right-hand-sides of Fronsdal's equations
in $AdS_4$.

That the currents do not contribute to the equations (\ref{Luv+1s})
is a manifestation of conformal invariance of the currents
which, being traceless, cannot contribute to the
trace part of the Fronsdal equations contained in Eq.~(\ref{Luv+1s}).

\renewcommand{\KKK}{ {\mathcal{J}}}
\renewcommand{\RRR}{\mathcal{I}}%

  \subsubsection{Half-integer spins}
Using  the decomposition   (\ref{jdecom}),
 from (\ref{CON1newh})
 we obtain for a half--integer $s$ 
\bee
  \label{level-0p.5}
D^L \go^{{[s]-1}\,,{[s]}}(y,\by|x)=
\gl e^{\ga\pb}{y}_\ga\f{\p}{\p \bar{y}^\pb }  \go^{{[s]}-2\,,{[s]}+1}(y,\by|x)+
\gl e^{\ga\pb} \by_\pb\f{\p}{\p y^\ga}  \go^{{[s]}\,,{[s]-1}}(y,\by|x)
+\\ \nn +
 {H}^{\ga \gb}\f{\p^2}{\p y_-{}^\ga \p y_-{}^\gb}
    \f{ \big(  \NNN_- \big)^{[s]-1 }   \,
  \big(\overline{\NNN}_-\big)^{  [s]  }}{ [s]\, [s]  ! }
      \KKK_{+{}}^{[s]+1,[s] }(y^\pm,\by^\pm|x)
  \,,\\
\label{level+0p.5}
D^L \go^{{[s]}\,,{[s]-1}}(y,\by|x)=
\gl e^{\ga\pb} \by_\pb\f{\p}{\p y^\ga}  \go^{{[s]}+1\,,{[s]}-2}(y,\by|x)+
\gl e^{\ga\pb}{y}_\ga\f{\p}{\p \bar{y}^\pb }  \go^{{[s]-1}\,,{[s]}}(y,\by|x)+
 \\ \nn + \overline{H}^{\pa \pb}\f{\p^2}{\p \by_-\p \by_-{}^\pb}
    \f{   \big(  \NNN_- \big)^{  [s] }   \,
 \big(\overline{\NNN}_-\big)^{ [s]  -1}}{ [s] \,[s] !}
 \KKK_{-{}}^{[s],[s]+1 }(y^\pm,\by^\pm|x)
\,,
\eee
where    $\overline{\KKK}_{+}=\KKK_{-{}}$.

 {}
 Hence (omitting the arguments)
\bee
  \label{Lovel-0p.5}
   D^L_{\ga\pmm} \go^{{[s]-1}\,,{[s]}}{}_{\ga}{}^{\pmm} =\gl {y}_\ga\f{\p}{\p \bar{y}^\pb }
 \go^{{[s]}-2\,,{[s]}+1} {}_{\ga}{}^{\pb}
+\gl   \by_\pb\f{\p}{\p y^\ga}            \go^{{[s]}\,,{[s]-1}}{}_{\ga}{}^{\pb}
 \\ \nn +
\f{\p^2}{\p {y}_-{}^{\ga} \p {y}_-{}^{\ga}}
\f{   \big(  \NNN_- \big)^{{[s]-1}}   \,
 \big(\overline{\NNN}_-\big)^{[s]}}{  [s]{[s]} !   }\,
 \KKK_{+{}}^{[s]+1,[s] }
  \,,\\   \nn\\   \nn
   D^L_{\mu\pb} \go^{{[s]-1}\,,{[s]}}{}^{\mu}{}_{\pb}=
   \gl  {y}_\ga\f{\p}{\p \bar{y}^\pb }
 \go^{{[s]}-2\,,{[s]}+1}{}^{\ga}{}_{\pb} +\gl   \by_\pb\f{\p}{\p y^\ga}
 \go^{{[s]}\,,{[s]-1}}{}^{\ga}{}_{\pb}
  \,,
  \eee

  \bee
\label{Lovel+0p.5}
 D^L_{\mu\pb} \go^{{[s] }\,,{[s]-1}}{}^{\mu}{}_{\pb}=
 \gl   \by_\pb\f{\p}{\p y^\ga}  \go^{{[s]}+1\,,{[s]}-2}{}^{\ga}{}_{\pb} +\gl
 {y}_\ga\f{\p}{\p \bar{y}^\pb }  \go^{{[s]-1}\,,{[s]}}{}^{\ga}{}_{\pb}
\\ \nn
+ \f{\p^2}{\p \by_-{}^{\pb} \p \by_-{}^{\pb}} \f{ \big(  \NNN_- \big)^{{[s]}}   \,
 \big(\overline{\NNN}_-\big)^{ [s]-1}}{[s]{[s]} !  }\,
  \KKK_{-{}}^{[s],[s]+1 }
\,,
\\
 \nn\\   \nn
  D^L_{\ga\pmm} \go^{{[s] }\,,{[s]-1}}{}_{\ga}{}^{\pmm} =
  \gl   \by_\pb\f{\p}{\p y^\ga}  \go^{{[s]}+1\,,{[s]}-2}{}_{\ga}{}^{\pb}+\gl
 {y}_\ga\f{\p}{\p \bar{y}^\pb }  \go^{{[s]-1}\,,{[s]}}{}_{\ga}{}^{\pb}
 \,.
\eee
Substitution of the bilinear $\KKK_{+}=
C _+{}  \big(  y_-+y_+ , \by_-+\by_+\big| x \big)
  \overline{C}_-{}\big(  y_+-y_- , \by_+-\by_-\big| x \big)\,$ and
  $\KKK_{-}= \overline{\KKK_{+}}$ (\ref{X})
gives
\bee
  \nn    D^L_{\ga\pmm} \go^{{[s]-1}\,,{[s]}}{}_{\ga}{}^{\pmm}
=\gl {y}_\ga\f{\p}{\p \bar{y}^\pb }
 \go^{{[s]}-2\,,{[s]}+1} {}_{\ga}{}^{\pb}
 +\gl   \by_\pb\f{\p}{\p y^\ga}\go^{{[s]}\,,{[s]-1}}{}_{\ga}{}^{\pb}
   +\qquad
   \\\nn+
  \f{\p^2}{\p {y}^-{}^{\ga} \p {y}^-{}^{\ga}}
\f{   \big(  \NNN_- \big)^{{[s]-1}}   \,
 \big(\overline{\NNN}_-\big)^{[s]}}{  [s] [s]!  { }  }\,
\sum_{p ,\,\,  n+m=[s]-p}    \!\!  \Big(C^{p+n+1, n }_+{}  \big(  y_- , \by_-\big| x \big)
   {C}^{m, p+m}_-{}  \big( -y_- , -\by_- \big| x \big)
  +\\+C^{m, p+m}_+{}  \big(  y_- , \by_-\big| x \big)
   {C}^{p+n+1, n }_-{}\big( -y_- , -\by_- \big| x \big) \nn\Big)\big|_{y_-=\by_-=0} \,\, ,
  \eee
  \bee\nn
   D^L_{\mu\pb} \go^{{[s] }\,,{[s]-1}}{}^{\mu}{}_{\pb}\, =
  \gl   \by_\pb\f{\p}{\p y^\ga}  \go^{{[s]}+1\,,{[s]}-2}{}^{\ga}{}_{\pb} +\gl
 {y}_\ga\f{\p}{\p \bar{y}^\pb }  \go^{{[s]-1}\,,{[s]}}{}^{\ga}{}_{\pb}
+\qquad  \\ \nn+
 \f{\p^2}{\p \by^-{}^{\pb} \p \by^-{}^{\pb}} \f{ \big(  \NNN_- \big)^{{[s]}}   \,
 \big(\overline{\NNN}_-\big)^{ [s]-1}}{[s][s]!{ }  }\,
\sum_{p ,\,\,  n+m=[s]-p}    \!\!  \Big(\overline{C}^{n, p+n+1 }_+{}  \big(  y_- , \by_-\big| x \big)
   \overline{{C}}^{p+m,  m}_-{}  \big( -y_- , -\by_- \big| x \big)
  +\\ +
  \overline{C}^{p+m,  m }_+{}  \big(  y_- , \by_-\big| x \big)
   \overline{{C}}^{n, p+n+1 }_-{}\big( -y_- , -\by_- \big| x \big)  \nn\Big)\big|_{y_-=\by_-=0} \,\, .
   \eee

Projecting out the terms, that contain the extra fields $\go^{{[s]}-2\,,{[s]}+1}$
and $\go^{{[s]}+1\,,{[s]}-21}$ by the contraction of free indices
with $y^\ga y^\ga$ and  $\bar{y}^\pb \bar{y}^\pb$, respectively,
we obtain the Fang-Fronsdal field equations \cite{FF} in $AdS_4$
with the conformal currents on the right-hand-sides.

\renewcommand{\KKK}{\mathcal{J}}

\section{Conclusion}
\label{conc}
In this paper, the unfolded  equations for free massless fields of
all spins are extended to  current interactions. The resulting equations
have  linear form where the currents are realized as rank-two linear fields
of \cite{tens2}.  More precisely, the construction of \cite{tens2} deals with conformal
currents built from $4d$ massless fields. Correspondingly, in this
paper, we describe interactions of massless fields with conformal
currents. We have checked in detail how  usual current interactions
for lower spins as well as their generalization to the HS sector are
reproduced. Remarkably, the same system reproduces
Yukawa interactions in the sector of spins zero and half.

More precisely, the set of currents, that results from the construction of \cite{gelcur}, is
infinitely degenerate with most of the currents being
exact, describing no charge conservation. However, the infinite set of currents of a given
spin contains one member that involves a minimal number of derivatives
of the constituent
fields and is not exact. In this respect the set of
currents resulting from our construction is analogous to that considered recently for the case
of any dimension in \cite{Bekaert:2010hk} which is also infinitely degenerate
(note, however, that our construction contains HS currents built from fields of different
integer and half-integer spins, while in the paper \cite{Bekaert:2010hk} only the HS currents
built from a scalar field were considered). Let us stress that exact currents may also play a
nontrivial role in the interaction theory: the difference is that nontrivial currents
(elements of the current cohomology) describe minimal HS interactions while the exact currents
(also known as improvements) describe non-minimal HS interactions of anomalous magnetic
moment type, that may also be important in the full interacting HS theory.

The analysis of this paper is performed in the $AdS_4$ background. The unfolded machinery
makes is  technically as simple as that in Minkowski case.
This should be compared to other approaches to the
analysis of HS conserved currents in $AdS$ background
\cite{Manvelyan:2004mb,Fotopoulos:2007yq,Manvelyan:2009tf,Fotopoulos:2009iw}.
(Note that the case of $AdS_3$ was considered in \cite{Prokushkin:1999xq,Prokushkin:1999ke}).

An interesting problem for the future is to see how the results of this paper are reproduced by
the full nonlinear system of equations of motion which is  known for HS fields
both  in $AdS_4$ \cite{more} and in $AdS_d$ \cite{non}
(see also reviews \cite{gol,solv}). This  may help to reach
better understanding of the full nonlinear problem allowing to interpret interactions as
the linear problem that involves fields that can either be interpreted as free fields
in higher dimensions or as currents in $AdS_4$. It should be noted however that
to proceed along this direction it is necessary to extend our results to the case of
non-gauge invariant HS currents built from HS gauge connection one-forms
rather than from the gauge invariant generalized Weyl zero-forms like the generalized
Bell-Robinson tensors of \cite{Berends:1985xx}. The complication is that currents of
this type, like, e.g., the stress tensor built from  HS gauge fields, are not gauge
invariant as was pointed out in \cite{Deser:2004rr}. In fact, it is this property that leads to
peculiarities of the HS interactions \cite{Aragone:1979hx}, that require additional interactions
with higher derivatives and non-zero cosmological constant to restore the gauge invariance
\cite{FV1}. It would be  interesting to see how this works within the approach presented in
this paper.

One of the conclusions of this paper is that, within the unfolded dynamics approach,
at least some of interactions can be interpreted in terms of free fields in higher
dimensions. The remarkable feature of the unfolded approach is that it  makes it easy to
put on the same footing field theories in different dimensions.
The only source of nonlinearity comes from the realization of higher-dimensional
fields as bilinears of the lower-dimensional ones as in Eq.~(\ref{bilinear}).
Let us note that from this perspective, the results of this paper are
somewhat reminiscent of the correspondence between
pairs of massless fields in two dimensions and sources of massless fields
in four dimensions observed in \cite{Ritus2}. It would be interesting to
reconsider the analysis of \cite{Ritus2} in the framework of the unfolded
machinery. Also it is interesting to extend our analysis to dynamical
systems in different dimensions. In particular, in accordance with
the results of \cite{Vasiliev:2001dc} $3d$ conformal currents should
identify with $4d$ massless fields and $6d$ conformal currents should
identify with $10d$ conformal fields.

More generally, it is tempting to elaborate further the interpretation of
 the obtained results in the context of $AdS/CFT$ correspondence. Moreover,
 we believe that the further analysis of HS gauge theories within the unfolded
 approach may help to understand the origin of the remarkable interplay between
 space-times of different dimensions suggested by $AdS/CFT$ correspondence
 \cite{Maldacena:1997re,Witten:1998qj,Gubser:1998bc} going beyond the standard
  $AdS/CFT$ interpretations  of HS theories
  \cite{Sundborg:2000wp,Witten,Sezgin:2002rt,Klebanov:2002ja,
  Giombi:2009wh,Henneaux:2010xg,Campoleoni:2010zq,Gaberdiel:2010ar,Gaberdiel:2010pz}.
The results of this paper indicate that
HS theories, that involve infinite towers of massless fields associated with
infinite dimensional HS symmetries, suggest that the usual space-time picture we are used
to work with results from  localization of an infinite dimensional space by virtue of
chosen dynamical systems as discussed in \cite{Vasiliev:2001dc}. Also we interpret the results of
this paper as a further evidence in favor of  the idea of an infinite chain of dualities
that relate the spaces $\M_M$ with different $M$, as suggested in \cite{BHS}.

 \section*{Acknowledgments}
We are grateful to O.~Shaynkman for useful discussions. M.V. is grateful for hospitality
Theory Division of CERN, where a considerable part of this work was done,
and acknowledges a partial support from the Alexander von Humboldt
Foundation Grant PHYS0167.
This research was supported in part by RFBR Grant No 08-02-00963.
The extension of the original version of the paper by the evaluation of
the symmetry parameters  of $AdS_4$ currents in Section \ref{symparam} and
the trivial gluings in Appendix D
was supported    by the Russian Science Foundation grant 14-42-00047.

\newcounter{appendix}
\setcounter{appendix}{1}
\renewcommand{\theequation}{\Alph{appendix}.\arabic{equation}}
\addtocounter{section}{1} \setcounter{equation}{0}
 \renewcommand{\thesection}{\Alph{appendix}.}

 \addtocounter{section}{1}
\addcontentsline{toc}{section}{\,\,\,\,\,\,\,Appendix A.  Weyl sector gluing operators}
\renewcommand{\TTT}{ {I}}
\renewcommand{\PPP}{ {J} }

 \section*{Appendix A.  Weyl sector gluing operators}

In Section \ref{Current deformation} we have introduced the
gluing operators, polynomial in the operators $\NNN_\pm$,
$\overline{\NNN}_\pm$ (\ref{Npm}). Here we present details of
the derivation.

 The following simple properties of an arbitrary   function $\mathcal{G}\big(\NNN_\pm ,  \overline{\NNN}_\pm\big)$ are used below
 \bee\label{properN}
\,\,\,\,\Big [     \mathcal{G}\big(\NNN_\pm ,  \overline{\NNN}_\pm\big) \,,  {y}^j{}^\mu \Big]=
 {y}^\mu\frac{\partial }{ \partial \NNN_j}\mathcal{G}\big(\NNN_\pm ,  \overline{\NNN}_\pm\big)
  \q\,\,
  \Big[\frac{\partial }{ \partial {y}^\mu} ,g{} \big(\NNN_\pm ,  \overline{\NNN}_\pm\big)\Big]
 =\frac{\partial }{ \partial \NNN_j}\mathcal{G}\big(\NNN_\pm ,  \overline{\NNN}_\pm\big)
\frac{\partial }{ \partial {y}^j{}^\mu}
 \q\quad
\\\nn
  \!\Big[   \mathcal{G}\big(\NNN_\pm ,  \overline{\NNN}_\pm\big) \,,  {\by}^j{}^\pmm  \Big]
={\by}^\pmm \frac{\partial }{ \partial \overline{\NNN}_j}\mathcal{G}\big(\NNN_\pm ,  \overline{\NNN}_\pm\big)
\q\!\!
\Big[\frac{\partial }{ \partial {\bar{y}}^\pmm }\,,\mathcal{G}\big(\NNN_\pm ,  \overline{\NNN}_\pm\big)\Big]
=\frac{\partial }{ \partial \overline{\NNN}_j}\mathcal{G}\big(\NNN_\pm ,  \overline{\NNN}_\pm\big)
\frac{\partial }{ \partial {\bar{y}}^j{}^\pmm } \q\,\,
\\ \label{idenNN}
  \!\! \mathcal{G}\big(\NNN_\pm ,  \overline{\NNN}_\pm\big)\,
  y^k{}_\ga F \Big|_{{y^\pm=\by^\pm=0}}
     =y_\ga\f{\p   }{\p \NNN_k}
  \,\mathcal{G}\big(\NNN_\pm ,  \overline{\NNN}_\pm\big)\, F
    \Big|_{{y^\pm=\by^\pm=0}}
    \quad \forall F(y^\pm)\,.\qquad\qquad\qquad\qquad  \eee
   For the future convenience  let us introduce  a set of functions 
\bee \label{GENF_K}
 \FFF_K{}^{n_+ ,n_-} \big( \NNN_\pm , \overline{\NNN}_\pm \big)=    \big( \NNN_+\big)^{ {n}_+ }
   \big( \NNN_-\big)^{ {n}_-   }
  \sum_{m\ge0\, }
\f{\big(\overline{\NNN}_+\,\NNN_-+ \overline{\NNN}_-\,\NNN_+\big)^{m}}
 { \,m! (m+ {n}_++{n}_-+K )!}\, ,\quad
 \eee
 which have useful properties
\bee \label{propF_K}
    \f{\p   }{  \p \overline{\NNN}_\pm}
\FFF_K{}^{n_+\,,n_-}{}&=& \NNN_\mp\FFF_{K+1}{}^{n_+\,,n_-}{}   \, ,\quad\\ \nn
\Big\{K  + \NNN_A \frac{\p }{\p  \NNN_A }
  \Big\}   \FFF_K{}^{n_+\,,n_-}{}&=&\FFF_{K -1}{}^{n_+\,,n_-}{}
\\ \nn
\Big(   \f{\p^2  }{\p \NNN_+ \p \overline{\NNN}_-}+
 \f{\p^2 }{\p \NNN_- \p \overline{\NNN}_+}\Big)\FFF_K{}^{n_+\,,n_-}{}
 &=& (1-K)\FFF{}_{K+1}{}^{n_+\,,n_-}+\FFF{}_{K}{}^{n_+\,,n_-} \,.
 \eee
Note that the function $\FFF{} {}^{n_+\,,n_-}$ used through the
paper coincides with $ \FFF{}_{1}{}^{n_+\,,n_-}$.
The functions (\ref{GENF_K}) are  related   to the
regular Bessel functions $I_{k}(x)$
 (see, e.g., \cite{bessel}) as follows
\bee \nn\f{\FFF_K{}^{n_+ ,n_-}  \big( \NNN_\pm , \overline{\NNN}_\pm \big)}{
\big( \NNN_+\big)^{  {n}_+ }  \big( \NNN_-\big)^{  {n}_-   }}=  f_{ {n}_++{n}_-+K } \big(\overline{\NNN}_+\,\NNN_-+ \overline{\NNN}_-\,\NNN_+\big)
\q  f_k (r)= r^{- \half k}I_{ k}(2r^{\half}).
\eee

The deformed conformal equations are of the form
\bee\label{GeneralDefC}
 D^{tw} C
 + e^{ \gm  }{}^{\pn}    G^k_j B_k^j{}_{ \gm  }{}_{\pn}
  \PPP\,\Big|_{{y^\pm=\by^\pm=0}}
  =0
 \,,
\eee
where  $B^k_j{}_{ \gm  }{}_{\pn}$
are  bilinear in $\p_k{}_\ga$,
$y^j{}^\ga$, $\bar{\p}_k{}_\pa$, $ {\by}^j{}^\pa$ with
 $j,k=\{+,-\}$ , namely,
\bee\label{Bili}
B^k_{j}{}_{\, \ga  }{}_{\pa}=y^k{}_\ga  {\bp}{}_j{}_\pa \q
\overline{B}^k_{j}{}_{\, \ga  }{}_{\pa}=\by^k{}_\pa  {\p}{}_j{}_\ga\q
B^k{}^{j}{}_{\, \ga  }{}_{\pa}=y^k{}_\ga \by^j {}_\pa \q
B_k{}_{j}{}_{\, \ga  }{}_{\pa}= {\p}{}_k{}_\ga{\bp}{}_j{}_\pa  \q \eee
 $G_a\big(\NNN_\pm\,,\overline{\NNN}_\pm\big)$\ are some
 gluing  operators,
$  D^{tw}  $ is the rank-one twisted covariant derivative
(\ref{tw}) and  the rank-two current field
$ \PPP(y^\pm\,,\by^\pm)$  satisfies  the current equation
  (\ref{newtwistjads}).
The system of equations (\ref{newtwistjads}) decomposes into a set
of subsystems
associated with different elements of $\slv$-modules realized by
bilinear operators
$B_a$ (\ref{Bili}).

 The consistency condition for the  equation (\ref{GeneralDefC})
\bee
\label{genercons}\!\!\!
\Big(  H^{ \gm  }{}^{ \ga  } \gvep^{\pn}{}^{\pb}+  H^{\pn}{}^{\pb} \gvep^{ \gm  }{}^{ \ga  }\Big)
 \Big\{\big(y_\ga \bar{y}_\pb+\p_\ga\bp{}_{\pb}\big)
 G^k_j B_k^j{}_{ \gm  }{}_{\pn}+\qquad\\ \nn
  - G^k_j B_k^j{}_{ \gm  }{}_{\pn}\big(y^+{}_\ga\,\bar{y}^-{}_\pb+
y^-{}_\ga\,\bar{y}^+{}_\pb\, +
\p_-{}_\ga\bar{\p}_+{}_\pb+\p_+{}_\ga\bar{\p}_-{}_\pb
\big)
  \Big\}\PPP (y^\pm \,,\bar{y}^\pm|x )  \Big|_{{y^\pm=\by^\pm=0}}
  =0
\,\quad
\eee
 imposes   restrictions on the gluing operators $G_a $ analyzed  below.
Evidently,  Eq.~(\ref{genercons}) decomposes into a set of subsystems
 characterized by different eigenvalues of the rank-two helicity
 operator  $\hv $  (\ref{slv}). We begin with the simpler Minkowski
 case then showing that the obtained solution also works in $AdS_4$.

 \subsection*{Minkowski case }
\label{Aflatlim}
 First consider  $\slv$ highest  element  $ B^k_{j}{}_\ga{}_\pa$ (\ref{Bili}), which   satisfies
$[\hv,B^k_{j}{}_\ga{}_\pa ] =2 B^k_{j}{}_\ga{}_\pa  $. In this case
 the flat limit of Eq.~(\ref{genercons})
   gives
along with (\ref{idenNN})
     \bee
\label{genercons1}
e^{\gm\pa}e^{\ga\pb}\Big( \p_\ga{}\bar{\p}_\pb
  y_\gm F^j\bp_j{}_{\pn}
 -  y_\gm F^j\bp_j{}_{\pn}
(
    \p_-{}_\ga\bar{\p}_+{}_\pb+\p_+{}_\ga\bar{\p}_-{}_\pb)
    \Big)
\PPP (y^\pm ,\bar{y}^\pm|x )  \Big|_{{y^\pm=\by^\pm=0}}=0
\,,\quad
\eee
where
\be
\label{ccc}
{F}^j=  \f{\p }{\p \NNN_i  }  G^{j }_i \,.
\ee
 Hence,   by virtue of (\ref{properN}), we have
\bee\label{Acontw1.1}  H^{\pmm \pb}
 \Big(\bar{\p}{}_\pmm\,
 \Big\{2 + \NNN_K \frac{\partial }{\partial \NNN_K }
 \Big\}F^j
- \NNN_+ F^j\bar{\p}_-{}_\pmm
 - \NNN_- F^j\bar{\p}_+{}_\pmm
    \Big)\bar{\p}_j{}_\pb
 &=&0\,, \qquad\, \eee
\bee\label{Acontw1.2}H^{\mu\ga}
y_\ga\bar{\p}_j{}^\pmm\,   \Big(
\bar{\p}_\pmm\,
\p_\mu F^j
  -  F^j    (\p_-{}_\mu\bar{\p}_+{}_\pmm+\p_+{}_\mu\bar{\p}_-{}_\pmm)
    \Big)
 &=&0 \,.\qquad\eee
   This gives the following  conditions on $F^\pm$ (\ref{ccc})
 \bee\label{Acontw++}
   \beee {rcl}%
\displaystyle{ \Big\{2 + \NNN_K \frac{\partial }{\partial \NNN_K }
  \Big\}  \frac{\partial {F}^+\!}{ \partial  \overline{\NNN}_+}\,
   - \NNN_- {F}^+\!}
 &=&0\,, \qquad\,
 \\
\\\displaystyle{
 \Big\{2 + \NNN_K \frac{\partial }{\partial \NNN_K }
  \Big\}  \frac{\partial {F}^-\!}{ \partial  \overline{\NNN}_-}\,
    - \NNN_+ {F}^-\!}
 &=&0\,, \qquad\, 
 \\ \\\displaystyle{
 \Big\{2 + \NNN_K \frac{\partial }{\partial \NNN_K }
 \Big\}\Big( \frac{\partial {F}^+\!}{ \partial  \overline{\NNN}_-}\,
 +
  \frac{\partial {F}^-\!}{ \partial  \overline{\NNN}_+}\,
 \Big)
  - \NNN_- {F}^-\!- \NNN_+ {F}^+\!}
  &=&0\,, \qquad\,
  \\ \\  \displaystyle{ \f{\p  }{\p \NNN_-  } {F}^+\! \,+  \f{\p  }{\p \NNN_+  }{F}^-\! }&=&0
    ,\quad
\eeee \eee\bee\nn  \beee{rcl}    
    \displaystyle{
\Big(   \f{\p^2  }{\p \NNN_+ \p \overline{\NNN}_-}+
 \f{\p^2 }{\p \NNN_- \p \overline{\NNN}_+}-1\Big){F}^+\!
  }&=&0
    ,\quad\\
    \\\displaystyle{\Big( \f{\p^2 }{\p \NNN_- \p \overline{\NNN}_+}+
 \f{\p^2 }{\p \NNN_+ \p \overline{\NNN}_-}-1\Big ){F}^-\!
     }&=&0.\quad\eeee
  \eee
Elementary straightforward analysis shows   that ${F}^\pm$ have the
form (\ref{ccc})\\ie, ${F}^\pm=  \f{\p }{\p \NNN_i  }  G^{\pm }_i \,,$ with
\bee   \label{resultG}
  G_+^+=-G_-^-=\sum_{n_+\,,\,\, n_-\ge0} a_{n_+\,,n_-}\FFF_1{}^{n_+\,,n_-}{ }  \q
  G_+^-=-G_-^+=0\,.\eee
 The respective   deformation, \ie the second term on the
 left-hand-side of Eq.~(\ref{GeneralDefC}),  is
\bee\label{dennK}
  e^{ \gm  }{}^{\pb}    \sum_{n_+\,,\,\, n_-\ge0} a_{n_+\,,n_-}
  \FFF_1 {}^{n_+\,,n_-}{ } \big(y^+{}_\gm\bar{\p}_+{}_\pb
-y^-{}_\gm\bar{\p}_-{}_\pb
\big)  \PPP\,\Big|_{{y^\pm=\by^\pm=0}}
  \, \q
\eee
where $a_{{n}_+\,,{n}_-}$ are arbitrary coefficients. Note that the ambiguity in the coefficients $a_{{n}_+\,,{n}_-}$
is in  accordance with the
ambiguity of contributions of different spin fields to the currents.

For the complex conjugate
   $\overline{B^k_{p}}{}_{ \ga  }{}_{\pa} $  satisfying
  $[\hv,\overline{B^k_{p}}{}_{ \ga  }{}_{\pa} ] =-2 \overline{B^k_{p}}{}_{ \ga  }{}_{\pa} $
  the respective  gluing operators  $\overline{G}_a$ are
\bee   \nn
\overline{G}_+^+=-\overline{G}_-^-=\sum_{\bar{n}_+\,,\,\, \bar{n}_-\ge0} \bar{a}_{\bar{n}_+\,,\bar{n}_-}
\overline{\FFF}_1{}^{\bar{n}_+\,,\overline{n}_-}{ }  \big( \NNN_\pm , \overline{\NNN}_\pm \big)\q
  \overline{G}_+^-=-\overline{G}_-^+=0\,. \eee
where $\bar{a}_{\bar{n}_+\,,\bar{n}_-}$ are arbitrary coefficients, and
\bee \label{GENF_Kcc}
\overline{\FFF}_K{}^{\bar{n}_+\,,\bn_-}{ }  \big( \NNN_\pm , \overline{\NNN}_\pm \big)=
  {\big(\overline{\NNN}_+ \big)^{\bn_+}  \big( \overline{\NNN}_- \big)^{\bn_-}}
 \sum_{m\ge0\, }
\f{\big(\overline{\NNN}_+\,\NNN_-+ \overline{\NNN}_-\,\NNN_+\big)^{m}}
 { \,m! (m+ {\bn_+}+{\bn_-}+K )!}\,  \quad
 \eee
is complex conjugate to $\FFF_K{}^{n_+\,,n_-}$ (\ref{GENF_K}).
 The respective   deformation is
\bee\label{dennKc}
  e^{ \gm  }{}^{\pn}    \sum_{\bn_+\,,\,\, \bn_-\ge0} \bar{a}_{\bar{n}_+\,,\bar{n}_-}
  \overline{\FFF}_1 {}^{\bar{n}_+\,,\bar{n}_-}{ }\, \by^j{}_\pn {\p}_j{}_\gm
  \PPP\,\Big|_{{y^\pm=\by^\pm=0}}
  \, \q
\eee
where  we use notations $a^j b_j=a^+ b_+ -a^- b_- $.
Note that the  operators $
 y^j{}_\gm\bar{\p}_j{}_\pb$ and $
 \by^j{}_\pn {\p}_j{}_\gm $ in the deformations
(\ref{dennK}) and (\ref{dennKc})
 are invariant under   $\slh$.

It is also not difficult to see, that the
deformation   (\ref{GeneralDefC})
with the remaining $B^a$ (\ref{Bili}) satisfying $[\hv,B^a ]   =0 $ is trivial, \ie
can be removed by a local field redefinition (in other words,
it is  $D_{fl}^{tw}$-exact on solutions of the current equation).

 \subsection*{AdS}
\label{AAdS}
In the  $AdS_4$ case
the gluing coefficients  remain the same as in Minkowski case.
For example consider $B^a{}^\gm{}_\pb$ of the form
$y^+{}^\gm\bar{\p}_+{}_\pb- y^-{}^\gm\bar{\p}_-{}_\pb$  found above.   Eq.(\ref{genercons})
   gives
  \bee  \!\!
 \label{Aadditads}\Big(H^{\mu\ga}\!\gvep^{\pmm \pb}\!+\! H^{\pmm \pb}\!\gvep^{\mu\ga}\Big)
\Big\{\!  \p_\gm{}\bar{\p}_\pmm
  y_\ga F^j\bp_j{}_{\pb}
 -  y_\ga F^j\bp_j{}_{\pb}
\big(\p_-{}_\mu\bar{\p}_+{}_\pmm+\p_+{}_\mu\bar{\p}_-{}_\pmm\big)
    +\qquad\qquad
\\ \nn
 +y_\gm \bar{y}_\pmm  y_\ga F^j\bp_j{}_{\pb}
- y_\ga F^j\bp_j{}_{\pb}   \Big(
y^+{}_\gm\,\bar{y}^-{}_\pmm+
y^-{}_\gm\,\bar{y}^+{}_\pmm\,\Big)  \,
    \Big\}\PPP\,   (y^\pm \,,\bar{y}^\pm|x )  \Big|_{{y^\pm=\by^\pm=0}}
   =0  \,.\quad \eee
One can see that (\ref{Aadditads}) is true provided that
  $F^\pm(\NNN_\pm\,,\overline{\NNN}_\pm)$ satisfy  the conditions (\ref{Acontw++})
along with
\bee    \label{Aadditads1}
 y_\ga y_\gm \bar{y}_\pmm  \Big\{1
- \Big( \f{\p}{\p \overline{\NNN}_-{} } \f{\p}{\p \NNN_+{} }
+  \f{\p}{\p \overline{\NNN}_+{} } \f{\p}{\p \NNN_-{} }\Big) \Big\} F^j\bp_j{}_{\pb}
    \PPP\,   (y^\pm \,,\bar{y}^\pm|x )  \Big|_{{y^\pm=\by^\pm=0}}+\\ \nn
    -
  y_\ga y_\gm\gvep_{\pb\pmm}\Big\{ \f{\p}{\p \NNN_-{} } F^+ +\f{\p}{\p \NNN_+{} } F^-
   \,
    \Big\}\PPP\,   (y^\pm \,,\bar{y}^\pm|x )  \Big|_{{y^\pm=\by^\pm=0}}=0\q
    \eee
which, however, is true by virtue of (\ref{Acontw++}).
 Hence, the deformation of the form   (\ref{dennK})  remains
 consistent in the $AdS_4$ case  as well.
The  complex conjugate case  is   analogous.

Analogously to the Minkowski case, it is not   difficult to see that
the consistent  deformed equations (\ref{GeneralDefC}) with  $B^a$ 
obeying
$[\hv,B^a ]   =0 $  are trivial ($D{}^{tw}$-exact) for any current field $J$.

 \renewcommand{\theequation}{\Alph{appendix}.\arabic{equation}}
\addtocounter{appendix}{1} \setcounter{equation}{0}

 \addtocounter{section}{1}
 \addcontentsline{toc}{section}{\,\,\,\,\,\,\,Appendix B.  Spin-s$\ge$2  one-form  sector}
 \section*{Appendix B. Spin-s$\ge$2  one-form  sector}
 \label{Appendix 2}

Since zero-forms contribute to the right-hand-sides of \,Eq.~(\ref{CON1}),
their formal consistency in presence of the deformation (\ref{newtwC})
requires an appropriate deformation in the one-form sector  
\bee
 \label{CON1newAP}
  D^{ad}\go(y,{\bar{y}}|x)=
     \overline{H}^{\pa\pb}
    \bp_{\pa} \bp_{\pb}
\overline{C}(0,\by\mid x) +    H^{\ga\gb}\p_{\ga}  \p_{\gb}C (y,0\mid x)\,+\qquad\qquad\\\nn
   \overline{H}^{\pa\pb}  \Gg{}_{\pa\pb}
   \big(\NNN_\pm\,,\overline{\NNN}_\pm\big)\,\RRR(y^\pm,\by^\pm|x) \big|_{y^\pm=\by^\pm=0}
+ H^{\ga\gb}
G_{\ga\gb}  \big(\NNN_\pm\,,\overline{\NNN}_\pm\big)\,\KKK(y^\pm,\by^\pm|x) \,\big|_{y^\pm=\by^\pm=0}
\,\quad\eee
for some gluing operators $G_{\ga\gb}$ and ${\Gg}_{\pa\pb}$
and current  fields $\RRR$ and $\KKK$ with  $\NNN_\pm$,
$\overline{\NNN}_\pm$ (\ref{Npm}).

Let  $s\ge2$.
 (The case of  $s=3/2$ is special  and is considered in Appendix C.)

Since the horizontal $\slh$ (\ref{slh}) acts on current fields
$\KKK$, $\RRR$ and, hence, on the gluing functions,
it is convenient to require
 $G_{\ga\gb} $ and ${\Gg}_{\pa\pb}$  be highest vectors with respect
  to $\slh$, setting
   \bee\label{comGG}
G_{\ga\gb}=\p_-{}_\ga \p_-{}_\gb G^{s-1}(\NNN_-\,,\overline{\NNN}_-)    \q
{\Gg}_{\pa\pb}=\bp_-{}_\pa \bp_-{}_\pb
\Gg^{s-1}(\NNN_-\,,\overline{\NNN}_-)
 \q  \eee
where $G^{s-1}$, $\Gg^{s-1}$ are
  some degree--$2(s-1)$  homogeneous polynomials of
$\NNN_-$ and $\overline{\NNN}_-$ to match the fact
that the one-forms $\go$   are degree-$2(s-1)$ homogeneous
polynomials of $y,\,\,\by$.

Taking into account the form of the
$\slh$ highest-weight deformation in the zero-forms sector,
namely (\ref{newtwCs}) and (\ref{newtwC-s}) with $\bar{a}_{m, 2s-m}=a_{m, 2s-m}=\gd_m^0 a_{0, 2s}$,
and  setting
for definiteness $a_{0, 2s}= 2s+1\,$,
 one can see that the consistency  condition
 for Eq. (\ref{CON1newAP})  imposes
\renewcommand{\TTT}{{ \overline{J}}}
\renewcommand{\TTT}{{ \overline{J}}}
 the following  conditions on the
 current fields $\PPP$ ,   $\TTT$, $\KKK$ and $\RRR$
 \bee \label{CON1newad}
   D^{ad}\Big(  \overline{H}^{\pa\pb}  \Gg{}_{\pa\pb}  \big(\NNN_-\,,\overline{\NNN}_-\big)\,\RRR(y^\pm,\by^\pm|x) \big|_{y^\pm=\by^\pm=0}
 +H^{\ga\gb}
G_{\ga\gb}  \big(\NNN_-\,,\overline{\NNN}_-\big)\,\KKK(y^\pm,\by^\pm|x) \,\big|_{y^\pm=\by^\pm=0}\Big)
= \quad \\ \nn
\!\!\!   =  \f{2}{(2s-2)!}
   {\Hh}^{\ga \pb}\p_-{}_\ga\overline{\p}_-{}_\pb\left\{
  \big(  \NNN_- \big)^{2s-2}   \PPP\,  \,
  -    \big(  \overline{\NNN}_- \big)^{2s-2}
\TTT{}\right\}      \Big|_{y^\pm=\by^\pm=0}.
  \qquad\qquad\eee
Substituting (\ref{comGG})   into   (\ref{CON1newad}),
and  using  (\ref{newtwistjads}), (\ref{iden2-3})
along with the evident identities
 $$
G (  {\NNN}_-  ,{\overline{\NNN}_-})  \Big( \p_-{}_\gga\p_+{}^\gga
   - f_- \Big)\KKK  \big|_{y^\pm=\by^\pm=0}\equiv0\,,
\quad
\Gg(  {\NNN}_-  ,{\overline{\NNN}_-})
\Big(\bar{\p}_-{}_\pga\bar{\p}_+{}^\pga
 -f_+ \Big){\RRR}\big|_{y^\pm=\by^\pm=0}
\equiv0\,,
$$
where
  $f_-$ and $f_+ $ are generators   of the vertical $\slh$  (\ref{slv}),
we obtain
\bee \label{DadGH}\!\!
     \gl  \p_-{}_\ga\bar{\p}_-{}_\pb
    \Big(
-\overline{\NNN}_-\f{\p}{\p  {\NNN}_-}
\Gg^{s-1}   \RRR
+\Gg^{s-1}    f_+
\RRR +
\NNN_-\f{\p}{\p \overline{\NNN}_-}
G^{s-1}     \KKK
-G^{s-1}    f_-
\KKK
\Big)\big|_{y^\pm=\by^\pm=0}
\quad\\ \nn \rule{0pt}{20pt}
   =  \f{1}{(2s-2)!}
    \p_-{}_\ga\overline{\p}_-{}_\pb\left\{
  \big(  \NNN_- \big)^{2s-2}   \PPP\,  \,
  -    \big(  \overline{\NNN}_- \big)^{2s-2}
\TTT{}\right\}      \Big|_{y^\pm=\by^\pm=0}.
\qquad\eee
This equation can be easily solved by the Ansatz
$$\Gg^{s-1}=\big(  \overline{\NNN}_- \big)^{2s-2}\q G^{s-1}=\big(   {\NNN}_- \big)^{2s-2}\,.$$
As shown in Appendix D, currents of the form
\bee   \label{triv1}
      \PPP=-\gl (2s-2)!
  f_-\,\KKK \q
\TTT{}=-\gl  (2s-2)!  \, f_+\, {\RRR}
\eee
that solve (\ref{DadGH}),
lead to  a trivial deformation in the zero-form sector
and, hence, to a trivial deformation
in the one-form sector.

The proper strategy is to  start with
some ``seed current field" $\LLL_{(l)}$ under the conditions
\be\label{eigenl-s}  \hv  \LLL_{(l)} =   2(l-s )  \LLL_{(l)} \ee
with some integer $l$ in the interval  $2\le l\le 2s-2$.
Setting
   \bee \label{Gs}
  G^{s-1} \KKK=G_{(l)}^{s-1} \KKK_{(l)}= \f{1}{(l-1)!}\sum_{k=0}^{l-2}  \f{ \big(  \NNN_- \big)^{l-k-2}   \,
  \big(\overline{\NNN}_-\big)^{ 2s-l+k}}{ (2s-l+k)! } \left( f_-\right)^k  \LLL_{(l)}\, \q
  \\ \label{Gs'}
    \Gg^{s-1}\RRR = \Gg_{(l)}^{s-1} \RRR_{(l)} =\f{1}{(2s-l-1)!}\sum_{k=0}^{(2s-l-2)}  \f{  \big(  \NNN_- \big)^{  l+k}   \,
 \big(\overline{\NNN}_-\big)^{2s-l-k-2}
}{  (l+k)!}\left( f_+\right)^k\LLL_{(l)}
  \qquad\eee
 in Eq.~(\ref{DadGH}) we obtain
\bee \label{DadGH2}\!\!
     \gl {\Hh}^{\ga\pb} \p_-{}_\ga\bar{\p}_-{}_\pb
    \Big(
\f{  \big(  \NNN_- \big)^{ 2s-2}   \,
  }{  (2s-2)!(2s-l-1)!}
\left( f_+\right)^{2s-l-1}\LLL_{(l)}
-\f{  \big(  \overline{\NNN}_- \big)^{ 2s-2}   \,
  }{  (2s-2)!( l-1)!}
\left( f_-\right)^{ l-1}\LLL_{(l)}
\Big)\big|_{y^\pm=\by^\pm=0}
\quad\\ \nn \rule{0pt}{20pt}
   =  \f{1}{(2s-2)!}
   {\Hh}^{\ga \pb}\p_-{}_\ga\overline{\p}_-{}_\pb\left\{
  \big(  \NNN_- \big)^{2s-2}   \PPP\,  \,
  -    \big(  \overline{\NNN}_- \big)^{2s-2}
\TTT{}\right\}      \Big|_{y^\pm=\by^\pm=0}.
\qquad\eee

Then         \bee   \nn
      \PPP= \gl \f{1}{ (2s-l-1)!} \left( f_+\right)^{2s-l-1}
   \,\LLL_{(l)}
     \q
\TTT{}= \gl \f{1}{ (l-1)!} \left( f_-\right)^{ l-1}\LLL_{(l)}
\eee
  solve (\ref{DadGH2}).
  Resulting   deformed equations    are
 \bee \label{ps_deformation}
 D^{ad}\go(y,{\bar{y}}|x) -   \overline{H}^{\pa\pb}
\f{\p^2}{\p \by^{\pa} \p \by^{\pb}}
\overline{C}(0,\by\mid x) -  H^{\ga\gb}
\f{\p^2}{\p {y}^{\ga} \p {y}^{\gb}}
C (y,0\mid x)\,=\qquad\\\nn=
  {H}^{\ga \gb}\f{\p}{\p y_-{}^\ga} \f{\p}{\p y_-{}^\gb}
 \sum_{k=0}^{l-2}  \f{ \big(  \NNN_- \big)^{l-k-2}   \,
  \big(\overline{\NNN}_-\big)^{ 2s-l+k}}{ (2s-l+k)! (l-1)!}  
 \left(   f_-\right)^k
  \LLL_{(l)}
  \,
  \big|_{y^\pm=\by^\pm=0}\qquad
 \\ \nn{}
+  \overline{H}^{\pa \pb}\f{\p}{\p \by_-{}^\pa} \f{\p}{\p \by_-{}^\pb}
 \sum_{k=0}^{(2s-l-2)}  \f{  \big(  \NNN_- \big)^{  l+k}   \,
 \big(\overline{\NNN}_-\big)^{2s-l-k-2}
}{  (l+k)!(2s-l-1)!}
 \left( f_+\right)^k
 \LLL_{(l)}\, \big|_{y^\pm=\by^\pm=0}  \,
 \eee
 and
  \bee
\label{defl0}
D^{tw}C(y,\by\mid x)
+ e^{ \gm  }{}^{\pb}
     \FFF_{1 }{}^{ 0\,,2s}{}   y^j{}_\gm\bar{\p}_j{}_\pb
         \f{ \gl (2s+1)  }{ (2s-l-1)! }
        \left(  f_+\right)^{2s-l-1}\LLL_{(l)}\big|_{y^\pm=\by^\pm=0}
  =0
 \,,\quad\\\nn
 D^{tw}\overline{C} (y,\by\mid x) + e^{ \gm  }{}^{\pb}
     \overline{\FFF}_{1 }{}^{ 0\,,2s}{}  \p_j{}_\gm\bar{y}^j{}_\pb
     \f{ \gl(2s+1)}{ ( l-1)! }       \,
 \left( f_-\right)^{l- 1} \LLL_{(l)} \,
 \big|_{y^\pm=\by^\pm=0} =0\,,\qquad
\eee
where  $\FFF_{1 }{}^{ 0\,,2s}{}  $ is  given by (\ref{GENF_K})  with $n_+=0$, $n_-=2s$.

As shown in Appendix D, the final result is independent of
the choice of   $\LLL_{(l)}$.
 Namely, up to $D^{ad}$-exact one-forms and
the $D^{tw}$-exact zero-forms, the final result remains the same
upon the identification $\dis{\LLL_{(l+1)}=\f{1}{(2s-l-1)}  f_+\LLL_{(l)}}.$
On the other hand, in the flat limit this procedure works
properly only for $|l-s|\le \half$.
 For this reason,
the formulae (\ref{CON1new}) and (\ref{CON1newh})  of
Section \ref{Current def1} were presented
 for the case of $|l-s|\le \half$
with the following identifications of  current fields $\LLLR_{\hel}$ (\ref{primfield})
 \be\label{primidenif}{ \KKK_{0,s}=\f{\LLL_{(s)}}{(s-1 )!}} \quad \mbox{ for integer }s\q
 {\KKK_{\pm1,s}=\f{\LLL_{(s\pm\half)}}{(s-\half)!}}
  \quad\mbox{ for  half-integer } s.\ee

Note  that  for any
$G( {\NNN}_-,\overline{\NNN}_-)$ with ${\NNN}_-,\overline{\NNN}_-$   (\ref{Npm}) for an
  arbitrary integer $m\ge0$
\bee\label{akthor1}
 \Big(
 \mathrm{ad}^m _{  \vv_-}\left( {\p}_-{}_\ga {\p}_-{}_\gb G( {\NNN}_-,\overline{\NNN}_-)
\left( f_-\right)^k  \right)
  -
  {\p}_-{}_\ga {\p}_-{}_\gb G( {\NNN}_-,\overline{\NNN}_-)
\big( f_-\big)^k \big(-\vv_-\big)^m\, \Big) \LLLL\big|_{y^\pm=\by^\pm=0}=0\qquad  \eee
   because   $g_{-}$    (\ref{slh}) is zero at
$y^\pm =\bar y^\pm=0$ and
 $[f_a,g_b]=0$  by virtue of (\ref{slv}), (\ref{slh})  (recall that $\mathrm{ad}_x(y)=[x,y]$).
 The complex conjugate formula  is analogous.

Since  $G_{\ga\gb} $ and ${\Gg}_{\pa\pb}$ in (\ref{comGG}) are $\slh$-highest vectors,
$\mathrm{ad}^m _{  \vv_-}\left(G_{\ga\gb}\right) $ and $
\mathrm{ad}^m _{  \vv_-}\left({\Gg}_{\pa\pb}\right)$
reproduce  the current deformations of
the dynamical equations in the zero-form sector,  associated with arbitrary gluing
coefficients in (\ref{resultFpm}) and (\ref{resultFpmcc}).

As an application of this mechanism, we observe that Eq.~(\ref{akthor1}) implies that
   the deformation
   \bee  \label{CON1newm}&&
 D^{ad}\go(y,{\bar{y}}|x) -   \overline{H}^{\pa\pb}
\bar{\p}{}_\pa\bar{\p}{}_\pb
\overline{C}(0,\by\mid x) -  H^{\ga\gb}
 {\p}{}_\ga {\p}{}_\gb
C (y,0\mid x)\,\quad\qquad\\\nn&=&
  (-1)^{m }  {H}^{\ga \gb}
\mathrm{ad}^m _{  \vv_-}  \left(   {\p}_-{}_\ga {\p}_-{}_\gb
\sum_{k=0}^{s-2}  \frac{ \big(  \NNN_- \big)^{s-k-2}   \,
  \big(\overline{\NNN}_-\big)^{ s+k}}{ (s+k)!  }\right)
 \left( f_-\right)^k\LLLL \big|_{y^\pm=\by^\pm=0} \quad
   \\ \nn&+&
   (-1)^{m }
  \overline{H}^{\pa \pb}\mathrm{ad}^m _{  \vv_-}  \left(
   \bar{\p}_-{}_\pa\bar{\p}_-{}_\pb
 \sum_{k=0}^{s-2}  \frac{   \big(  \NNN_- \big)^{  s+k}   \,
 \big(\overline{\NNN}_-\big)^{s-k-2}
}{  (s+k)! } \right)
 \left( f_+\right)^k  \LLLL \big|_{y^\pm=\by^\pm=0}\qquad\,
  \eee
is    consistent  for any $m \ge0$. By virtue of
  (\ref{CON2new}), the associated
  deformations in the zero-form sector  are
 \bee\nn  \label{CON2newg-m}
 D^{tw}C(y,\by\mid x)
  +\gl(2s+1)e^{ \ga  }{}^{\pb}
       \FFF {}^{0 ,2s } y^j{}_\ga\bar{\p}_j{}_\pb
    \Big( f_+\Big)^{s- 1} \big(\vv_-\big)^m\,\LLLL \,\big|_{y^\pm=\by^\pm=0}
     &=&0\q
 \,\\\nn
 D^{tw}\overline{C} (y,\by\mid x)
+  \gl(2s+1) e^{ \ga  }{}^{\pb}
     \overline{\FFF }{\,}^{0 ,2s }  {\p}_j{}_\ga \by^j{}_\pb
 \Big( f_-\Big)^{s- 1} \big(\vv_-\big)^m\,\LLLL\,\big|_{y^\pm=\by^\pm=0}
  &=&0\, .
\eee
Since $\overline{\vv_\pm}=-{\vv_\pm}$, the reality conditions require  to consider the
horizontal algebra $\mathfrak{sl}_2$ spanned by
$$
\vv_+=:i\vv_+\q \vv_-=:-i\vv_-\q \uu\,.
$$
Therefore,
according to (\ref{actingdlhF}) and  (\ref{akthor0}),
 the  deformed equations in the zero-form sector   can be rewritten as
 \bee\label{CON2newm}
 D^{tw}C(y,\by\mid x)
   + e^{ \ga  }{}^{\pb}
     \f{ \gl(-i)^m \, (2s+1)!} { \,m!} \FFF {}^{m ,2s-m } y^j{}_\ga\bar{\p}_j{}_\pb
 \Big( f_+\Big)^{s- 1}  \LLLL \,\big|_{y^\pm=\by^\pm=0}
   &=&0\q
 \,\\\nn
 D^{tw}\overline{C} (y,\by\mid x)
   + e^{ \ga  }{}^{\pb}
    \f{\gl( i)^m\,( 2s+1)!} { \,m!}\overline{   \FFF} {}^{m ,2s-m }  {\p}_j{}_\ga \by^j{}_\pb
      \Big( f_-\Big)^{s- 1}  \LLLL
    \,\big|_{y^\pm=\by^\pm=0}
  &=&0\,,
\eee
that gives the general result
that all
 zero-form  gluing operators (\ref{resultFpm}) and  (\ref{resultFpmcc})    are in the game, that  allows us to
conclude that the formulae (\ref{CON1newm}) contain all possible
 nontrivial  current deformations  of integer-spin fields
in the one-form sector.

The case
of half-integer spins is  analogous.

 \renewcommand{\theequation}{\Alph{appendix}.\arabic{equation}}
\addtocounter{appendix}{1} \setcounter{equation}{0}
  \addtocounter{section}{1}
\addcontentsline{toc}{section}{\,\,\,\,\,\,\,Appendix C. Spin-$3/2$ one-form  sector}
 \section*{Appendix C. Spin-$3/2$ one-form  sector}
\label{Appendix 3}
The case of  $s=3/2$ is special.
Let us look for solution of (\ref{CON1newad})
in the form
\bee\label{KKK32}
  \KKK=      \LLL_{ (1)}\q 
  \RRR \,
=
  \LLL_{ (-1)}\,\q  \LLL_{ (1)}=\overline{\LLL_{(-1)}}\,,
   \eee where
 $$
 \hv \LLL_{ (\pm 1)}=  \pm \LLL_{ (\pm1)}\,, 
 $$
Setting
\bee\label{Gss32}
G^{\half}_{\ga\gb}  = \p_-{}_\ga{\p}_-{}_\gb  \overline{\NNN}_-\q
{\Gg}^{\half}_{\pa\pb}  = \bp_-{}_\pa{\bp}_-{}_\pb
   \NNN_-   \q 
   \eee
 and plugging (\ref{KKK32}) into
    (\ref{DadGH})
 we obtain
    \bee   \label{newEINconcon333}
           \gl  \p_-{}_\ga\bar{\p}_-{}_\pb
    \Big(
-\overline{\NNN}_-   \LLL_{ (-1)}
+ {\NNN}_-    f_+
 \LLL_{ (-1)} +
\NNN_-      \LLL_{ (1)}
-\overline{\NNN}_-    f_-
\LLL_{ (1)}
\Big)\big|_{y^\pm=\by^\pm=0}
\quad\\ \nn \rule{0pt}{20pt}
   =
    \p_-{}_\ga\overline{\p}_-{}_\pb\left\{
    \NNN_-    \PPP\,  \,
  -     \overline{\NNN}_-
\TTT{}\right\}      \Big|_{y^\pm=\by^\pm=0}.
\qquad
\eee
As a result,
  \bee    \label{PPP32}
     \PPP\,   =    \gl      \LLL_{(1)} +\gl   f_+   \LLL_{ (-1)} \,  \q
  \TTT{}  =  \gl    \LLL_{ (-1)}   +\gl  f_-   \LLL_{ (1)}   \,
\qquad\eee
solve (\ref{newEINconcon333}) and the deformed equation is
 \bee \label{def32}
 D^{ad}\go(y,{\bar{y}}|x) =   \overline{H}^{\pa\pb}
\bp{}_\pa\bp {}_\pb
\overline{C}(0,\by\mid x) +  H^{\ga\gb}
\p{}_\ga\p{}_\gb
C (y,0\mid x)\,+\qquad\\\nn+
 {H}^{\ga \gb}
\p_-{}_\ga{\p}_-{}_\gb  \overline{\NNN}_-
     \LLL_{ (1)}
  \,
  \big|_{y^\pm=\by^\pm=0}+
  \overline{H}^{\pa \pb}\bp_-{}_\pa{\bp}_-{}_\pb
\NNN_- \LLL_{ (-1)} \big|_{y^\pm=\by^\pm=0} \,.
 \eee
This result coincides with (\ref{CON1newh}) at $s=3/2$
under convention
that all terms containing  $\sum_{k=0}^{-1}(...)$ or $\sum_{k=2}^{1}(...)$
are  zero.

\addtocounter{appendix}{1} \setcounter{equation}{0}
  \addtocounter{section}{1}
\addcontentsline{toc}{section}{\,\,\,\,\,\,\,Appendix D. Trivial gluings}

 \section*{Appendix D. Trivial gluings}

Here we identify a class of currents that upon substitution into the
equations (\ref{newtwC})   do not lead to a nontrivial deformation of the
massless field equations being removable by a local field redefinition. Also
the deformation (\ref{CON1new}) in the one-form sector is shown to be
insensitive to a particular choice of the
 seed current field $\LLL_{(l)}$ (\ref{eigenl-s}).

\subsection*{Trivial gluings in the zero-form sector}

 Using  that
  \be\nn
 \NNN_+\p_-{}_\ga-\NNN_-\p_+{}_\ga=y^\gga\p_+{}_\gga\p_-{}_\ga-y^\gga\p_-{}_\gga\p_+{}_\ga
=y^\gga \gvep_{\gga\ga}\p_+{}_\gb\p_-{}^\gb
=y_\ga\,\p_+{}_\gb \p_-{}^\gb
\,  \ee
and
taking into account the properties (\ref{propF_K}) of   $\FFF_K{}^{n_+\,,n_-}{}(\NNN_\pm,\overline{\NNN}_\pm)$
for any Minkowski current field $J_{fl}$
 we obtain
\bee\nn
  D^{tw}_{fl} \FFF{}_{0}{}^{n_+,n_-}{} \,
J_{fl}\, \Big|_{{y^\pm=\by^\pm=0}}
=
-e^{\gm\pb} \FFF{}_{1}{}^{n_+,n_-}{}
\big(y^+{}_\gm\,\bp_+{}_\pb- y^-{}_\gm\,\bp_-{}_\pb\big)  f{}_-{}_{fl}
J_{fl}\, \Big|_{{y^\pm=\by^\pm=0}}\q
\eee
where $f{}_-{}_{fl}=-\p_+{}_\gga \p_-{}^\gga  $ (\ref{slvflat}).
Analogously,  for any   $AdS$ current field $\PPP$
\bee\label{glu000}
\gl^{-1} D^{tw}  \FFF{}_{0}{}^{n_+,n_-}{} \,
\PPP\, \Big|_{{y^\pm=\by^\pm=0}}
=
-e^{\gm\pb} \FFF{}_{1}{}^{n_+,n_-}{}
\big(y^+{}_\gm\,\bp_+{}_\pb- y^-{}_\gm\,\bp_-{}_\pb\big)\,f_-\, \PPP\, \Big|_{{y^\pm=\by^\pm=0}}\,,
\eee
where $f_-=-\p_+{}_\gga\p_-{}^\gga+\by^+{}^\pga \by^-{}_\pga\,$ (\ref{slv}).

 Therefore the equation
 \bee\label{newtwC+2}
D^{tw} C(y,\by|x) +  e^{ \gm  }{}^{\pb}
\FFF_1{}^{n_+\,,n_-}{ }\big(y^+{}_\gm\bar{\p}_+{}_\pb-y^-{}_\gm\bar{\p}_-{}_\pb\big)
  \,f_-\, \PPP\, \Big|_{{y^\pm=\by^\pm=0}}=0\,
\eee
 follows from a local field redefinition of the  twisted
 equation
\be\label{Dtwexactdef}
D^{tw} (C(y,\by|x)-\gl^{-1} \FFF{}_{0}{}^{n_+,n_-}{} \, J (y^\pm \,,\bar{y}^\pm|x )
) \Big|_{{y^\pm=\by^\pm=0}}
 =0\,.
 \ee
 The same is true in the flat limit.
Complex conjugate formulae are analogous.

 \subsection*{Trivial gluings in the one-form sector}

Denoting the deformation term
in the \rhs of (\ref{ps_deformation}) $\Delta_{s,l}(   J_{(l)})$
let us show that the following deformation
\bee
\label{di}
 D^{ad}\go(y,{\bar{y}}|x) -   \overline{H}^{\pa\pb}
\f{\p^2}{\p \by^{\pa} \p \by^{\pb}}
\overline{C}(0,\by\mid x) -  H^{\ga\gb}
\f{\p^2}{\p {y}^{\ga} \p {y}^{\gb}}
C (y,0\mid x)\,=\qquad\\\nn=
\Delta_{s,l+1}\left(f_+\LLL_{(l )}\right)
- (2s-l -1)\Delta_{s,l }\left( \LLL_{(l )}\right)\q\qquad
\eee
where $\LLL_{(l)}$ with $l\ge2$  satisfies (\ref{eigenl-s}),
is trivial.
Consider  \bee\nn \Omega=
  \gl^{-1} e^{\ga\pb}\p{}_-{}_\ga  \bp{}_-{}_\pb
  \sum_{k=-1}^{l-2}  \f{ \big(  \NNN_- \big)^{l-k-2}   \,
  \big(\overline{\NNN}_-\big)^{ 2s-l+k }}{ (2s-l +k)! (l)!}  
     \left(   f_-\right)^{k+1}\LLL_{(l )}\,.
\eee
Straightforwardly
one can show that
\bee\label{raznica}
D^{ad}\Omega-
\Delta_{s,l+1}   \left(f_+\LLL_{(l )}\right)=
- (2s-l -1)\Delta_{s,l }\left(\LLL_{(l )}\right)
+\\ \nn+
\overline{H}^{\pa \pb}\bp{}_-{}_\pa  \bp{}_-{}_\pb
   \f{     \,
  \big(\overline{\NNN}_-\big)^{ 2s-2 }}{  (2s-2)! (l)!}     \left(f_-\right)^{l}
 \LLL_{(l)}\big|_{y^\pm=\by^\pm=0}\,.\qquad 
\eee
Proceeding as in appendix B, one can see that the respective deformation in the zero-form sector is proportional to
\be\label{trivdef0}
\overline{\FFF}{}_{0}{}^{ 0,2s}f_+ \left(f_-\right)^{l}
 \KKK_{(l)}{}\big|_{y^\pm=\by^\pm=0}\,.
\ee
Since $l>1$,   (\ref{trivdef0})  can be  rewritten  as
\be\label{trivdef00}
\overline{\FFF}{}_{0}{}^{ 0,2s}f_-\,
 \widetilde{\KKK_{(l)}}{}\big|_{y^\pm=\by^\pm=0}\,
\ee
for some current field $\widetilde{\KKK_{(l)}}$. By virtue of Eq.~(\ref{glu000}) this
 implies that the zero-form deformation (\ref{trivdef0}) is trivial
 resulting  from a local field redefinition.
It remains to observe that the one-form deformation is
indeed  trivial by virtue of  (\ref{raznica}).

\addtocounter{section}{1}
\addcontentsline{toc}{section}{\,\,\,\,\,\,\,References}

\section*{$\rule{0pt}{1pt}$}


\begin{thebibliography}{99}

\baselineskip=18pt
\parindent=0pt
\parskip=-4.5pt
     \bibitem{Maldacena:1997re}
J.~M.~Maldacena,
  Adv.\ Theor.\ Math.\ Phys.\  {\bf 2} (1998) 231
  [Int.\ J.\ Theor.\ Phys.\  {\bf 38} (1999) 1113]
  [arXiv:hep-th/9711200].
\bibitem{Gubser:1998bc}
S.~S.~Gubser, I.~R.~Klebanov and A.~M.~Polyakov,
  Phys.\ Lett.\  B {\bf 428}, 105 (1998)
  [arXiv:hep-th/9802109].
\bibitem{Witten:1998qj}
E.~Witten,
  Adv.\ Theor.\ Math.\ Phys.\  {\bf 2}, 253 (1998)
  [arXiv:hep-th/9802150].

\bibitem{more} M.A.Vasiliev, {\it Phys. Lett.}  {\bf B285} (1992) 225.

\bibitem{non}
M.A.Vasiliev,
{\it Phys.\ Lett.\ } {\bf B567} (2003) 139, [{\tt hep-th/0304049}].

\bibitem{solv}
 X.~Bekaert, S.~Cnockaert, C.~Iazeolla and M.~A.~Vasiliev, {\it ``Nonlinear
higher spin theories in various dimensions}, arXiv:hep-th/0503128.



\bibitem{tens2} O.~A.~Gelfond and M.~A.~Vasiliev,
    {\it Theor.Math.Phys.} {\bf 145} N1 (2005) 35, {\tt hep-th/0304020}.



\bibitem{F}C.Fronsdal, \emph{``Massless Particles, Ortosymplectic
Symmetry and Another Type of Kaluza-Klein Theory"}, Preprint
UCLA/85/TEP/10, in Essays on Supersymmetry, Reidel, 1986
(Mathematical Physics Studies, v.8).

\bibitem{Bandos:1999qf}
  I.~A.~Bandos, J.~Lukierski and D.~P.~Sorokin,
  Phys.\ Rev.\  D {\bf 61} (2000) 045002
  [arXiv:hep-th/9904109].




 \bibitem{BHS}
M.~A.~Vasiliev, \textit{Phys.Rev.} {\bf D66} (2002) 066006, {\tt
[hep-th/0106149]}.

      \bibitem{Vasiliev:2001dc}
        M.~A.~Vasiliev,
        arXiv:hep-th/0111119.


\bibitem{Vasiliev:1990bu}
  M.~A.~Vasiliev,
  Phys.\ Lett.\  B {\bf 257} (1991) 111.

\bibitem{gelcur} O.~A.~Gelfond and M.~A.~Vasiliev,
 {\it JHEP} {\bf 03} (2009) 125;
     [arXiv:0801.2191v4 [hep-th]].


\bibitem{GSV} O.~A.~Gelfond, E.~D.~Skvortsov and M.~A.~Vasiliev,  Theor.Math.Phys. 154 (2008) 294-302,
 {\tt  hep-th/0601106}.


\bibitem{Berends:1985xx}
  F.~A.~Berends, G.~J.~H.~Burgers and H.~van Dam,
  Nucl.\ Phys.\  B {\bf 271} (1986) 429.

\bibitem{Bandos:2005mb}
  I.~Bandos, X.~Bekaert, J.~A.~de Azcarraga, D.~Sorokin and M.~Tsulaia,
  JHEP {\bf 0505} (2005) 031
  [arXiv:hep-th/0501113].



\bibitem{Siegel} W.~Siegel, {\it Int.\ J.\ Mod.\ Phys.} {\bf A4} (1989)
2015.

\bibitem{Mets} R.~R.~Metsaev, {\it Mod.\ Phys.\ Lett.\ } {\bf A10} (1995) 1719.

\bibitem{Das:2003vw}
  S.~R.~Das and A.~Jevicki,
  Phys.\ Rev.\  D {\bf 68} (2003) 044011
  [arXiv:hep-th/0304093].

\bibitem{Koch:2010cy}
  R.~d.~M.~Koch, A.~Jevicki, K.~Jin and J.~P.~Rodrigues, Phys.Rev. D83 (2011) 025006,
  arXiv:1008.0633 [hep-th].

%
\bibitem{4dun} M.A.Vasiliev,
{\it Phys.\ Lett.}  {\bf B209} (1988) 491.


\bibitem{Ann}   M.A.Vasiliev, {\it Ann. Phys.} (N.Y.) {\bf 190} (1989) 59.


\bibitem{frame}
 M.A.~Vasiliev, {\it Sov. J. Nucl. Phys.\/} {\bf 32} (1980) 855
               $(p.\,439$ in English translation).



\bibitem{Fort1}M.A.~Vasiliev, {\it Fortschr. Phys.\/} {\bf 35} (1987) 741.

\bibitem{gol}
  M.~A.~Vasiliev,
  arXiv:hep-th/9910096.
\bibitem{33}
M.A.~Vasiliev, {\it Nucl.Phys.} {\bf B793} (2008) 469, {\tt
arXiv:0707.1085 [hep-th]}.



\bibitem{fronsdal_flat}
C.~Fronsdal, Phys.\ Rev.\ D {\bf 18} (1978) 3624.

\bibitem{FF}
 J.~Fang and C.~Fronsdal, Phys.\ Rev.\ D {\bf 18} (1978) 3630.

\bibitem{FV1}
  E.~S.~Fradkin and M.~A.~Vasiliev,
  Phys.\ Lett.\  B {\bf 189} (1987) 89.



\bibitem{Didenko:2003aa}
  V.~E.~Didenko and M.~A.~Vasiliev,
  J.\ Math.\ Phys.\  {\bf 45} (2004) 197
  [arXiv:hep-th/0301054].

\bibitem{fut}
  O.~A.~Gelfond and M.~A.~Vasiliev, arXiv:1412.7147 [hep-th].


\bibitem{bessel} H.~Beitman and A.~Erdelyi, ``Higher transcendental
functions'', New-York Toronto London MC Graw-Hill Book Company, inc. (1953).




\bibitem{Bekaert:2010hk}
  X.~Bekaert and E.~Meunier,
  JHEP {\bf 1011} (2010) 116
  [arXiv:1007.4384 [hep-th]].



\bibitem{Manvelyan:2004mb}
  R.~Manvelyan and W.~Ruhl,
  Phys.\ Lett.\  B {\bf 593} (2004) 253
  [arXiv:hep-th/0403241].



\bibitem{Fotopoulos:2007yq}
  A.~Fotopoulos, N.~Irges, A.~C.~Petkou and M.~Tsulaia,
  JHEP {\bf 0710} (2007) 021
  [arXiv:0708.1399 [hep-th]].

\bibitem{Manvelyan:2009tf}
  R.~Manvelyan and K.~Mkrtchyan,
  Mod.\ Phys.\ Lett.\  A {\bf 25} (2010) 1333
  [arXiv:0903.0058 [hep-th]].


\bibitem{Fotopoulos:2009iw}
  A.~Fotopoulos and M.~Tsulaia,
  JHEP {\bf 0910} (2009) 050
  [arXiv:0907.4061 [hep-th]].

\bibitem{Prokushkin:1999xq}
  S.~F.~Prokushkin and M.~A.~Vasiliev,
  Theor.\ Math.\ Phys.\  {\bf 123} (2000) 415
  [Teor.\ Mat.\ Fiz.\  {\bf 123} (2000) 3]
  [arXiv:hep-th/9907020].

\bibitem{Prokushkin:1999ke}
  S.~F.~Prokushkin and M.~A.~Vasiliev,
  Phys.\ Lett.\  B {\bf 464} (1999) 53
  [arXiv:hep-th/9906149].

\bibitem{Deser:2004rr}
  S.~Deser and A.~Waldron,
  arXiv:hep-th/0403059.

\bibitem{Aragone:1979hx}
  C.~Aragone and S.~Deser,
  Phys.\ Lett.\  B {\bf 86} (1979) 161.

\bibitem{Ritus2}
V.~I.~Ritus, J. Exp. Teor. Phys. {\bf 124} 1(7) (2003) p.14-27 [arXiv:hep-th/0309181];
V.~I.~Ritus, J. Exp. Teor. Phys. {\bf 129} 4 (2006) p.664-683 [arXiv:hep-th/0509209].

\bibitem{Sundborg:2000wp}
  B.~Sundborg,
  Nucl.\ Phys.\ Proc.\ Suppl.\  {\bf 102} (2001) 113
  [arXiv:hep-th/0103247].

\bibitem{Witten}
E. Witten, talk at the John Schwarz 60-th birthday symposium,
http://theory.caltech.edu/jhs60/witten/1.html

\bibitem{Sezgin:2002rt}
  E.~Sezgin and P.~Sundell,
  Nucl.\ Phys.\  B {\bf 644} (2002) 303
  [Erratum-ibid.\  B {\bf 660} (2003) 403]
  [arXiv:hep-th/0205131].


\bibitem{Klebanov:2002ja}
  I.~R.~Klebanov and A.~M.~Polyakov,
  Phys.\ Lett.\  B {\bf 550} (2002) 213
  [arXiv:hep-th/0210114].


\bibitem{Giombi:2009wh}
  S.~Giombi and X.~Yin,
  JHEP {\bf 1009} (2010) 115
  [arXiv:0912.3462 [hep-th]]; arXiv:1004.3736 [hep-th].

\bibitem{Henneaux:2010xg}
  M.~Henneaux and S.~J.~Rey,
  JHEP {\bf 1012} (2010) 007
  [arXiv:1008.4579 [hep-th]].

\bibitem{Campoleoni:2010zq}
  A.~Campoleoni, S.~Fredenhagen, S.~Pfenninger and S.~Theisen,
  JHEP {\bf 1011} (2010) 007
  [arXiv:1008.4744 [hep-th]].

\bibitem{Gaberdiel:2010ar}
  M.~R.~Gaberdiel, R.~Gopakumar and A.~Saha,
JHEP 1102 (2011) 004,
  arXiv:1009.6087 [hep-th].


\bibitem{Gaberdiel:2010pz}
  M.~R.~Gaberdiel and R.~Gopakumar, Phys.Rev. D83 (2011) 066007,
  arXiv:1011.2986 [hep-th].





\end{thebibliography}
\end{document}